%% file: continG191_OJAwithSM.tex
\documentclass[twocolumn,iop,numberedappendix,appendixfloats]{openjournal}
\usepackage[utf8]{inputenc}
\usepackage{amsmath,amssymb,amstext,amsfonts}
\usepackage{natbib}
\usepackage{fancyvrb}
\usepackage[colorlinks,allcolors=blue]{hyperref}
\usepackage{dsfont}
\usepackage{xcolor}
\usepackage{fontawesome}
\usepackage{tikz}
\usepackage{longtable}
\usepackage{graphicx}
\graphicspath{ {./images/} }
\usepackage{listings}

\usetikzlibrary{shapes,arrows}
\newcommand{\daa}{\ensuremath \Delta\alpha/\alpha}
\newcommand{\fdaa}{\frac{\Delta\alpha}{\alpha_0}}

\newcommand{\fev}{{Fe\sc\,v}}
\newcommand{\niv}{{Ni\sc\,v}}

\begin{document}
\journalinfo{The Open Journal of Astrophysics}
\submitted{Submitted 1 October 2025; accepted 5 April 2025}

\shorttitle{Searching for new physics using high precision absorption spectroscopy}
\shortauthors{Lee et al.}
\title{Searching for new physics using high precision absorption spectroscopy; continuum placement uncertainties and the fine structure constant in strong gravity}

\author{Chung-Chi Lee$^*$}
\affiliation{Big Questions Institute, Level 4, 55 Holt St., Surry Hills, Sydney, NSW 2010, Australia}
\email{$^*$lee.chungchi16@gmail.com}

\author{John K. Webb$^\dagger$}
\affiliation{Institute of Astronomy, University of Cambridge, Madingley Road, Cambridge, CB3 0HA, UK}
\affiliation{Clare Hall, University of Cambridge, Herschel Rd, Cambridge, CB3 9AL, UK}
\affiliation{Big Questions Institute, Level 4, 55 Holt St., Surry Hills, Sydney, NSW 2010, Australia}
\email{$^\dagger$jw978@cam.ac.uk}

\author{Darren Dougan}
\affiliation{Big Questions Institute, Level 4, 55 Holt St., Surry Hills, Sydney, NSW 2010, Australia}

\author{Vladimir A. Dzuba, Victor V. Flambaum}
\affiliation{School of Physics, University of New South Wales, Sydney, NSW 2052, Australia}

\author{Dinko Milakovi{\'c}}
\affiliation{Institute for Fundamental Physics of the Universe, Via Beirut, 2, 34151 Trieste, Italy}
\affiliation{INAF - Osservatorio Astronomico di Trieste, via Tiepolo 11, 34131, Trieste, Italy\\}

\usetikzlibrary{arrows.meta,chains,positioning,quotes,shapes.geometric}

\tikzset{FlowChart2/.style={
startstop/.style = {rectangle, rounded corners, draw, fill=yellow!10,
                    minimum width=3cm, minimum height=1cm, align=center, thick,
                    on chain, join=by arrow},
  process/.style = {rectangle, rounded corners, draw, fill=green!10,
                    text width=4cm, minimum height=1cm, align=center, thick,
                    on chain, join=by arrow},
 decision/.style = {diamond, aspect=1.5, draw, fill=orange!10,
                    minimum width=3cm, minimum height=1cm, align=center, thick,
                    on chain, join=by arrow},
       io/.style = {trapezium, trapezium stretches body, 
                    trapezium left angle=70, trapezium right angle=110, thick,
                    draw, fill=blue!30,
                    minimum width=3cm, minimum height=1cm,
                    text width =\pgfkeysvalueof{/pgf/minimum width}-2*\pgfkeysvalueof{/pgf/inner xsep},
                    align=center,
                    on chain, join=by arrow},
    arrow/.style = {thick,-latex'}
                        }
        }

\begin{abstract}
Searches for variations of fundamental constants require a comprehensive understanding of measurement errors. There are several potential sources of error and quantifying each one accurately is essential. This paper examines a source of error that is usually overlooked: the impact of continuum placement error. We investigate the problem using a high resolution, high signal to noise spectrum of the white dwarf G191$-$B2B. Narrow photospheric absorption lines allow us to search for new physics in the presence of a gravitational field approximately $10^4$ times that on Earth. Modelling photospheric lines requires knowing the underlying spectral continuum level. We describe the development of a fully automated, objective, and reproducible continuum estimation method. Measurements of the fine structure constant are produced using several continuum models. The results show that continuum placement variations result in small systematic shifts in the centroids of narrow photospheric absorption lines which impact significantly on fine structure constant measurements. This effect should therefore be included in the error budgets of future measurements. Our results suggest that continuum placement variations should be investigated in other contexts, including fine structure constant measurements in stars other than white dwarfs. The analysis presented here is based on {\niv} absorption lines in the photosphere of G191$-$B2B. Curiously, the inferred measurement of the fine structure constant obtained in this paper using {\niv} (the least negative of our measurements is $\daa = -1.462 \pm 1.121 \times 10^{-5}$) is inconsistent with the most recent previous G191$-$B2B photospheric measurement using {\fev} ($\daa = 6.36 \pm 0.35_{stat} \pm 1.84_{sys} \times 10^{-5}$). Given both measurements are derived from the same spectrum, we presume (but in this work are unable to check) that this 3.2$\sigma$ difference results from unknown laboratory wavelength systematics.
\end{abstract}

\maketitle

\section{Introduction} \label{sec:Intro}

This is the first paper in a companion series of two. The second paper applies the methods developed here to a high resolution quasar spectrum. 

Light scalar fields naturally arise in modern cosmological models and high-energy physics theories, influencing fundamental parameters of the Standard Model, such as coupling constants and mass ratios. Similar to gravitational charge, scalar charge is purely additive, meaning the effects of a scalar field can vary near massive objects like white dwarfs. For non-relativistic objects, both total mass and total scalar charge scale proportionally with the number of nucleons. However, different forms of coupling between the scalar field and other fields can enhance or suppress scalar interactions near massive gravitating bodies \citep{Magueijo2002}. For small variations, the scalar field at a distance \(r\) from an object of mass \(M\) is proportional to the change in the dimensionless gravitational potential \(\phi = GM/rc^2\). This proportionality is captured by the sensitivity parameter \(k_{\alpha}\) \citep{Flambaum2008}. Specifically, for variations in the fine-structure constant \(\alpha\), we define
\begin{equation}
\frac{\Delta \alpha}{\alpha} \equiv \frac{\alpha(r) - \alpha_0}{\alpha} = k_{\alpha} \Delta \phi = k_{\alpha} \frac{GM}{rc^2}.
\end{equation}
This dependence is explicitly demonstrated in theories of varying \(\alpha\), such as those proposed by \cite{Bekenstein1982, Sandvik2002, Barrow2002b}, as well as their generalizations e.g. \cite{Barrow2012}. Depending on the balance between electrostatic and magnetic energy in the surrounding matter fields, \(\alpha\) can either increase (\(\Delta \alpha/\alpha > 0\)) or decrease (\(\Delta \alpha/\alpha < 0\)) near a massive object \citep{Magueijo2002}.

Precise limits on \(k_{\alpha}\) come from comparisons of two Earth-based atomic clocks over the course of a year \citep{Flambaum2008, Bauch2002, Ferrell2007, Fortier2007, Blatt2008, Barrow2008, Guena2012, Leefer2013, Dzuba2017}. These constraints arise from the ellipticity of Earth's orbit, which induces a 3\% seasonal variation in the Sun's gravitational potential at Earth. The peak-to-trough sinusoidal change in potential has a magnitude of $\Delta \phi = 3 \times 10^{-10}$. Since different clocks exhibit varying sensitivities to changes in \(\alpha\), measurements of \(\Delta \alpha/\alpha\) allow for the extraction of \(k_{\alpha}\).

In searches for spacetime variations of fundamental constants, it is essential to understand all possible sources of uncertainty, irrespective of a positive detection or upper limit. For absorption line measurements, one source of uncertainty that is often ignored in this context is that associated with imperfections in the continuum level against which absorption takes place. Existing continuum fitting methods, at least those we looked at, do not have the level of automation, reproducibility, and accuracy required for high precision $\daa$ measurements. Therefore, in this paper, we present a new, fully automated, spectral continuum fitting algorithm. This iterative method identifies and removes regions containing spectral features and then fits cubic splines to the remaining data points. The method is applied to a white dwarf spectrum obtained using the Space Telescope Imaging Spectrograph (STIS) on the Hubble Space Telescope (Section \ref{sec:G191}). By varying the continuum algorithm fitting parameters, i.e. generating different continuum models, we explore how variations in the adopted continuum model impact on measurements of the fine structure constant $\daa=(\alpha_z - \alpha_0)/\alpha_0$, where the subscripts $z, 0$ indicate redshift (resulting from gravitational and peculiar velocity terms) and the terrestrial value, and where the fine structure constant $\alpha = e^2/4\pi\epsilon_0\hbar c$ in SI units.

The analysis in this paper applies specifically to measurements along the lines of \cite{Berengut2013}, in which the initial continuum that is fitted to the spectrum to be analysed is never subsequently modified i.e. any potential $\daa$ systematic associated with continuum placement uncertainty is not explicitly taken into account. Many previous quasar absorption system $\daa$ measurements have also ignored this source of uncertainty. As we discuss shortly, $\daa$ measurements in which profile modelling is carried using codes like {\sc vpfit} and {\sc ai-vpfit} can allow the introduction of additional parameters for fine-tuning the local continuum for each spectral segment fitted, as done in, for example, \cite{Hu2021}. In that kind of analysis, the impact of initial continuum placement errors may be much smaller than the effects we report in the present paper.

\section{Line shift vs. continuum slope variation and potential impact on \texorpdfstring{$\daa$}{daa}} \label{sec:shifts}

Discussions on different methods, in various contexts, for spectral line centroiding are given in \cite{Garnir1987, Guo2011, Ipsen2017, Kandel2017, Teague2018, Rodgers2021}. Consider a single absorption line measured relative to a model continuum that is tilted with respect to the true underlying continuum. The measured centroid will be slightly different to that derived using the true continuum, potentially emulating $\alpha$ variation.  Here we develop a simple model to illustrate this effect.

The measured rest-frame frequency $\omega_z$ of an atomic transition at redshift $z$ is related to a change in the fine structure constant by
\begin{equation}
        \omega_z = \omega_0 + q\left( \frac{\alpha_z^2}{\alpha_0^2} -1 \right) \label{eq:wqda}
\end{equation}
where $\omega_0$ is the terrestrial value and $q$ is the sensitivity coefficient, \cite{Dzuba1999b, Webb1999}. Eq.\,\eqref{eq:wqda} can be used to obtain
\begin{equation}
    \Delta\lambda \approx -\frac{2q\lambda_0^2}{10^8} \frac{\Delta\alpha}{\alpha} \label{eq:lamshift},
\end{equation}
where $\lambda_0$ is in\,{\AA}. Taking a representative transition with $|q_{Ni\,v}| \approx 2000$\,cm$^{-1}$ and a rest-frame wavelength $\lambda_0 = 1500$\,{\AA}, Eq.~\eqref{eq:lamshift} gives
\begin{equation}
    \left|\Delta\lambda\right| = 90 \left| \frac{\Delta\alpha}{\alpha} \right|, \label{eq:dlamdalph}
\end{equation}
from which we see that for a $\daa$ measurement uncertainty contribution of $\sim$$10^{-6}$, any wavelength shift $\gtrsim$$10^{-4}$\,{\AA} is of concern. 
To see if this is possible, approximate optical depth using a Gaussian model,
\begin{equation}
    \tau_x = \tau_0 \exp\left(-\frac{x^2 \ln 2}{g^2} \right)
\end{equation}
where $x = \lambda-\lambda_0$, $\lambda_0$ is the central wavelength, $\tau_0$ is the optical depth at $x=0$, $g = b \lambda_0/c$, $b = \sqrt{kT/m}$, $k$ is Boltzmann's constant, $T$ is temperature, $m$ is atomic mass, and $c$ is the speed of light in vacuum. 

The profile intensity is $I_x = I_0 \exp\left(-\tau_x\right)$, where $I_0$ is the true continuum. Now introduce a small change on the true continuum, emulating a continuum-fitting uncertainty, parameterised using a multiplicative term, such that the profile intensity becomes
\begin{equation}
    I^{\prime}_x = I_x \left(c_l + c_s \left(\frac{\lambda}{\lambda_0} -1 \right) \right), \label{eq:gausstilt}
\end{equation}
where the parameters $c_l$ and $c_s$ allow for a level and slope change. This method for continuum adjustment was introduced in {\sc vpfit} \citep{web:VPFIT}. 
Then, using
\begin{equation}
    e^{-\tau_x} = \sum_{n=0}^{\infty} \left(-\frac{\tau_x^n}{n !} \right),
\end{equation}
and taking only only the first two terms in the expansion such that we are assuming the optically thin case (sufficient for our illustrative purposes here),
\begin{equation}
    I_x \simeq I_0 (1 - \tau_x) \,.  \label{eq:gaussapproach}
\end{equation}
The line centroids using the true continuum, and the continuum after adjustment according to Eq.\,\eqref{eq:gausstilt}, are given by
\begin{equation}
    C = \frac{ \int^{\infty}_{-\infty} A_\lambda \lambda d \lambda}{\int^{\infty}_{-\infty} A_\lambda d \lambda} \,, \quad
    \mathrm{and}
    \quad \Tilde{C} = \frac{ \int^{\infty}_{-\infty} \Tilde{A}_\lambda \lambda d \lambda}{\int^{\infty}_{-\infty} \Tilde{A}_\lambda d \lambda} 
\end{equation}
where $A_x = I_0 - I_x$ and where
\begin{equation}
    \Tilde{A}_\lambda = \frac{A_\lambda}{c_l + c_s x/\lambda_0} \simeq \frac{A_\lambda}{c_l} \left( 1- \frac{c_s x}{c_l \lambda_0} \right) \,.
\end{equation}
From the equations above, one can derive $C = \lambda_0$ and
\begin{align}
    \Tilde{C} &= \lambda_0 + \frac{\int^{\infty}_{-\infty} A_\lambda \left( 1 - \frac{c_s x}{c_l \lambda_0} \right) x dx}{\int^{\infty}_{-\infty} A_\lambda \left( 1 - \frac{c_s x}{c_l \lambda_0} \right) dx} \nonumber \\ &\simeq \lambda_0 - \frac{g^2}{2 \ln 2} \frac{c_s }{c_l  \lambda_0 } \,,
\end{align}
where we have put $A_\lambda \simeq \tau_\lambda$. Then, the wavelength shift due to the continuum variation is
\begin{equation}
    \Delta \lambda = \Tilde{C} - \lambda_0 = - \frac{b^2 \lambda_0}{(2 \ln 2)c^2} \frac{c_s}{c_l} \,.
    \label{eq:dlambda}
\end{equation}
If we now consider 2 transitions, with sensitivity coefficients $q_1$ and $q_2$, Eqs.~\eqref{eq:lamshift} and \eqref{eq:dlambda} can be used to obtain 
\begin{equation}
    \frac{\Delta\alpha}{\alpha} = \frac{10^8}{4 \ln 2 } \frac{b^2}{c^2 \lambda_0} \frac{1}{(q_1 - q_2)} \frac{c_s}{c_l} \approx \frac{0.04 c_s}{\lambda_0 (q_1 - q_2)}. \label{eq:davscs}
\end{equation}
where in the last term we have taken an illustrative $b=10$ km\,s$^{-1}$ and used $c_l \approx 1$ (adjustments to $c_l$ should not impact on the line position measurement). Recalling that $c_s$ describes a local continuum tilt with respect to the true value, Eq.\,\eqref{eq:gausstilt} can be used (by taking representative values for $\lambda/\lambda_0$) to show that a reasonable expected range for the slope parameter is $0 \lesssim |c_s| \lesssim 100$. 

\begin{figure}
\centering
\includegraphics[width=0.8\linewidth]{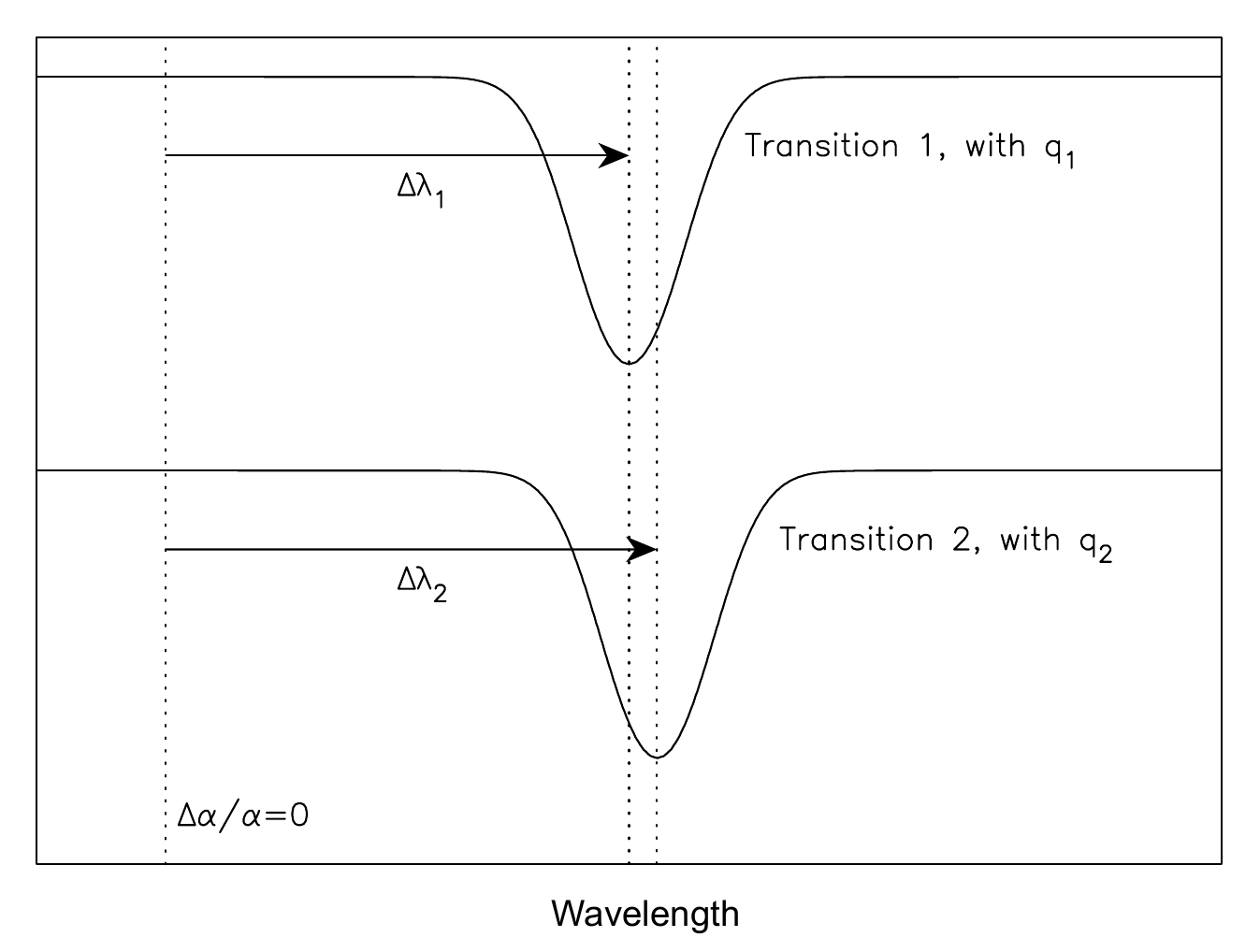}
\caption{Illustration of 2 transitions used to measure $\daa$, where $\Delta q \ll |q|$. Both transitions shift far from their $\daa=0$ positions, but since $\Delta q$ is small, a pair of lines with this property have (by themselves) little sensitivity to a change in $\alpha$. In this case, the denominator in Eq.\,\eqref{eq:davscs} is small and the measured value of $\daa$ is poorly constrained.}
\label{fig:shiftcompare}
\end{figure}

For a pair of transitions with $q_1 \approx q_2$, as illustrated in Fig.\,\ref{fig:shiftcompare}, constraints on $\daa$ are poor. However, taking the upper limit of $c_s \sim 100$, we can use Eq.\,\eqref{eq:davscs} to approximate the impact on the measured $\daa$ associated with a continuum slope uncertainty, of $\sim$$3 \times 10^{-6}$ (representative of best-case $\daa$ uncertainties to date, ignoring continuum slope errors). The simple analytic analysis above is only illustrative, because a real measurement involves the simultaneous use of a large number of transitions. Nevertheless, the above suggests that continuum uncertainty may contribute significantly to the overall $\daa$ error budget, for this kind of linear regression analysis, motivating the more detailed numerical modelling described in the following Sections.

\section{Continuum fitting procedure}\label{sec:method}

Several continuum fitting algorithms have been reported previously in the literature, mostly for quasar spectroscopy, although none are suitable for our requirement here of fitting the HST/STIS G191$-$B2B continuum. The particular problem we face is that the G191$-$B2B HST data exhibit many small-scale undulations against which absorption lines of interest must be measured, as can be seen in e.g. the Appendix spectral plots.

We did not entertain the possibility of using a stellar synthesis code as a means to estimate an underlying continuum because: (i) whilst some undulations are presumably due to blends of weak absorption lines, others (possibly the majority) are caused by imperfect spectral order flattening prior to combination into a co-added one-dimensional spectrum. In any case, incomplete input atomic data may mean that a stellar synthesis code may not identify all weak blends. Comparisons (and discrepancies) between synthetic stellar spectra and observations are given in e.g. \cite{Coelho2014, Coelho2020}; (ii) our aim was to produce a generalised continuum fitting method, useful for various kinds of astronomical spectra.

Examples of previous continuum fitting procedures include: \cite{Eilers2017} apply a Bayesian method to estimate continuum levels for Lyman alpha forest data. \cite{Sanchez2018} describe Python software {\sc statcont}. \cite{Ciollaro2014} develop methods applied to Hubble Space Telescope Faint Object Spectrograph and Baryon Oscillation Spectroscopic Survey spectra. \cite{Lee2012,Davies2018} discuss principal component analysis methods to estimate continua in the Lyman-$\alpha$ forest, for SDSS data i.e. at lower spectral resolution. \cite{Paris2011} use a PCA method to estimate quasar continua over the full wavelength range. \cite{Suzuki2005} also uses principal component analysis methods to estimate continua in high resolution Lyman-$\alpha$ forest data. \cite{Aghaee2009} compare various Lyman-$\alpha$ forest continuum fitting approaches. Many other methods exist, but we are not aware of any that could provide the level of detail needed for absorption complex modelling in high resolution spectra (and do so in an objective, automated, and reproducible way). Given the lack of any existing suitable method (for our application), we therefore developed the new continuum modelling procedure described next.

The first step towards obtaining the final estimated continuum is to make a preliminary model using cubic splines. To do this we first identify significant absorption and emission profiles in order to be able to fit cubic splines to feature-free spectral segments. The method for identifying features is discussed shortly. One problem to avoid is allowing the continuum model to follow an absorption profile, thereby effectively removing it. For example, consider a shallow absorption profile, spread across $\sim$2\,{\AA} such that it is reasonably well fitted by a cubic spline with a knot interval of 0.5\,{\AA}, such that the absorption line can be accidentally removed. This problem can be avoided by flagging the data points within the absorption line prior to fitting the cubic spine model. Doing this interactively prior to the continuum fitting stage would be both time consuming and subjective. Therefore, we instead iteratively increase the spline order: the initial model is a single cubic spline, after which the number of knots is doubled at each subsequent iteration. Figure \ref{fig:flowchart} shows two loops: loop A (iterating over stages 2-5) and loop B (iterating over stages 2-6). Each loop has its own specific stopping criterion. These are discussed shortly (Sections \ref{sec:stg5} and \ref{sec:stg6}).

\subsection{Stage 1: Re-bin and smooth spectrum}\label{sec:stage1}

The optical depth associated with absorption features in the spectrum is given by
\begin{equation}
\tau_{tot}(\lambda) = \sum_i \tau_{abs}^i(\lambda)
\end{equation}
where $\tau_{abs}^i$ is the optical depth associated with an individual $i^{th}$ absorption line and where the sum is taken over all absorption (or emission) lines present in the model. The observed spectrum is the sum of three terms,
\begin{equation}
\label{eq:I_tot}
    I(\lambda) = I_{0}(\lambda) e^{-\tau_{tot}(\lambda)} + I_{em}(\lambda) + I_{n}(\lambda) \,,
\end{equation}
where $I_{0}$, $I_{em}$, and $I_{n}$ are the underlying continuum from the object being observed, the emission line spectrum, and noise, respectively. The noise term has several contributions, e.g. photon counting from the object, photon counting from the sky, assuming the spectrum is sky-subtracted, detector read-out noise, detector dark current, cosmic rays, and any detector defects that have not been fully removed in the data pre-processing. 

In Stage 1 we smooth the data by convolving the spectrum with a Gaussian function. We first re-bin the spectrum onto a finer grid (with pixel size 1/10 the original value). The re-binned spectrum is then smoothed using a Gaussian. The FWHM of the smoothing Gaussian is taken to be a multiple of the mean original pixel size $\bar{x}$ (over the entire spectrum). The default pixel space is in\,{\AA} and in practice a value of 3$\bar{x}$ for the FWHM was found to work well, but this is a user-defined quantity. The observed pixel size is approximately constant in velocity units, so convolution is done in velocity space. This initial smoothing is helpful in the next Stage where we calculate the derivative of the spectrum. Stage 1 is only carried out once - it is not included in subsequent loops (as illustrated in Fig.\,\ref{fig:flowchart}).

\subsection{Stage 2: Preliminary feature identification}\label{sec:stage2}

In order to determine whether a pixel in the spectrum falls within an absorption (or emission) feature, or whether it can be considered as a continuum pixel, two properties are examined: the closeness of the pixel intensity to the current continuum estimate and the derivative of the smoothed spectrum. The derivative spectrum is obtained using the re-binned smoothed spectrum from Stage 1. 

The first step is to make a preliminary identification of all absorption (or emission) features in the data. To do this we make simple use of the rebinned, smoothed derivative spectrum from Stage 1; the largest $x$\% of the data points in that spectrum (across the entire spectrum) are identified (using a simple array sort) and excluded . In practice we found $x=30$\% works well. This process of course identifies line edges where the derivative is high (but also, necessarily) line centres are {\it not} selected.

Having selected the points as described above, an initial straight line continuum is fitted. This initial continuum will of course be only a very crude representation of the true continuum. Then, now having selected low-slope data points (the remaining $100-x$\%) and with a preliminary (lowest order) continuum fit, we calculate the normalised chi-squared, $\chi^2_\nu$ over the selected data points, 
\begin{align}
&\chi^2_\nu \approx \frac{\chi^2}{M_s} \label{eq:chisqn} \\
&\chi^2 = \sum_{i=1}^{M_s} \left( \frac{I_i - C_i}{\sigma_i}\,\right)^2 \,, \label{eq:chisq}
\end{align}
where $I_i$ is the intensity of the $i^{th}$ pixel in the original spectrum, with uncertainty $\sigma_i$, $C_i$ is the current estimated continuum value at the $i^{th}$ pixel, and $M_s$ is the total number of selected data points. Prior to refinement, with only a crude continuum fit, the initial value of $\chi_\nu^2 \gg 1$. To reject data points at line centres we search for pixel values having $(I_i - C_i)^2/\sigma_i^2 > \zeta^2 \chi^2_\nu$, using a default value of $\zeta=3$ to correspond approximately to $3\sigma$ deviations.

\subsection{Stage 3: Adjust number of knots}\label{sec:stage3}

At first pass through Stage 3 (when $K_i=0$), the continuum model is very simple (initially only a straight line) so Stage 3 applies only to higher iterations. During the development of a continuum model, the number of cubic splines increases. It is possible that at certain positions along the spectrum, only a few continuum data points are available to fit. If this happens, that segment of the continuum will of course be poorly determined. Therefore, Stage 3 checks that a sufficiently large number of data points is present between each knot pair. The minimum acceptable number of continuum data points between each knot pair is a user-defined parameter, having a default value of $k_{merge}=1/3$, this parameter being the fraction of pixels available within the range defined by the knot pair. When there are too few continuum data points left between any particular knot pair, the two flanking regions are checked and the test region merged into whichever side region has the least continuum data points (such that the knot spacing is no longer necessarily regular).

\subsection{Stage 4: Solve for search direction}

By the end of Stage 3, the model continuum can comprise a large number of free parameters. The goal is to minimise $\chi^2$ between the cubic spline model and the set of continuum data points. We do this using a standard non-linear least-squares procedure. The components of the gradient vector $g(a,b)$ and of the Hessian matrix $H_{ab}$ are
\begin{align}
&g_a = \frac{\partial \chi^2}{\partial y_a} \,, \\
&H_{ab} = \frac{\partial^2 \chi^2}{  \partial y_a \partial y_b} \,,
\end{align}
where $y_a$, $y_b$ are the knot values (i.e. intensities) at positions $a$ and $b$. Then we apply a standard Gauss-Newton minimisation procedure,
\begin{equation}
    H_{ab} p_b = - g_a \,,
\end{equation}
where $p_b$ is the search direction, providing the best-fit set of knot intensities. Detailed descriptions of optimisation methods, including this one, can be found in many books, an excellent example being \cite{GMW81}.

\subsection{Stage 5: Update parameters}\label{sec:stg5}

The model continuum parameters are updated (using univariate minimisation) by finding the scalar $d$ which minimises $\chi^2$. The updated parameters are then
\begin{equation}
    \bar{y}_b = y_b + p_b \times d \,.
\end{equation}
The values of $\chi_\nu^2$ for the current and previous iterations are compared and if they differ by more than a value of $\chi_{thres}^2$, the algorithm returns to an earlier stage (analogous to Stage 2) and re-identifies all absorption and emission features using the newly updated continuum model. This is illustrated in Figure \ref{fig:flowchart} by the ``Refining'' box. If the difference is smaller than $\chi_{thres}^2$, the algorithm moves to Stage 6. The default value of $\chi_{thres}^2$ is $10^{-3}$ (although this can be user-defined). Final results are rather insensitive to this parameter.

\subsection{Stage 6: Increase order of fit}\label{sec:stg6}

At this stage, the number of cubic spline knots is $K_i$. $K_f$ is the full data range divided by the smallest permitted knot spacing and is a user input, specified to terminate the algorithm. If $K_i < K_f$, the algorithm thus iterates further as indicated in Figure \ref{fig:flowchart}. We choose the form for the evolution of the number of knots $K_i \rightarrow K_{i+1}$ to be
\begin{equation}
    K_{i+1} = \mathrm{Int} \left( \frac{K_f}{2^{n_i}\,} \right) \,,
\end{equation}
where $n_{i}\,= \mathrm{Int} (\log_2 K_f -i+1)$, such that the number of knots approximately doubles at each successive iteration.

However, the number of knots used has to be carefully considered. For example, to avoid over-fitting, one should avoid making the knot spacing comparable to or smaller than the width of an absorption line or blended feature. Also, the STIS echelle format means that many spectral orders are pieced together to form a final one dimensional spectrum. The order flattening process is imperfect and small wiggles may remain in the final one dimensional spectrum on scales corresponding to the order separation. Approximately matching the knot density to that scale enables the continuum model to follow these undulations and hence can help remove this effect. Detailed descriptions of STIS data reduction procedures are given in e.g. \cite{Ayres2010, Ayres2022}.

\subsection{Refining: Re-identify Absorption and Emission Lines}\label{sec:refining}

\begin{figure*}
\centering
\tikzstyle{decision} = [diamond, draw, thick, fill=red!10, text width=7em, text badly centered, node distance=3cm, inner sep=0pt]
\tikzstyle{block} = [rectangle, draw, thick, fill=blue!10, text width=12em, text centered, rounded corners, minimum height=4em]
\tikzstyle{line} = [draw, thick, -latex']
\tikzstyle{cloud} = [draw, ellipse,fill=red!20, node distance=3cm, minimum height=2em]
\tikzstyle{end} = [trapezium, draw, thick, trapezium right angle=110, text width=7em, text centered, rounded corners, fill=yellow!10, node distance=1.9cm, minimum height=3.2em]
\begin{tikzpicture}[node distance = 2.5cm, auto]
\node [block] (stage1) {{\bf Stage 1} \\ Re-bin and smooth spectrum};
\node [block, below of=stage1, node distance=2.0cm] (stage2) {{\bf Stage 2} \\ Preliminary feature identification};
\node [block, below of=stage2, node distance=2.0cm] (stage3) {{\bf Stage 3} \\ Adjust number of knots};
\node [block, right of=stage3, node distance=4.5cm] (refining) {{\bf Refining} \\ Re-identify features};
\node [block, below of=stage3, node distance=2.0cm] (stage4) {{\bf Stage 4} \\ Solve for search direction};
\node [decision, below of=stage4, node distance=3.0cm] (stage5) {{\bf Stage 5} \\ Update parameters};
\node [decision, below of=stage5, node distance=3.7cm] (stage6) {{\bf Stage 6} \\ Increase order of fit};
\node [end, below of=stage6, node distance=2.7cm] (final) {{\bf FINAL MODEL}};
\path [line] (stage1) -- node[right]{$K_{i=1}=0$}(stage2);
\path [line] (stage2) -- (stage3);
\path [line] (refining) -- (stage3);
\path [line] (stage3) -- (stage4);
\path [line] (stage4) -- (stage5);
\path [line] (stage5) -| node[near start,above]{$\Delta \chi^2_\nu > \chi^2_{thres}$} (refining);
\path [line] (stage5) -- node[right]{$\Delta \chi^2_\nu \le \chi^2_{thres}$} (stage6);
\path [line] (stage6) -| node[near start,above]{$K_i<K_f$} (refining);
\path [line] (stage6) -- node[right]{$K_i=K_f$} (final);
\end{tikzpicture}
\vspace{0.3in}
\caption{This flow chart summarises the continuum fitting model based on spline fitting with automated line removal. $K_i$ indicates the number of cubic spline knots used at the $i^{th}$ iteration. $K_f$ is the number of knots used for the final best-fit continuum model. See Section \ref{sec:method} for details.
\label{fig:flowchart}
}
\end{figure*}
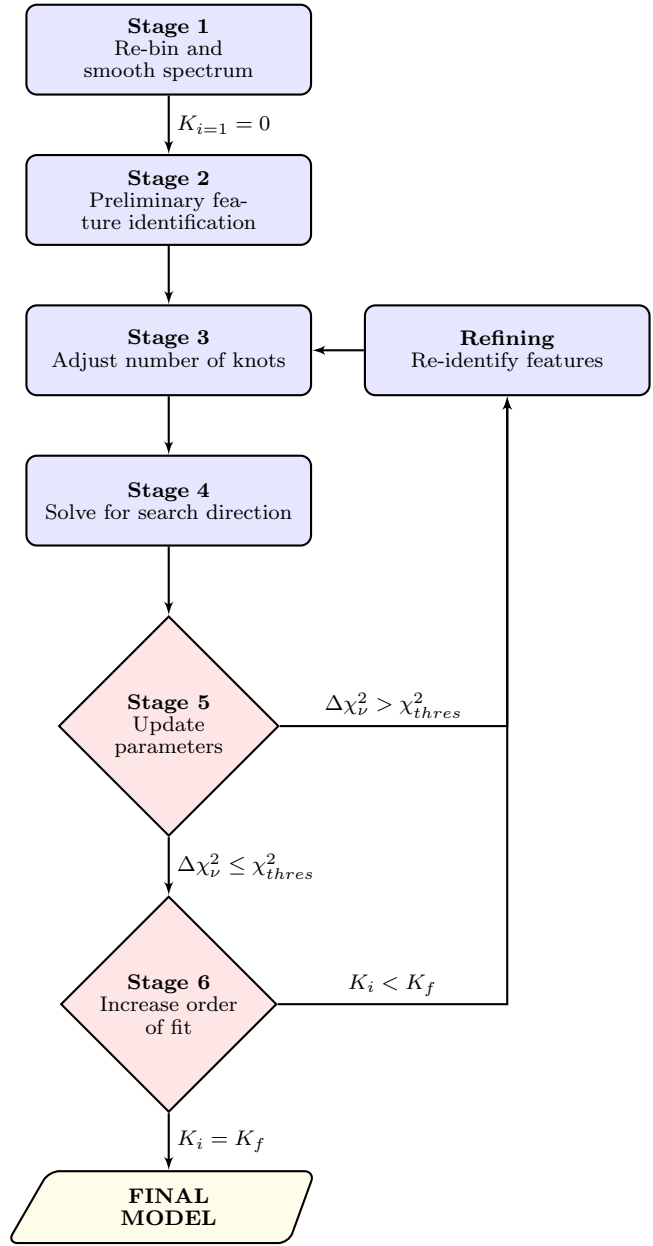

If the conditions for further iterations specified in Stages 5 and 6 are satisfied (see Figure \ref{fig:flowchart}), the algorithm return to re-identify all absorption and emission features, relative to the current continuum fit. The procedure during refinement is similar to Stage 2: 

\begin{enumerate}

\item Moving out from an absorption line centre, one can specify a point at which the observed intensity recovers to the underlying continuum level (within some tolerance). Also, once the intensity recovers to the local continuum level, noise fluctuations mean that the derivative is likely to change sign away from the absorption line centre. Thus we locate the point at which the derivative of the smoothed spectrum changes sign, moving outwards from the line centre. These points (left and right of each spectral line centre) provide a {\it preliminary} estimate of the feature boundaries and hence define an initial set of continuum pixels. However, several further steps refine these initial estimates.

\item Having identified feature edges, i.e. discarded pixels within and near to absorption (or emission) features, we now attempt to put previously excluded pixels back into the set used for continuum fitting. For each pixel in the unsmoothed spectrum, we calculate the intensity difference between it and the current continuum estimate. If the pixel in the unsmoothed spectrum is very close to the current continuum, it can be ``re-assigned'' as a pixel to be used the continuum fit, whether or not it formed part of the preliminary continuum pixel set. However, to be re-assigned, it must satisfy a second condition: the derivative of the smoothed spectrum at that point must be below a specified threshold. In this way, additional pixels are picked up to be used in fitting a continuum function.

To express the above more rigorously, let the intensity of the $i^{th}$ pixel in the original unsmoothed spectrum be $I_i$, with uncertainty $\sigma_i$. Also let the intensity of the $i^{th}$ pixel in the current continuum estimate be $C_i$. Then let
\begin{equation}
 \Delta_i = \lvert (I_i - C_i)/\sigma_i \rvert,
\end{equation}
and $d_i$ be the derivative of the smoothed original spectrum. We then require both $\Delta_i$ and $d_i$ to be smaller than threshold values. The threshold values are defined as follows. For all continuum pixels, we compute the distribution of $\Delta_i$ and determine its standard deviation, $\sigma_D$. The same is done for $d_i$ to determine $\sigma_d$. For a pixel to be ``re-assigned'' as a continuum pixel, we then require, simultaneously,
\begin{equation}
\Delta_i < \sigma_D \,\,\, \textrm{and} \,\,\, d_i < \sigma_d
\end{equation}

\item A further check is now made on pixels that are deemed to be free from absorption or emission features i.e. pixels that are used to fit a continuum fit to. For this test, we use a normalised spectrum i.e. the original spectrum has been divided by the continuum fit. For each contiguous continuum segment i.e. a set of $M$ pixels that are all deemed to be continuum pixels, linear regression is used to determine that segment's slope $s$ and its uncertainty $\sigma_s$. Provided $s- k \sigma_s \le 0 \le s+ k \sigma_s$, that segment remains identified as a continuum segment, where $k$ is a user defined parameter with a default value of 4.0.

The first two conditions above generally succeed in identifying and excluding strong and relatively narrow absorption lines whilst the third condition helps to identify shallow and weaker lines. These conditions above apply equally the emission lines (with a minus sign on the tests). However, at this stage there may nevertheless still be true continuum pixels that have been discarded. Therefore one further test is done:

\item Where a contiguous set of $n$ or more pixels (default $n=10$) lies {\it above} the continuum fit, if $|d_i| < \sigma_d$, those pixels are re-assigned as being pixels to include in the continuum fit.
\end{enumerate}

The whole process described above is illustrated as a flowchart in Figure \ref{fig:flowchart}.

\section{Application to a high resolution spectrum of the white dwarf G191--B2B} \label{sec:G191} 

\subsection{Astronomical data} \label{sec:wddata}

The astronomical data used is the Hubble Space Telescope STIS spectrum of the well known white dwarf G191$-$B2B. The data were obtained using the E140H grating, with resolution $R\approx114,000$ (the highest currently available for UV astronomical spectroscopy), and have a signal to noise $\sim$100 per resolution element. A comprehensive description covering both the observational data and the data reduction processes is given in Appendix A of \cite{Hu2019}. This object/spectrum was chosen because: (a) several detailed studies of it already exist, including measurements of $\daa$, \cite{Berengut2013, Hu2021}; (b) the STIS echelle instrument format means individual spectral orders are extracted from the detector individually before combination to form a continuous 1-d spectrum. This means the data are fairly ``challenging'' in that weak undulations in the apparent continuum are present due to imperfect order flattening prior to combination (as seen in the illustrations later in this paper); (c) the spectrum is riddled with hundreds of photospheric absorption lines, \cite{Preval2013}, that must be detected and removed prior to (or in conjunction with) estimating the underlying continuum. Nevertheless, the G140H STIS spectrum of G191$-$B2B may be considered atypical due to its high signal to noise such that the settings used in the continuum fitting code (Section \ref{sec:method}) for this particular spectrum may of course need tuning for other applications.\\

\subsection{Atomic data} \label{sec:wdatom}

In this analysis we opted to use {\niv} transitions and atomic data. {\niv} wavelengths and energy level measurements have been published in \cite{Raassen1976a, Raassen1976b, Ward2019}. Errors in the published values from \cite{Ward2019} were found and corrected by \cite{NIST_ASD}. The {\niv} wavelengths used in this analysis are those given in the {\it NIST Atomic Spectra Database database v10}.

In order to assess the impact of variations in the fitted continuum level on $\daa$ measurements, we broadly follow the procedures described in \cite{Berengut2013}. To do so we require input rest-frame wavelengths and sensitivity coefficients describing wavelength shifts as a function of $\daa$ ($q$-coefficients), calculated using Eq.\,\eqref{eq:wqda}.

Calculations of sensitivity coefficients applied in an astronomical context were first described in \cite{Dzuba1999a, Dzuba1999b}. {\niv} (and {\fev}) $q$-coefficients for UV transitions were first reported in \cite{Ong2013}. Improved $q$ calculations are provided in \cite{Hu2021}, but only for {\fev} and not for {\niv}. For the calculations described in the present paper, we have revised the {\niv} $q$-coefficients given in \cite{Ong2013} using the same methods described in \cite{Hu2021}. We therefore do not re-iterate technical details provided in those papers and provide the relevant atomic data tables as online supplementary material, in which the transition frequencies and  $q$-coefficients are presented in units of cm$^{-1}$. \\

\section{$\daa$ measurements using different continuum models} \label{sec:results}

We now show how continuum uncertainty impacts on $\daa$ measurements. Our $\daa$ analysis for white dwarfs follows that of \cite{Berengut2013}, and not the analysis of \cite{Hu2021}. The former is far simpler because observed line wavelengths are measured using a simple centroiding procedure -- no absorption profile modelling is carried out. In the analysis of \cite{Hu2021}, each absorption feature is fitted using one or more Voigt profiles and in this case small variations in the continuum level can give rise to changes in the number of absorption components required, which in turn changes relative positions. The procedure of \cite{Hu2021} has another advantage over that of \cite{Berengut2013}; the {\sc vpfit} and {\sc ai-vpfit} modelling procedures allow for the inclusion of additional free parameters, to refine the local continuum of each spectral segment in the model, as shown in Eq.\,\eqref{eq:gausstilt} and discussed in Section \ref{sec:shifts}. We address this in a companion paper, in the same volume of this journal.

\begin{figure*}
\centering
\includegraphics[width=0.95\linewidth]{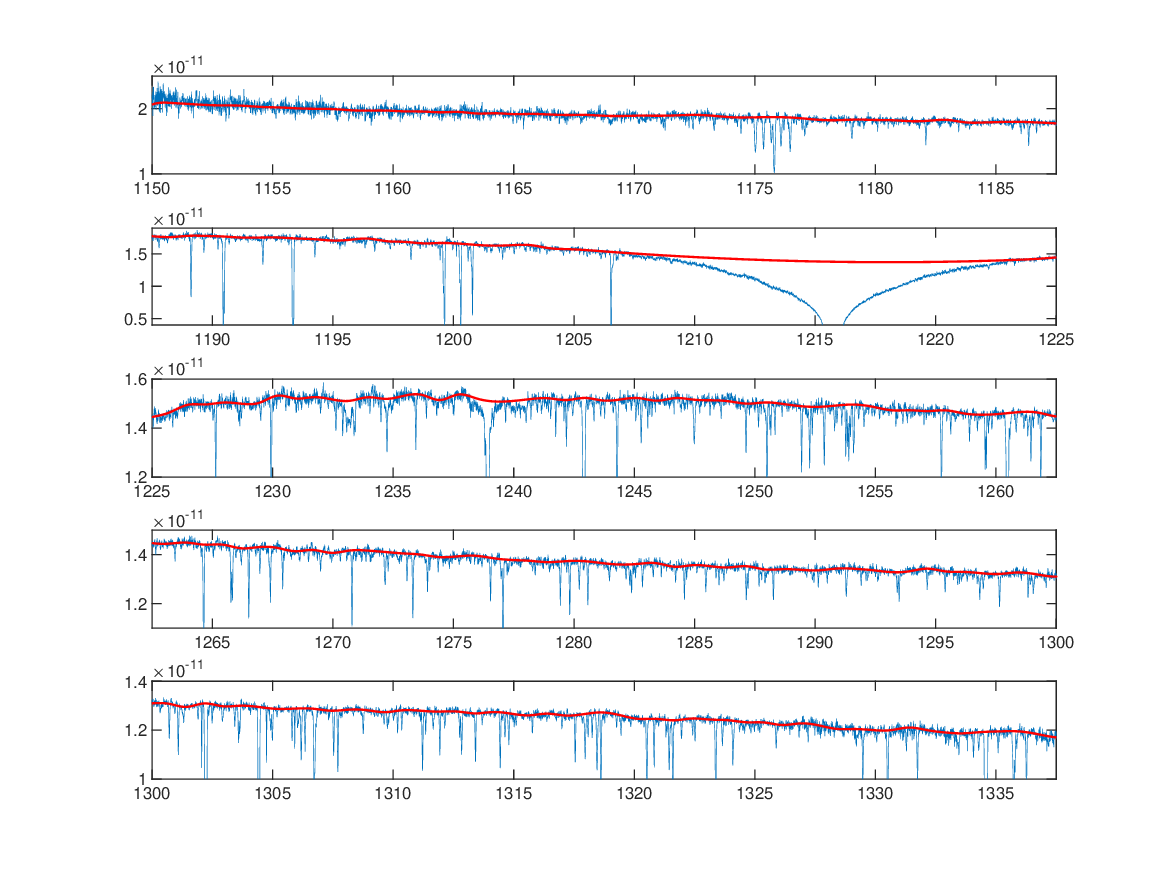}\\
\vspace{-1cm}
\includegraphics[width=0.95\linewidth]{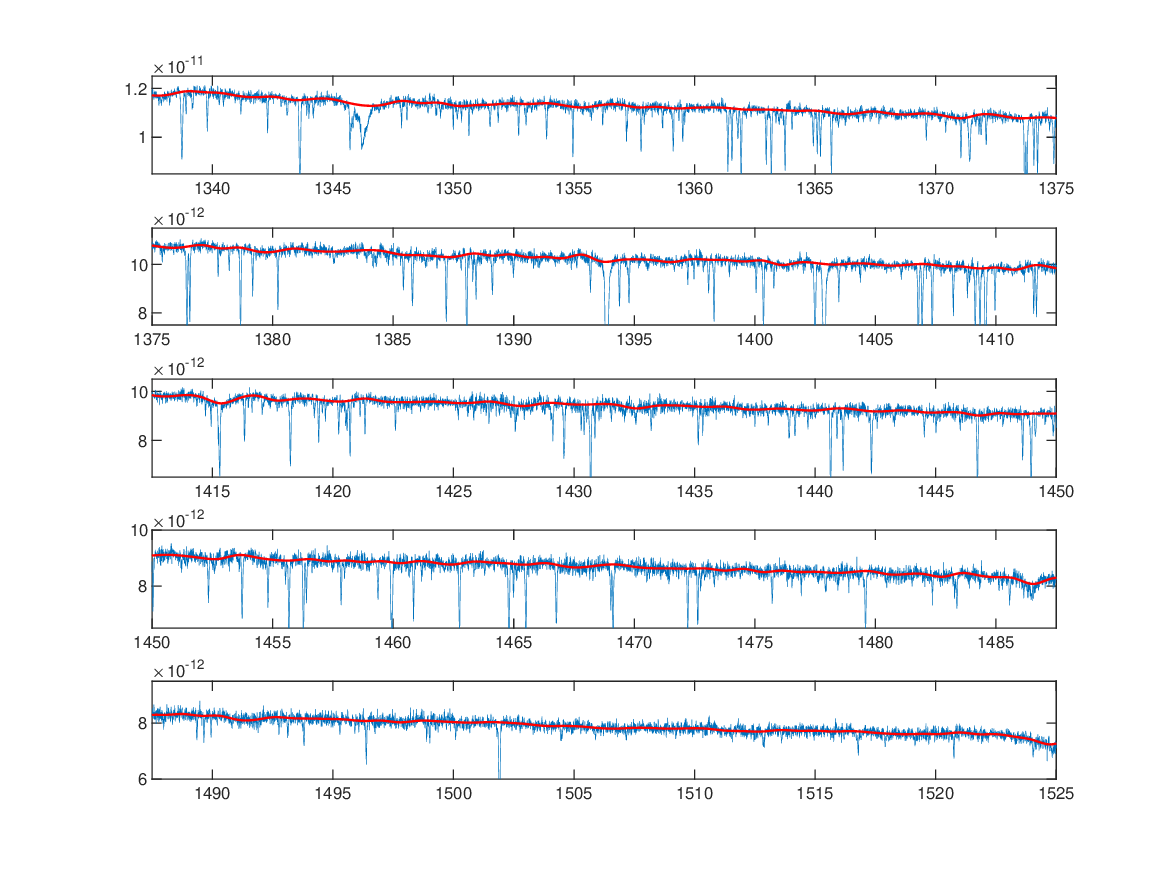}
\caption{G191$-$B2B continuum model with 0.75\,{\AA} knot spacing.
\label{fig:full075-1}}
\end{figure*}

Measuring the relative positions of a large number of narrow photospheric absorption lines allows us to explore whether the fine structure constant $\alpha$ varies in the presence of strong gravity. For discussions on this and general overviews see e.g. \cite{magueijo02, Barrow2005, flambaum08, Uzan2011, Berengut2013, Bainbridge2017, Hu2021}. Repeating measurements using different continuum models quantifies the sensitivity of the measured $\alpha$ to small variations in the continuum. We therefore generated four example model continua, using knot spacings of 0.5, 0.75, 1.0, and 1.25\,{\AA}. The range is intended to bracket (for our specific spectrum i.e. the STIS G191$-$B2B data) a plausible range of values within which the returned continuum fits look reasonable. Figure \ref{fig:full075-1} illustrates one continuum model derived using the method described in Section \ref{sec:method} using a knot spacing of 0.75{\AA}. Plots using other settings are provided as online supplementary material. As can be seen in Figure \ref{fig:full075-1}, the data reveal many slight small-scale undulations (discussed earlier in Section \ref{sec:method}) that seem to be followed fairly well by this model.

Using these four continua, we then detect absorption lines which are significant at or above 5$\sigma$ (see Table \ref{tab:resultsWD}). This number of lines identified at or above this significance limit, in the range 1150\,{\AA} to 1894\,{\AA}, for each of the four knot spacings, is also given in Table \ref{tab:resultsWD}). Using custom Python code, the observed feature wavelengths were then shifted to the rest-frame using a G191$-$B2B redshift of 23.8 $\pm$ 0.03 km s$^{-1}$ from \cite{Preval2013}). The rest-frame wavelengths were then matched to recently updated {\niv} laboratory wavelengths (provided as online supplementary material), where a match was accepted for agreement equal to or better than 3$\sigma$, allowing for both observational and laboratory wavelength uncertainties. For each knot spacing, a total of 299, 316, 325 and 320 potential {\niv} matches were made respectively, of which 258, 270, 281 and 278 have associated $q$-coefficients (see Table \ref{tab:resultsWD}). However, as can be seen in Table \ref{a:ids}, there are many cases where multiple identifications are possible. All multiple IDs are removed from the linear regression procedure to measure $\daa$. Finally, a visual inspection of Figure \ref{fig:full075-1} shows that there are a few spectral regions that need to be avoided: [1205, 1227], [1238,1241], [1242,1243.5], [1345,1347] {\AA}. All {\niv} transition within these 4 windows are discarded.

Having identified the lines to be used, we follow a similar procedure to that used in \cite{Berengut2013}. The observed-frame wavelength $\lambda_{obs,i}$ of the $i^{th}$ absorption line in the white dwarf photosphere, and the corresponding rest-frame laboratory wavelength $\lambda_{0,i}$, are related by
\begin{align}
\frac{\lambda_{obs,i} - \lambda_{0,i}}{\lambda_{0,i}} = \frac{\Delta\lambda_i}{\lambda_{0,i}} &= z - \frac{2q_i \lambda_{obs,i}}{10^8} \fdaa \nonumber \\ &= z - Q_i \fdaa \label{eq:mm3}
\end{align}
where wavelengths are in {\AA}, $q_i$ (in cm$^{-1}$) is the relative sensitivity of the transition frequency to variation in $\alpha$, $z$ is the white dwarf redshift (comprising gravitational and kinematic contributions), and $\daa = (\alpha_{WD} - \alpha_0)/\alpha_0$.

\begin{figure*}
\centering
\includegraphics[width=\linewidth]{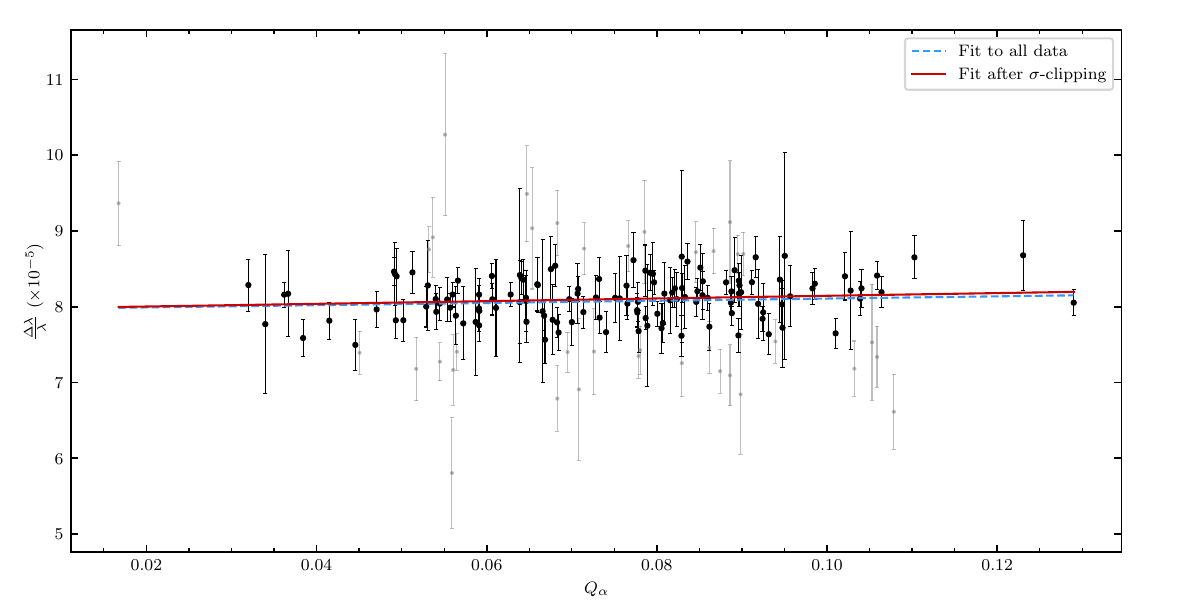}
\caption{Eq.\,\eqref{eq:mm3} for one example trial continuum (0.75{\AA} knot spacing). Grey points were removed when $\sigma$-clipping was applied. See Section \ref{sec:results} for details.
\label{fig:linregress}
}
\end{figure*}

\begin{table*}
\renewcommand*{\arraystretch}{1.4}
\centering
\begin{tabular}{|l|r|r|r|r|}
\hline
\textbf{Knot spacing (\AA)} & \textbf{0.5} & \textbf{0.75} & \textbf{1.0} & \textbf{1.25} \\
\hline
\textbf{5$\sigma$ detections} & 1566 & 1678 & 1789 & 1752 \\
\textbf{NiV matches total} & 299 & 316 & 325 & 320 \\
\hline
\textbf{NiV matches having q (no $\sigma$-clipping)} & 167 & 165 & 163 & 164 \\
\textbf{$\daa \times 10^{-5}$} & $-1.530 \pm 1.101$ & $-1.462 \pm 1.121 $ & $-1.624 \pm 1.156$ & $-1.980 \pm 1.150$ \\
\textbf{$z \times 10^{-5}$} & $7.957 \pm 0.085$ & $7.962 \pm 0.087$ & $7.940 \pm 0.090$ & $7.920 \pm 0.090$ \\
\textbf{$\chi^2_{\nu}$} & 1.950 & 1.869 & 1.943 & 1.918 \\
\hline
\textbf{NiV matches having q (after $\sigma$-clipping)} & 125 & 128 & 125 & 127 \\
\textbf{$\daa \times 10^{-5}$} & $-2.153 \pm 1.193$ & $-1.771 \pm 1.187$ & $-2.856 \pm 1.229$ & $-2.625 \pm 1.215$ \\
\textbf{$z \times 10^{-5}$} & $7.955 \pm 0.092$ & $7.968 \pm 0.092$ & $7.888 \pm 0.095$ & $7.906 \pm 0.095$ \\
\textbf{$\chi^2_{\nu}$} & 0.995 & 0.980 & 0.982 & 0.997 \\
\hline
\end{tabular}
\vspace{1em}
\caption{\begin{flushleft}
The impact of using different white dwarf continuum models on measurements of $\daa$. Four model continua, each corresponding to a different knot spacing. The results show that the additional uncertainty on $\daa$ associated by continuum placement variation is comparable to or even larger than the statistical uncertainty. The first row gives the total number of 5$\sigma$ NiV detections in the HST STIS spectrum. The second row shows the total number of lines identified as potential NiV, matched within 3$\sigma$ of the Ward et al.\ laboratory list (allowing for the white dwarf redshift), including multiple potential IDs. Where more than 1 possible match exists (see Appendix \ref{a:ids}), we discard that line entirely from further analysis. Observed and laboratory wavelength matches were made using the newly calculated {\niv} $q$-coefficients and the \protect\cite{Ward2019} {\niv} laboratory wavelengths updated as part of this work by A. Kramida. The second block in the table shows measurements for all lines that have $q$ coefficient calculations (given in Appendix \ref{sec:atomic}). The third block shows the same thing as the second block, except deviant points (measured in units of $\sigma$) were iteratively clipped during linear regression until $\chi^2_{\nu} \le 1$. $\chi^2_{\nu}$ is defined by Eq.\eqref{eq:chisqnu}. The other continuum fitting parameter settings are: smoothing FWHM $3\bar{x}$ (Section \ref{sec:stage1}), $\zeta=3$ (Section \ref{sec:stage2}), $k_{merge}=1/3$ (Section \ref{sec:stage3}), $n=10$ (Section \ref{sec:refining}).
\end{flushleft}}
\label{tab:resultsWD}
\end{table*}

A linear regression analysis on the matched {\niv} lines is then used to solve for the parameters $z$ and $\daa$, for each of the four model continua. The results of this are given in Table \ref{tab:resultsWD} and, for one example knot spacing, illustrated in Figure \ref{fig:linregress} (details for all knot spacings are given in the Supplementary Material. The table also gives the reduced chi-squared, which in this case is
\begin{equation}
\chi^2_{\nu} = \frac{1}{N-2} \sum_{i=1}^N \left( \frac{y_i - (\Delta\lambda/\lambda_0)_i}{\sigma(\Delta\lambda/\lambda_0)_i)} \right)^2 \label{eq:chisqnu}
\end{equation}
Clearly, the $\chi^2_{\nu}$ values are larger than the expected value of unity for a good fit to the data. This is unsurprising: the scatter in the ``continuum'' points fitted will be increased by the presence of weak unidentified absorption lines that were not detected using the 5$\sigma$ detection threshold. Further, the continuum model is likely to be inadequate in places for the instrumental/data processing reasons explained previously.

\subsection{Potential bias}

The impact of {\it assuming} $\daa=0$ when matching laboratory and white dwarf lines was discussed in \cite{Hu2021} (see section 5.2 and figure 2 in that paper). That numerical experiment was carried out using {\fev}, altering the input assumption multiple times, using a range $\daa = \pm 10^3$, in steps of $\daa=10^{-5}$. No evidence was seen for bias in the \cite{Hu2021} analysis. Nevertheless, it should not necessarily be assumed that there is no bias in the analysis carried out here using {\niv}. It is possible that the assumption of $\daa=0$ when performing line identifications may result in some fraction of false identifications, potentially biasing the slope in Fig.\,\ref{fig:linregress}. However, our focus in this paper is to explore the impact of continuum placement. A detailed study of this kind of bias, in the sense described above, is beyond the scope of the present work, but we note the potential issue here to motivate explicit checks in future measurements.

\section{Discussion and future work}\label{sec:discussion}

In this work we have developed a new algorithm for continuum fitting high resolution spectra. The method is automated, reproducible, and objective. Here we have applied it to a high-resolution, high signal-to-noise HST/STIS spectrum of the white dwarf G191$-$B2B. In a companion article we apply it to quasar spectroscopy. Using this algorithm we obtained four different estimates for the underlying G191$-$B2B continuum in order to examine the impact of continuum placement uncertainties on measurements of the fine structure constant in the strong gravitational field at the G191$-$B2B photosphere.

An earlier analysis \citep{Berengut2013} found anomalous results when comparing $\daa$ measurements derived using {\fev} and {\niv} transitions detected in the G191$-$B2B photosphere. We draw attention to figure 1 in that paper, where different slopes were found using the linear relationship expressed in Eq.\eqref{eq:mm3}; they found $\daa = (4.2 \pm 1.6) \times 10^{-5}$ and $(-6.1 \pm 5.8) \times 10^{-5}$ for {\fev} and {\niv} respectively. The interpretation suggested in \citep{Berengut2013} was that systematics were present in the available {\niv} laboratory wavelength measurements. 

More recently, \cite{Hu2021} carried out a more detailed analysis using {\fev} transitions, using the same astronomical data. The best result found by \cite{Hu2021} is $\daa = (6.36 \pm 0.35_{stat} \pm 1.84_{sys}) \times 10^{-5}$, consistent with the earlier {\fev} result from \cite{Berengut2013}. 

New, independent {\niv} laboratory wavelength measurements have been made \citep{Ward2019}. Also, new sensitivity coefficient ($q$) calculations have been carried out and reported in this paper (Appendix \ref{sec:atomic}). These new data thus allow us to re-examine the apparent {\fev}/{\niv} inconsistency previously reported. Comparing the best available {\fev} measurement \citep{Hu2021} with the most positive result obtained in the present work (no $\sigma$-clipping, 0.75{\AA} knot spacing, see Table \ref{tab:resultsWD}), a significant (3.2$\sigma$) discrepancy remains (more significant if we take any of the other {\niv} measurements obtained in the present work). Note that Table \ref{tab:resultsWD} quotes two sets of four ``final'' $\daa$ measurements derived using the {\niv} lines. We opt to avoid selecting a ``preferred'' value since there are no obvious criteria for selecting one continuum model over another. Nevertheless, it can be seen that there is reasonable internal agreement within each set of four measurements, and there is some justification for preferring the $\sigma$-clipped results, given the high $\chi^2_{\nu}$ results obtained without $\sigma$-clipping.

Our main results from this work therefore are:
\begin{enumerate}
    \item Slight changes in the underlying continuum can impact on $\daa$ measurements, as illustrated in Table \ref{tab:resultsWD}; the additional uncertainty on $\daa$ associated with continuum placement variations is not negligible compared to the statistical (i.e. covariance matrix) uncertainty on $\daa$. 
    \item Curiously, the $\daa$ measurements we have made, using entirely independent {\niv} laboratory wavelengths and new $q$ coefficients, are still of opposite sign to the {\fev} measurements reported in both \cite{Berengut2013} and \cite{Hu2021}. The explanation for this continuing significant discrepancy has yet to be found.
\end{enumerate}

There is an important caveat to the first result listed above: the analysis in the present work is based on absorption line centroiding and linear regression method, as for \cite{Berengut2013}. A different approach is that of \cite{Hu2021}, in which absorption lines are modelled using Voigt profiles. In the latter, the fits to local continuum segments can be refined using additional free parameters included in the non-linear least squares fitting procedure (see \citealt{web:VPFIT} for details). That process may substantially reduce the error contribution found here. We have not explored this here for white dwarf spectroscopy, although a companion paper to this one, describing the analysis of the spectrum of quasar PHL957, suggests local continuum refinement dramatically reduces the impact of variations in the initial continuum model.

The results presented in this paper have important implications for the claimed uncertainties in some previous published $\daa$ measurements, including the recent measurements of, for example, \cite{Hees2020, Murphy2022}. There are likely to also be significant implications for the achievable precision in forthcoming redshift drift studies \citep{Sandage1962, Liske2008}; the results reported here motivate a more careful exploration of the potential systematics that may arise from continuum estimation across the Lyman alpha forest.

\section*{Acknowledgments}
We are most grateful to Alexander Kramida and Tom Ayres for useful communications.

\section*{Data Availability}
Based on observations made with the NASA/ESA Hubble Space Telescope, obtained from the data archive at the Space Telescope Science Institute. The STIS spectra of G191$-$B2B are available from the Barbara A. Mikulski archive. The continuum fitting code described in this paper is available from the authors on request.

\bibliographystyle{aasjournal}
\bibliography{contin}

\clearpage

\appendix
\section{Solving for $\daa$ using linear regression}

Let $x = (\alpha/\alpha_0)^2 - 1$. 
For the $i^{th}$ absorption line,
\begin{equation}
    \omega_{rest,i} = \omega_{0,i} + q_i x \label{eq:mmm}
\end{equation}
where $\omega_{rest,i}$ is the observed rest-frame frequency of the $i^{th}$ line measured in the white dwarf spectrum (i.e. $\omega_{rest,i}$ does not include the white dwarf redshift, the combined effects of gravitational redshift and line of sight stellar velocity). $\omega_{0,i}$ is the terrestrial laboratory rest-frame frequency. $q_i$ is the sensitivity coefficient for that line. $\omega$ and $q$ are both in units of cm$^{-1}$. The observed-frame and rest-frame frequencies of the $i^{th}$ line measured in the white dwarf spectrum are related by
\begin{equation}
    \omega_{rest,i} = \omega_{obs,i} (1+z) \label{eq:obs2rest}
\end{equation}
so Eq.\,\ref{eq:mmm} becomes
\begin{equation}
    \omega_{obs,i} + z\omega_{obs,i} = \omega_{0,i} + q_i x \label{eq:mmm2}
\end{equation}
Using the approximation $x \approx 2\Delta\alpha/\alpha_0$, where $\Delta\alpha = \alpha - \alpha_0$, Eq.\,\ref{eq:mmm2} becomes
\begin{equation}
    \omega_{obs,i} - \omega_{0,i} = -z\omega_{obs,i} + 2 q_i \fdaa
\end{equation}
so
\begin{equation}
    \frac{\omega_{obs,i} - \omega_{0,i}}{\omega_{obs,i}} = -z + \frac{2q_i}{\omega_{obs,i}} \fdaa
\end{equation}
and with $\lambda$ in {\AA}, $\omega = 10^8/\lambda$, 
\begin{equation}
    \frac{\lambda_{obs,i} - \lambda_{0,i}}{\lambda_{0,i}} = \frac{\Delta\lambda_i}{\lambda_{0,i}} = z - \frac{2q_i \lambda_{obs,i}}{10^8} \fdaa \label{eq:mm3b}
\end{equation}
\vspace{0.25cm}

\noindent The version of Eq.\,\ref{eq:mm3b} used in Berengut et al 2013 is,
\begin{equation}
    \frac{\Delta\lambda_i}{\lambda_{0,i}} = z - \frac{2q_i \lambda_{0,i}}{10^8} (1+z) \fdaa. \label{eq:b13}
\end{equation}
Eq.\,\ref{eq:b13} thus assumes $\lambda_{obs,i}/\lambda_{0,i} = (1+z)$, which is correct if $\daa_0=0$, but an approximation otherwise. The exact expression is $\lambda_{obs,i}/\lambda_{rest,i} = (1+z_i)$. Whilst both equations give very close results ($\daa$ discrepancies are at the $10^{-8}$ level, well below any realistic experimental error), in this work we use Eq.\,\ref{eq:mm3b} to solve for the slope and intercept $\daa$ and $z$.\\

\section{Complete set of figures}

In the main paper we illustrate only one continuum (0.75{\AA} knot spacing). Here we illustrate the complete set used in these measurements, i.e. continua computed for all four knot spacings: 0.5, 0.75, 1.0, 1.25{\AA}.

\begin{figure}
\centering
\includegraphics[width=0.95\linewidth]{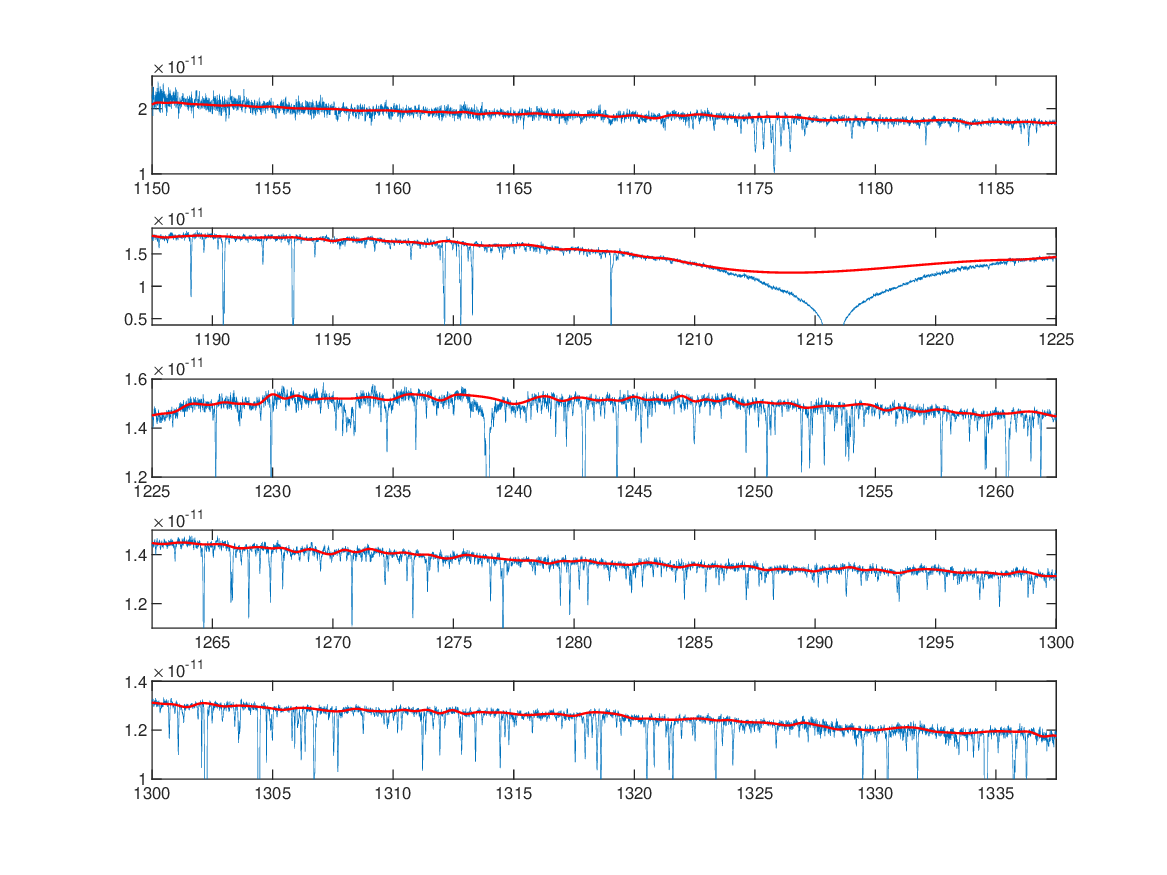}\\
\vspace{-1cm}
\includegraphics[width=0.95\linewidth]{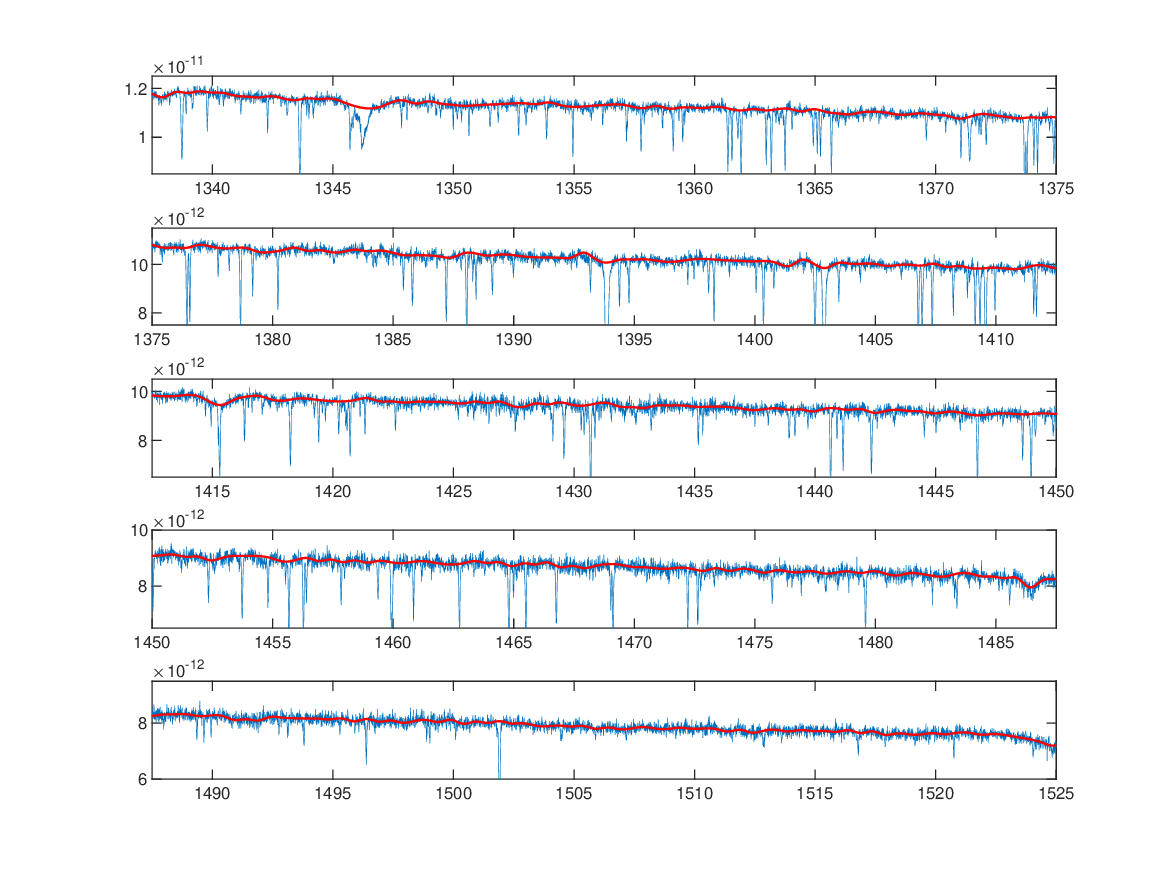}
\caption{G191$-$B2B continuum model with 0.5\,{\AA} knot spacing.
\label{fig:full05-1}
}
\end{figure}

\begin{figure}
\centering
\includegraphics[width=0.95\linewidth]{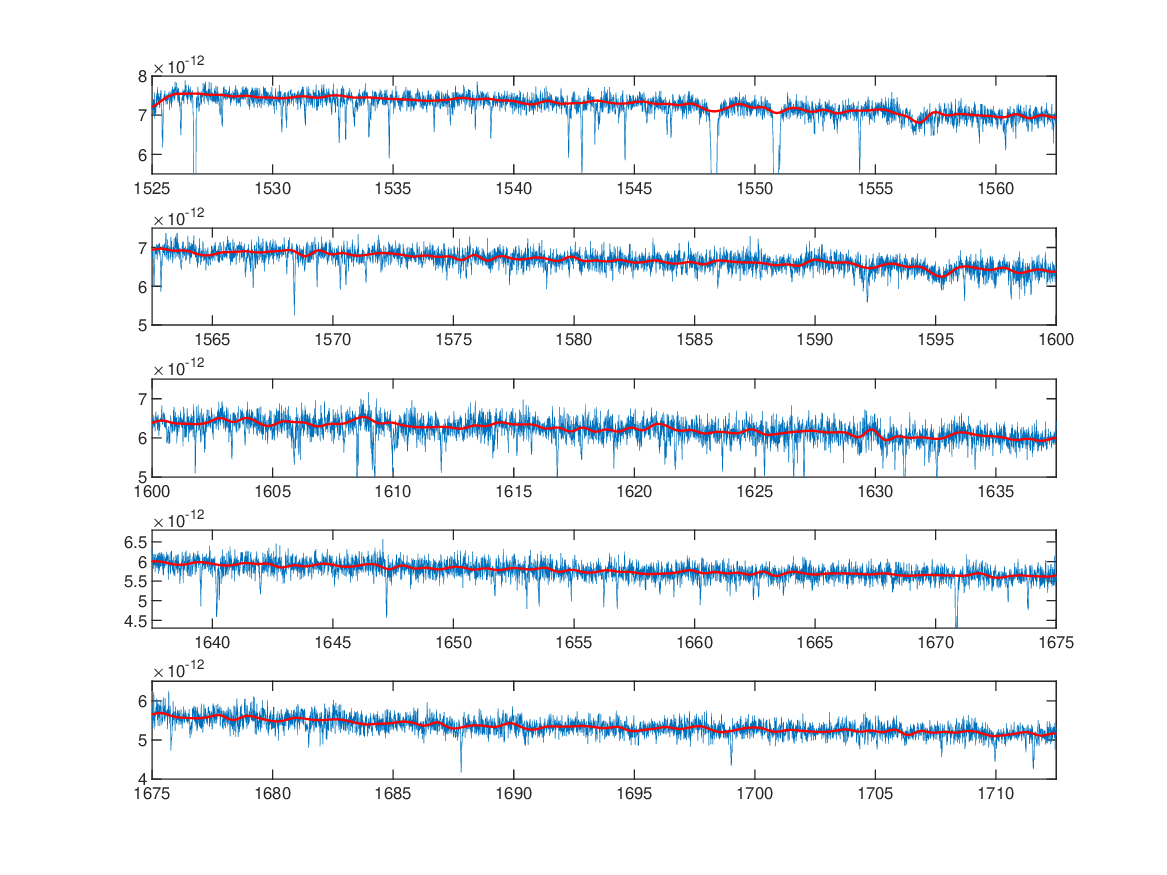}\\
\vspace{-1cm}
\includegraphics[width=0.95\linewidth]{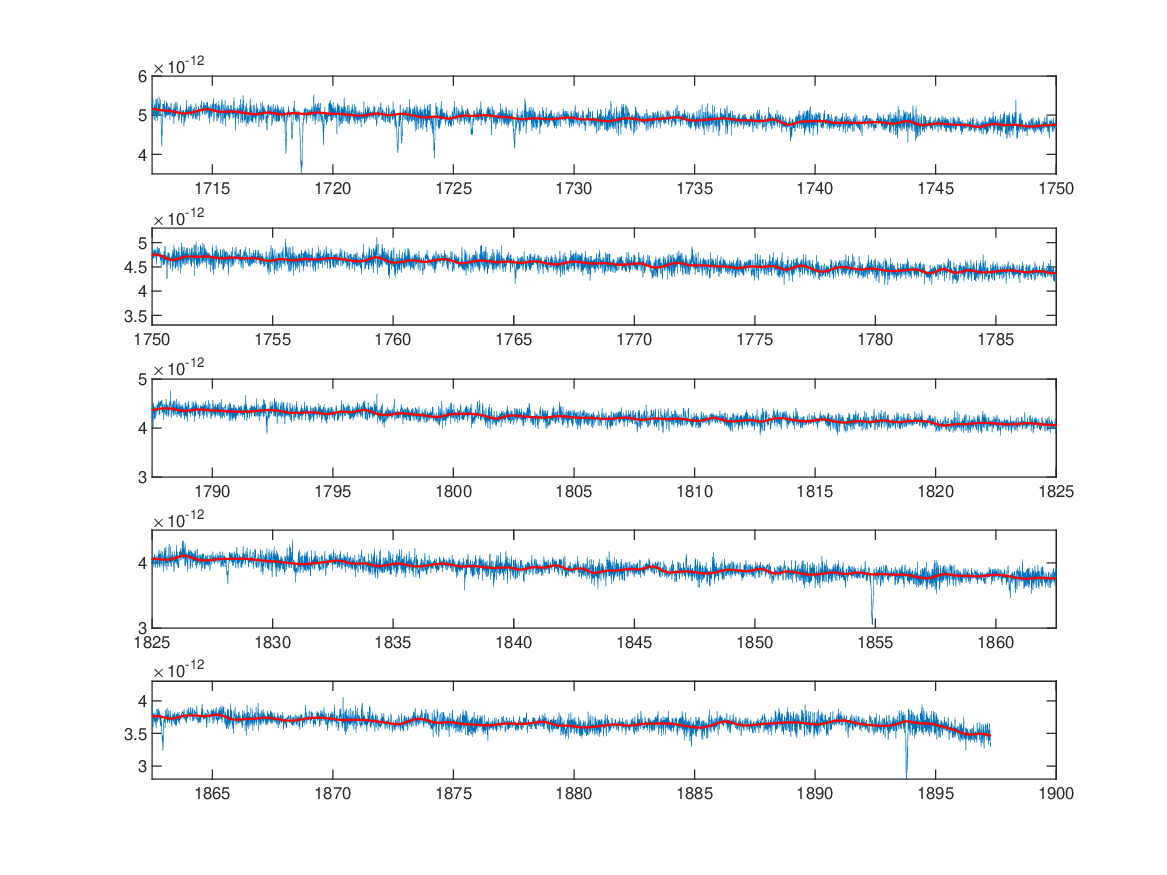}
\caption{G191$-$B2B continuum model with 0.5\,{\AA} knot spacing.
\label{fig:full05-2}
}
\end{figure}

\begin{figure}
\centering
\includegraphics[width=0.95\linewidth]{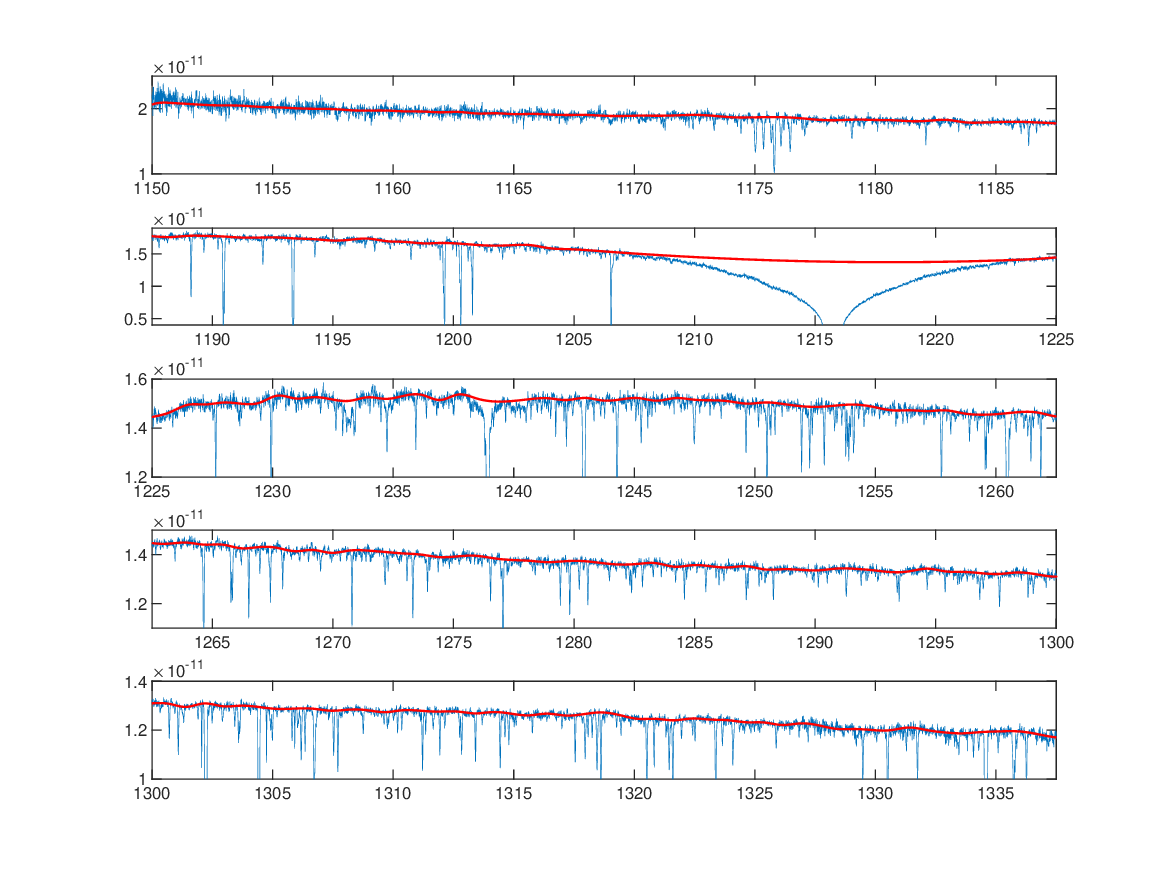}\\
\vspace{-1cm}
\includegraphics[width=0.95\linewidth]{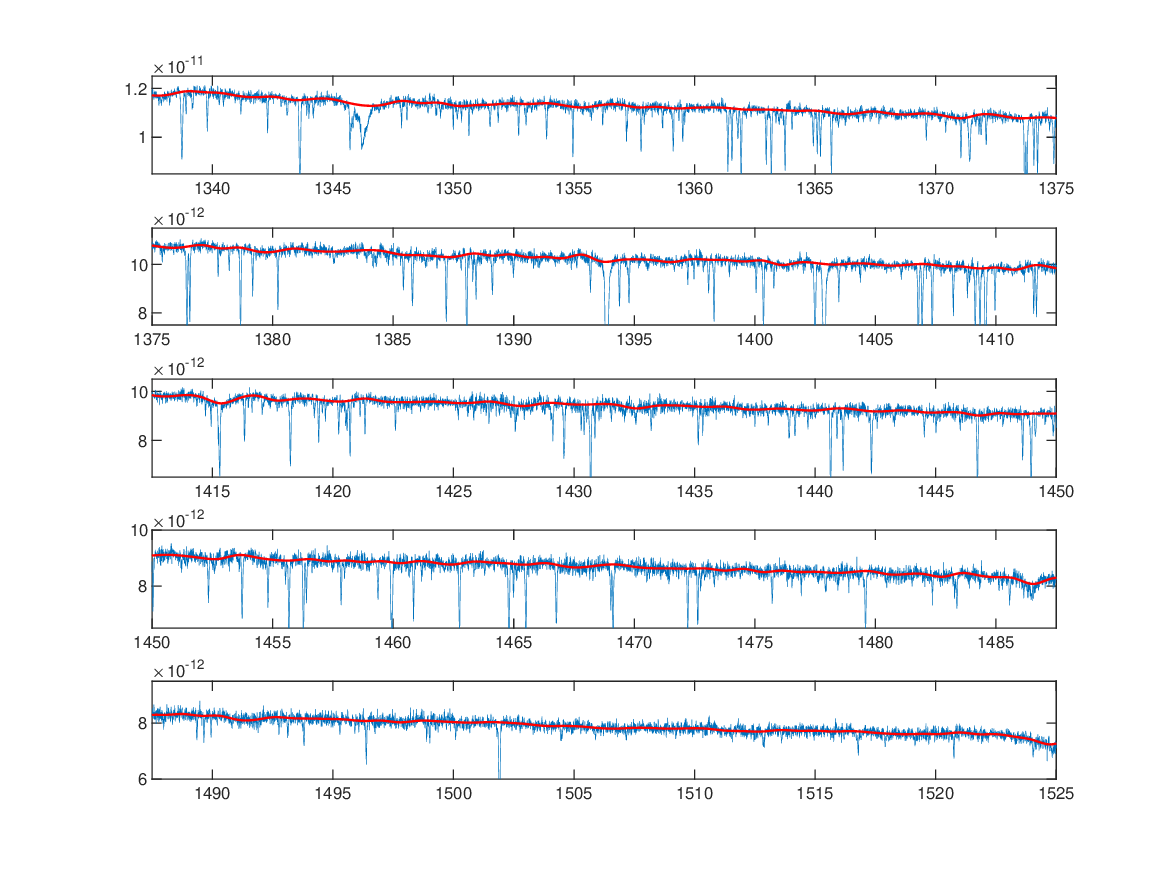}
\caption{G191$-$B2B continuum model with 0.75\,{\AA} knot spacing.
\label{fig:full075-1b}
}
\end{figure}

\begin{figure}
\centering
\includegraphics[width=0.95\linewidth]{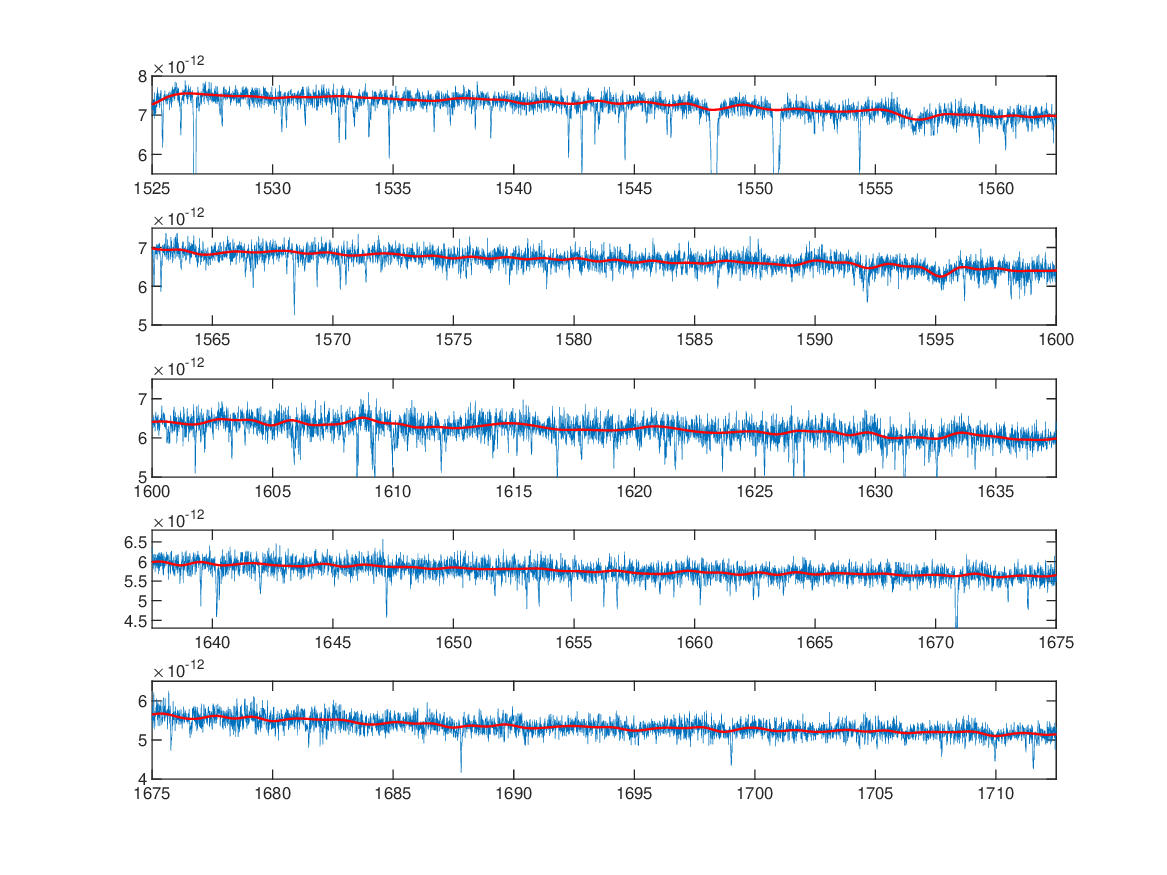}\\
\vspace{-1cm}
\includegraphics[width=0.95\linewidth]{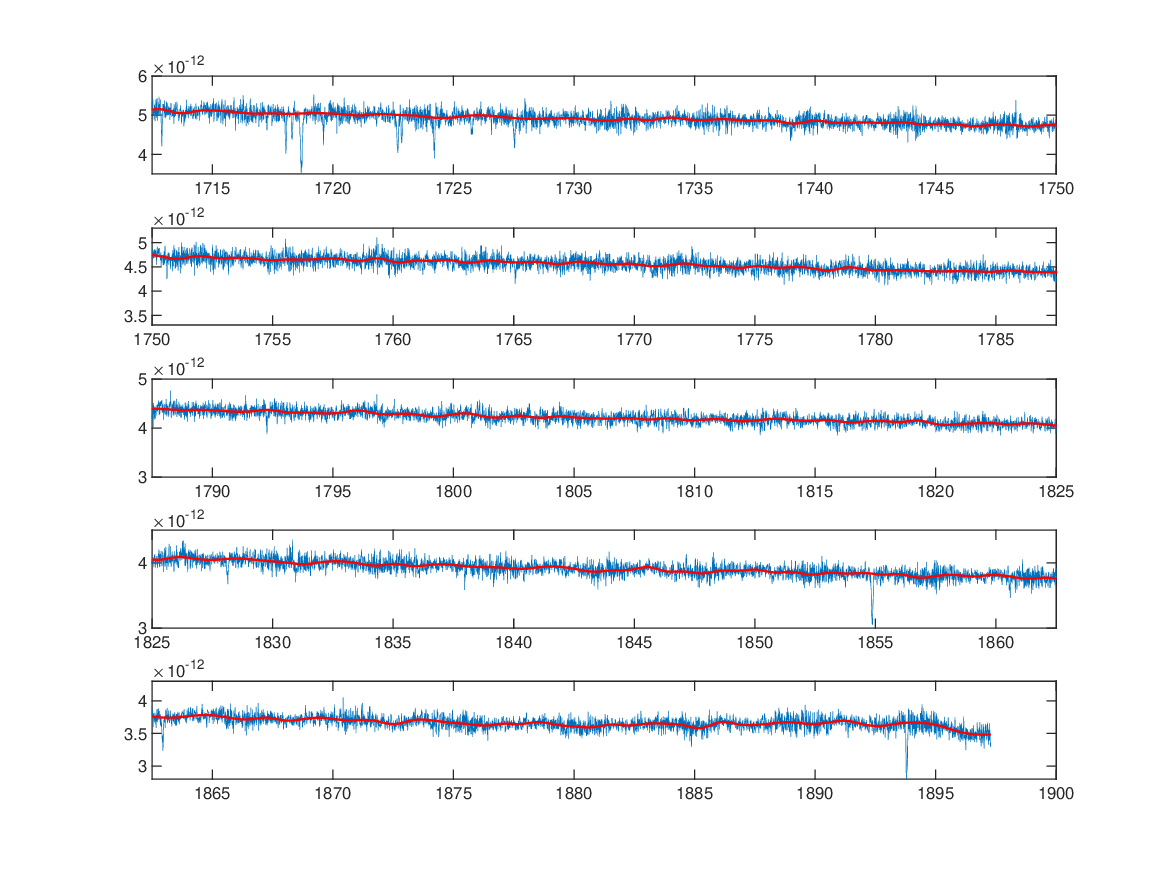}
\caption{G191$-$B2B continuum model with 0.75\,{\AA} knot spacing.
\label{fig:full075-2}
}
\end{figure}

\begin{figure}
\centering
\includegraphics[width=0.95\linewidth]{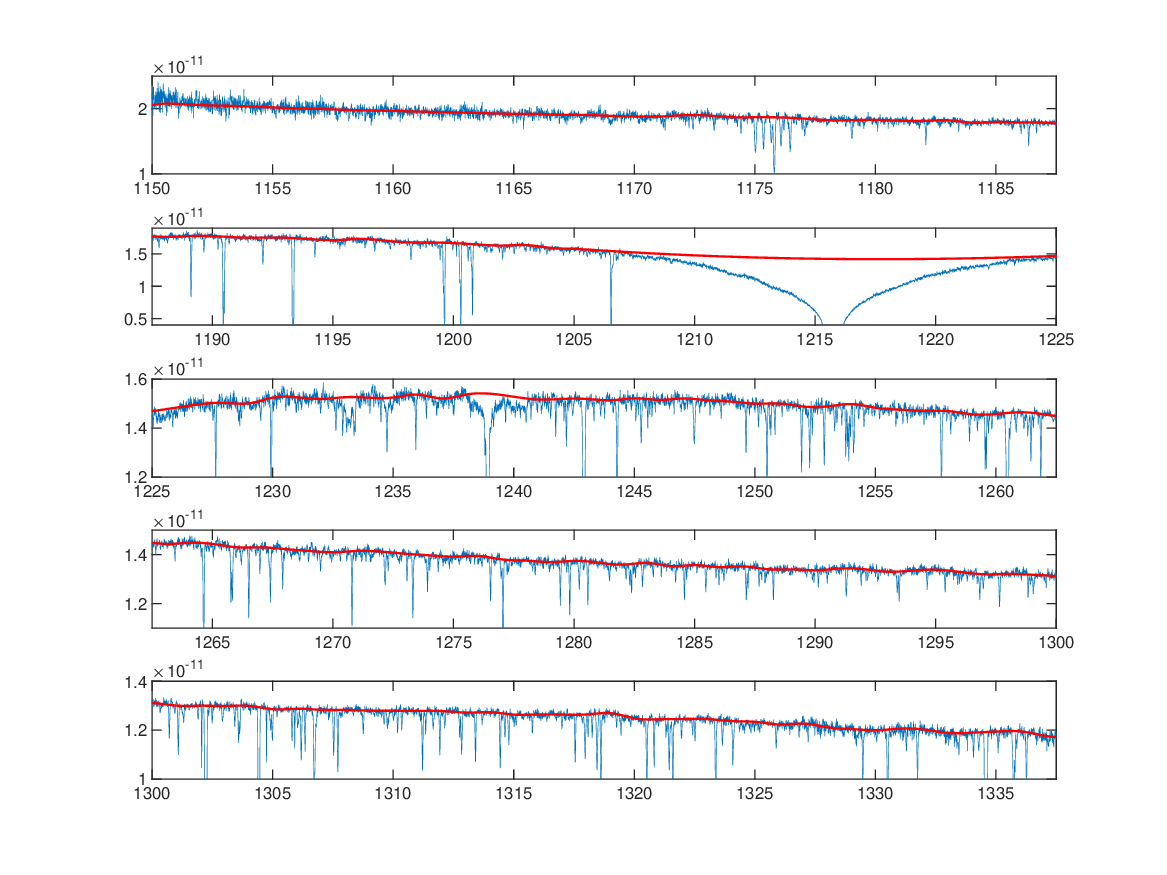}\\
\vspace{-1cm}
\includegraphics[width=0.95\linewidth]{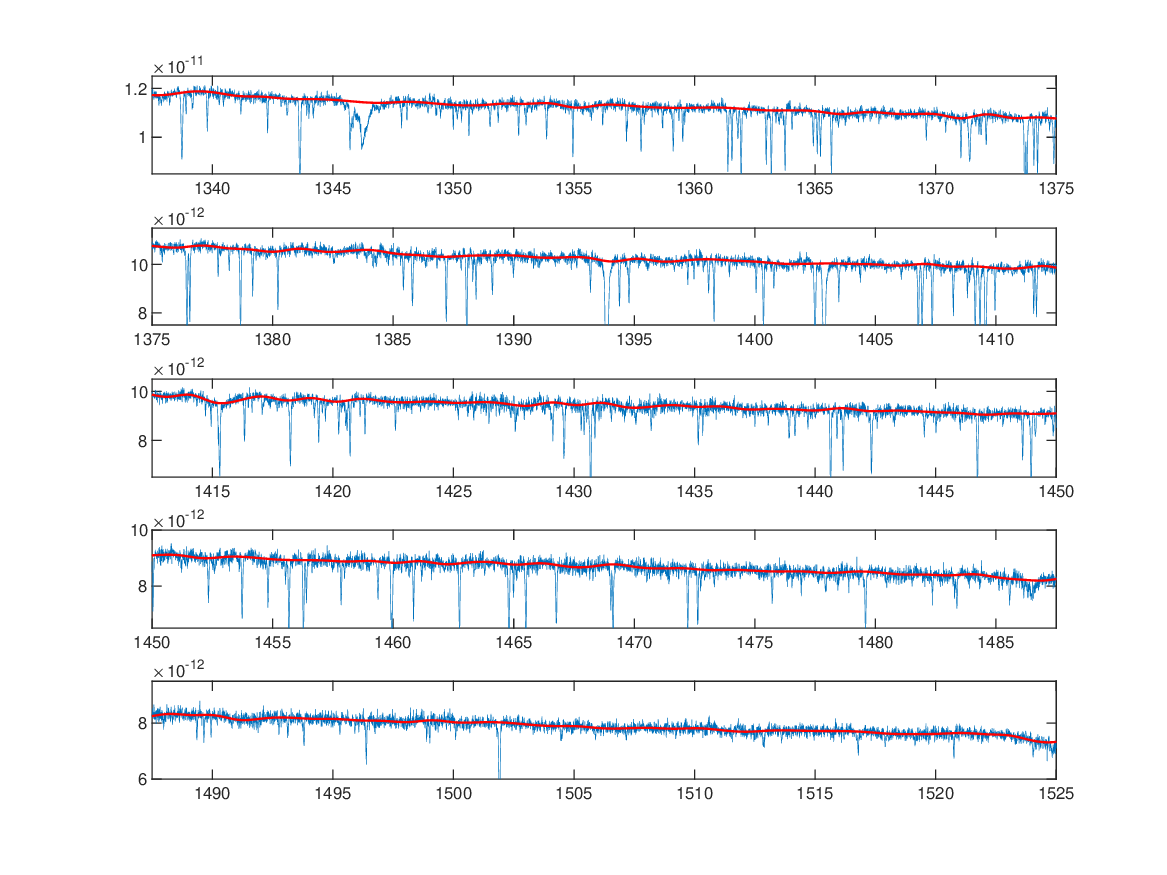}
\caption{G191$-$B2B continuum model with 1.0\,{\AA} knot spacing.
\label{fig:full100-1}
}
\end{figure}

\begin{figure}
\centering
\includegraphics[width=0.95\linewidth]{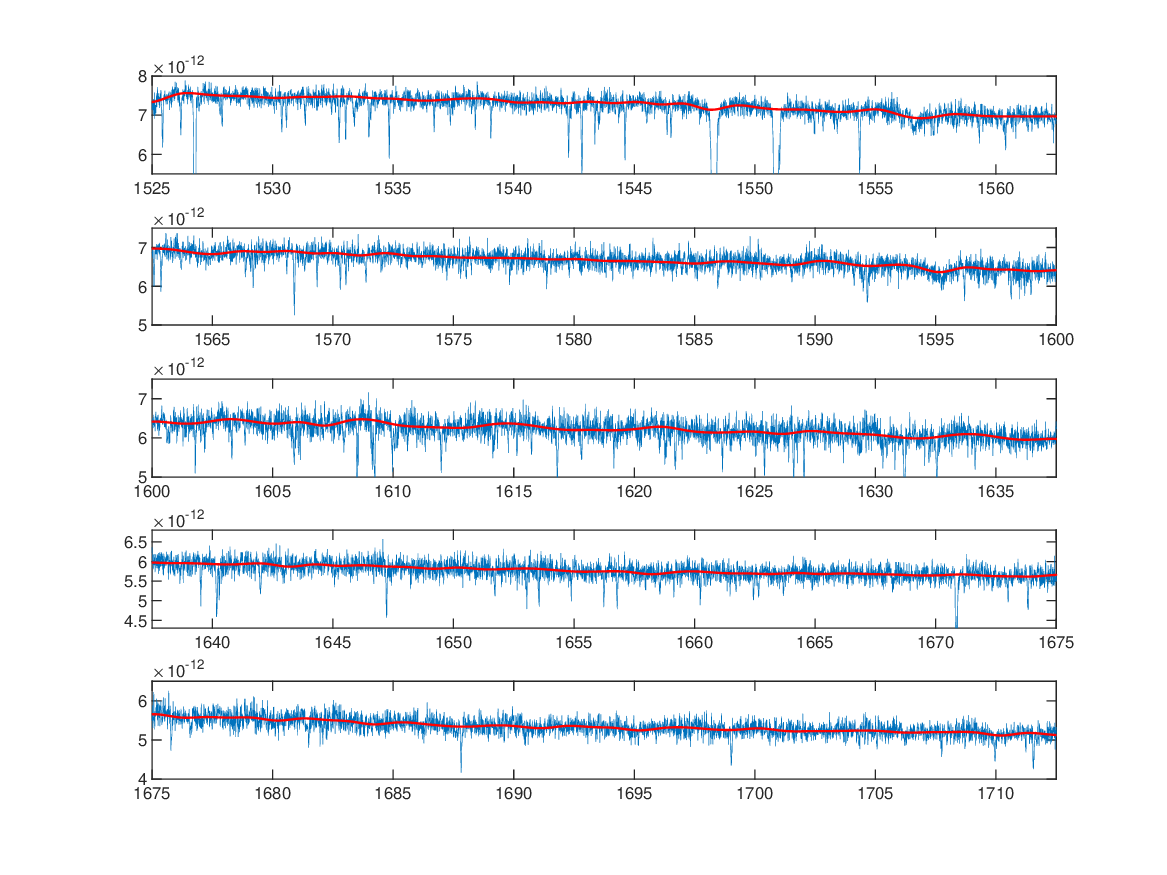}\\
\vspace{-1cm}
\includegraphics[width=0.95\linewidth]{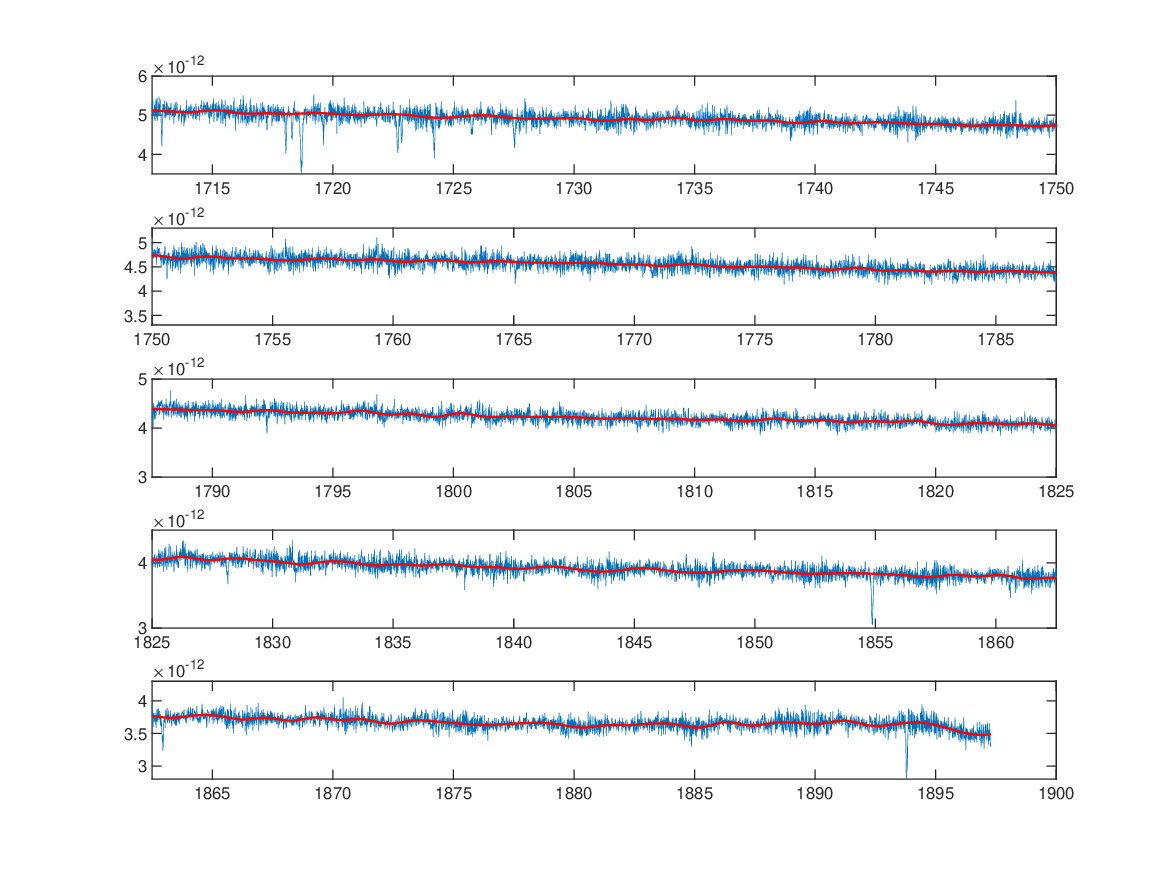}
\caption{G191$-$B2B continuum model with 1.0\,{\AA} knot spacing.
\label{fig:full100-2}
}
\end{figure}

\begin{figure}
\centering
\includegraphics[width=0.95\linewidth]{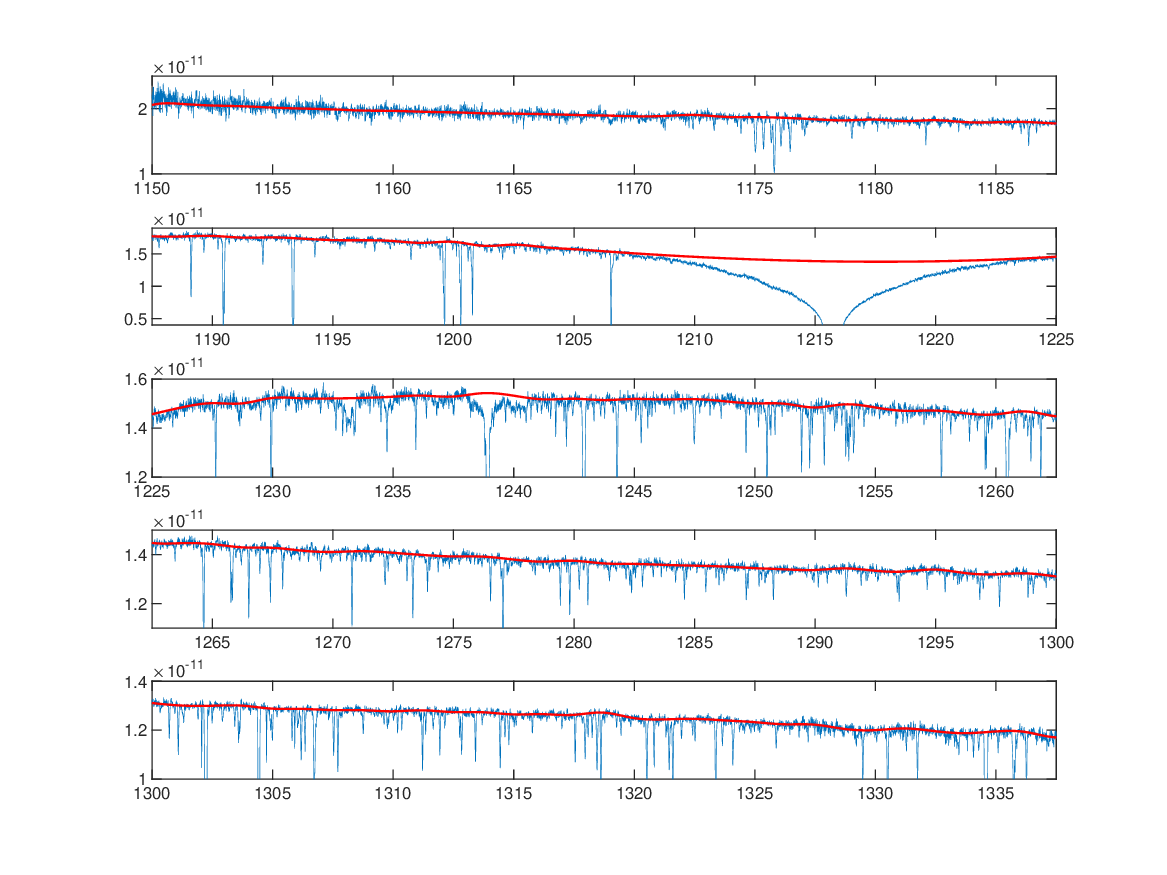}\\
\vspace{-1cm}
\includegraphics[width=0.95\linewidth]{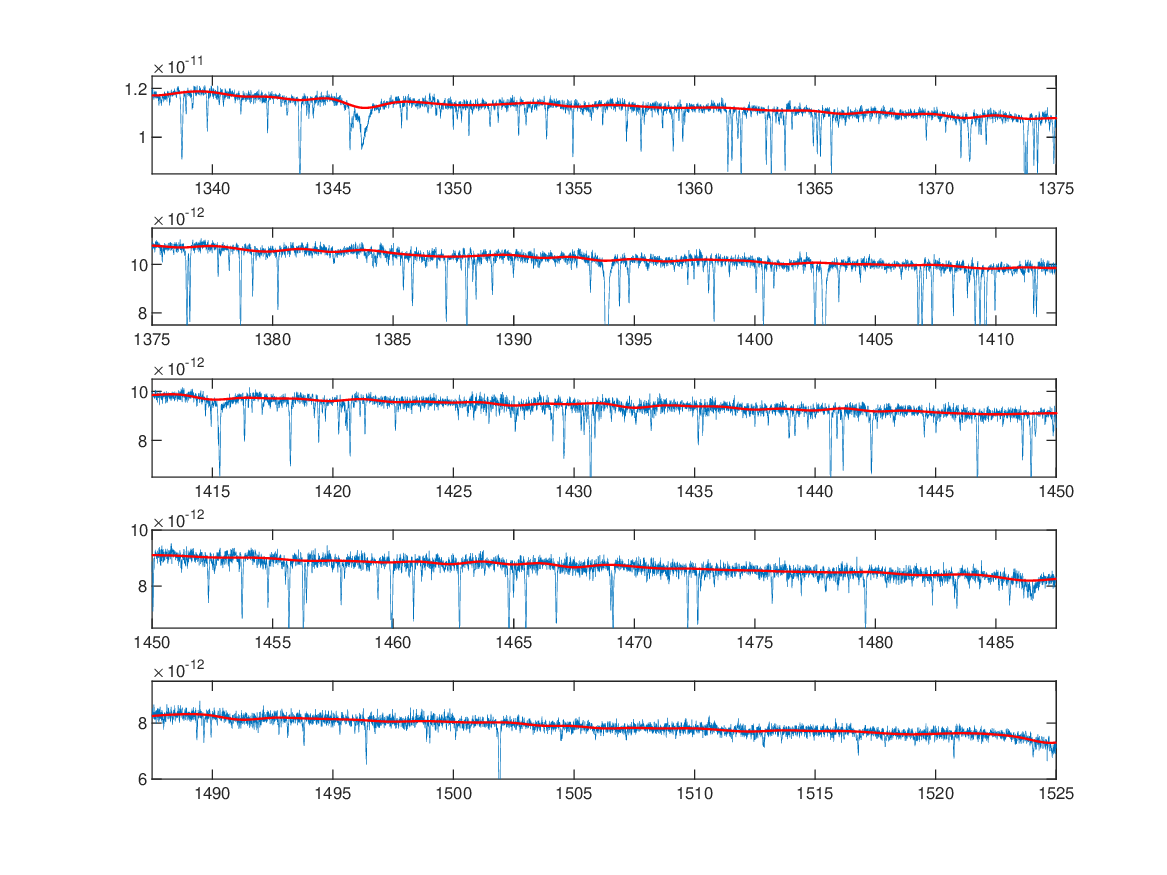}
\caption{G191$-$B2B continuum model with 1.25\,{\AA} knot spacing.
\label{fig:full125-1}
}
\end{figure}

\begin{figure}
\centering
\includegraphics[width=0.95\linewidth]{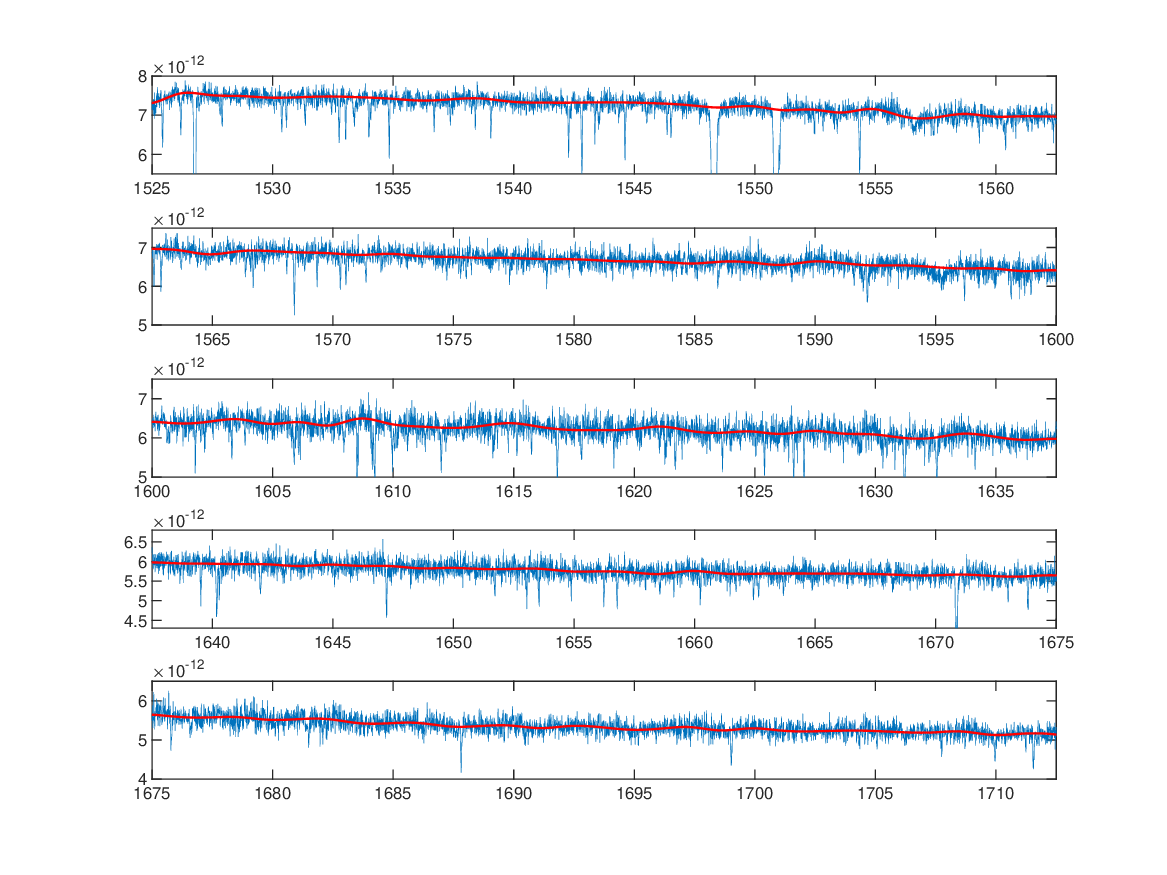}\\
\vspace{-1cm}
\includegraphics[width=0.95\linewidth]{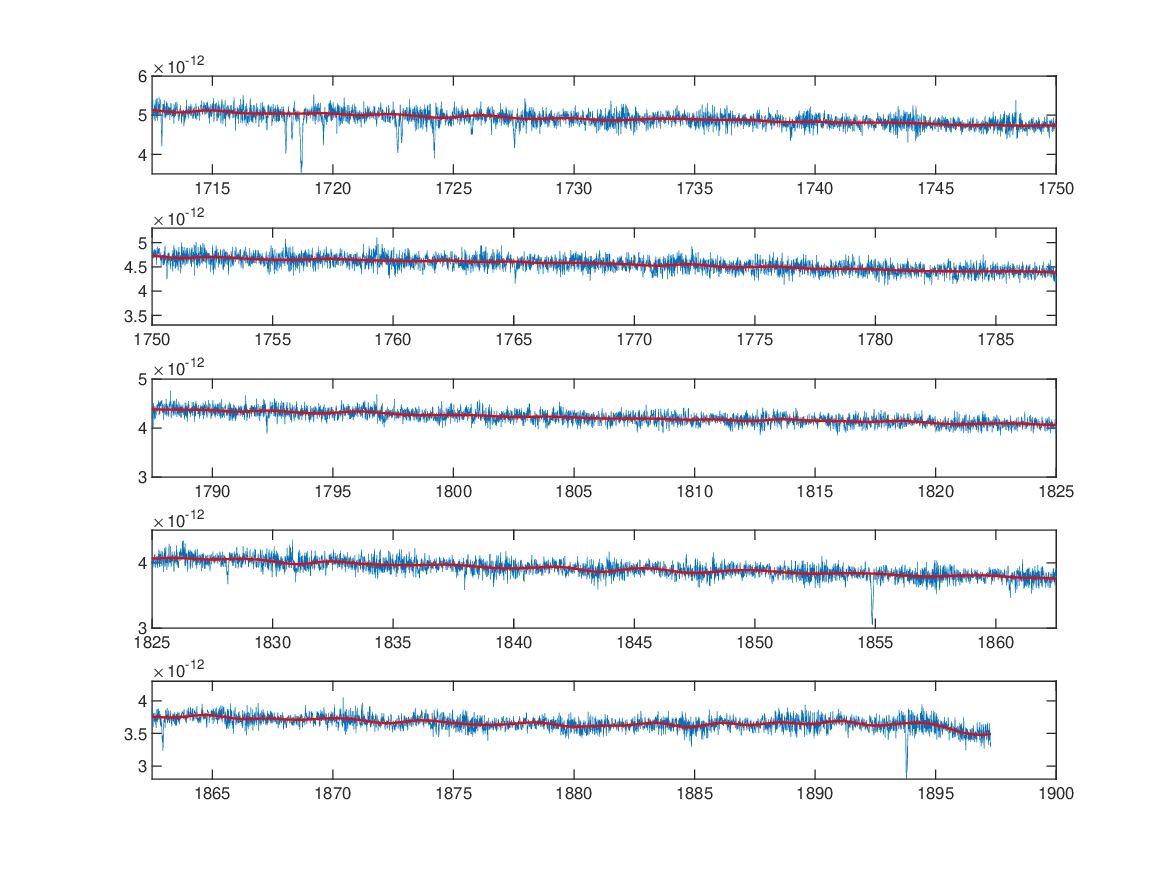}
\caption{G191$-$B2B continuum model with 1.25\,{\AA} knot spacing.
\label{fig:full125-2}
}
\end{figure}
\newpage

\section{Atomic data NiV} \label{sec:atomic}
Experimental and theoretical data on energy levels of Ni~V. For further details about the experimental data see \cite{NIST_ASD}. Apart from the energies, the theoretical data include calculated Land\'{e} $g$-factors and sensitivity coefficients $q$. The atomic data provided here for NiV is analogous to similar files for FeV provided as supplementary material in \cite{Hu2021}. For enquiries about the theoretical data, contact Vladimir Dzuba: \href{mailto:v.dzuba@unsw.edu.au}{v.dzuba@unsw.edu.au}.
\input{SMtables/niv}

\newpage

\section{Line IDs for each continuum model} \label{a:ids}

The columns in the tables in this Section are as follows: `lam\_K' is the laboratory wavelength in {\AA} from \cite{NIST_ASD} and `error' is its standard deviation in m{\AA}. `lam\_obs' is the observed wavelength in {\AA} and `error' is its standard deviation in m{\AA}. `q' is the sensitivity coefficient in cm$^{-1}$ (Appendix \ref{sec:atomic}). `lam\_rest' is the rest-frame observed wavelength in {\AA} (`lam\_obs' shifted to the rest frame using the G191-B2B redshift from \cite{Preval2013}). `match' is the absolute difference between the observed and laboratory rest-frame wavelengths in units of 1 standard deviation, allowing for uncertainties on both wavelengths in most cases, except where the uncertainty on the observed wavelength is very small (below 1m{\AA}, indicated as a '-' in the tables); in those cases, the observed wavelength uncertainty contribution has been ignored. Note that there are multiple potential IDs in many cases. These are left in the Table for illustration, but all multiples are discarded in the linear regression estimates. Further, by visual inspection of the data, the following four spectral ranges are excluded from the tables below and from the linear regression calculations: [1205, 1227], [1238,1241], [1242,1243.5], [1345,1347] {\AA}. 

We associate lower and upper energy levels and sensitivity coefficients using the tabulations provided in \cite{NIST_ASD} together with the $q$ values in Appendix \ref{sec:atomic} in the present work. As an illustration, the transition at Ritz wavelength 1202.427{\AA} \citep{NIST_ASD} has lower level configuration, term, and $J$: $3d^5\,(^4G)\,4s$, $^5G$, $4$, and upper level configuration, term, and $J$: $3d^5\,(^4G)\,4p$, $3F^{\circ}$, $3$. The associated sensitivities (from Tables \ref{t:NiVeven} and \ref{t:NiVodd}) are -6829 and -3939, such that the $q$ applicable in Eq.\,\ref{eq:mm3} is $q=+2890$ \citep{Dzuba1999a, Dzuba1999b}.

\subsection{0.5{\AA}}
\input{SMtables/G191_0.5.txt}
\newpage

\subsection{0.75{\AA}}
\input{SMtables/G191_0.75.txt}
\newpage

\subsection{1.0{\AA}}
\input{SMtables/G191_1.0.txt}
\newpage

\subsection{1.25{\AA}}
\input{SMtables/G191_1.25.txt}

\widetext

\end{document}

%% file: SMtables/niv.tex
\begingroup
\renewcommand*\arraystretch{2}
\begin{longtable}{|llcr | rrr|}
\caption{\raggedright Even energy levels.
\label{t:NiVeven}}\\
\hline
 \multicolumn{4}{|c|}{Experimental data} & \multicolumn{3}{c|}{Theoretical data} \\
Configuration & Term & $J$ & Energy      & Energy      & $g$~~~ & $q$~~~ \\
           &         &        & ~[cm$^{-1}$]~ & ~[cm$^{-1}$]~  &        & ~[cm$^{-1}$]~ \\
\hline
\endfirsthead
 \multicolumn{7}{c}%
 {\tablename\ \thetable\ -- \textit{Continued from previous page}} \\
\hline

\multicolumn{4}{|c|}{Experimental data \cite{NIST_ASD}} & \multicolumn{3}{c|}{Theoretical data} \\
Configuration & Term & $J$ & Energy      & Energy      & $g$~~~ & $q$~~~ \\
           &         &        & ~[cm$^{-1}$]~ & ~[cm$^{-1}$]~  &        & ~[cm$^{-1}$]~ \\
\hline
\endhead
 \hline \multicolumn{7}{r}{\textit{Continued on next page}} \\
 \endfoot
 \hline
 \endlastfoot
\phantom{()} &&&&&& \\
3d6\phantom{()}         &  5D  &  0 &     2057.6  &     1396 &  0.0000 &   1829 \\
3d6\phantom{()}         &  3P2 &  0 &    29640.0  &	   34626 &  0.0000 &   2557 \\
3d6\phantom{()}         &  1S2 &  0 &    47699.7  &	   51554 &  0.0000 &   1487 \\
3d6\phantom{()}         &  3P1 &  0 &    66737.8  &	   73326 &  0.0000 &   -940 \\
3d6\phantom{()}         &      &  0 &             &	  148275 &  0.0000 &    324 \\
3d5.(4D).4s &  5D  &  0 &   216305.7  &   211486 &  0.0000 &  -6971 \\
\phantom{()} 	    &      &  0 &  	      &	  215959 &  0.0000 &  -6250 \\
\phantom{()} 	    &      &  0 &  	      &   259128 &  0.0000 &  -6666 \\
\phantom{()} 	    &      &  0 &  	      &   302100 &  0.0000 &  -6966 \\
				          
3d6\phantom{()}          &  5D  &  1 &     1871.5 &    2516 &  1.4998 &   1691 \\
3d6\phantom{()}          &  3P2 &  1 &    28697.6 &   34535 &  1.4962 &   2209 \\
3d6\phantom{()}          &  3D  &  1 &    41701.1 &   46758 &  0.5049 &    629 \\
3d6\phantom{()}          &  3P1 &  1 &    67547.9 &   74931 &  1.4991 &   -168 \\
3d5.(4P).4s &  5P  &  1 &   212455.7 &  207163 &  2.4590 &  -7116 \\
3d5.(4D).4s &  5D  &  1 &   216434.7 &  211590 &  1.5371 &  -6657 \\
3d5.(4P).4s &  3P  &  1 &   221429.0 &  215767 &  1.4491 &  -6758 \\
3d5.(4D).4s &  3D  &  1 &   225545.1 &  220256 &  0.5474 &  -6088 \\
3d5.(2D3).4s&  3D  &  1 &   232910.8 &  229753 &  0.2174 &  -7827 \\
3d5.(4F).4s &  5F  &  1 &   235116.5 &  232189 &  0.2897 &  -5611 \\
3d5.(2S).4s &  3S  &  1 &   253905.2 &  254966 &  1.9996 &  -6825 \\
3d5.(2D2).4s&  3D  &  1 &   263700.9 &  264761 &  0.5003 &  -6878 \\
3d5.(2D1).4s&  3D  &  1 &   307105.1 &  302035 &  1.4955 &  -7063 \\
\phantom{()} 	    &      &  1 &  	     &  306249 &  0.9939 &  -7038 \\
\phantom{()} 	    &      &  1 & 	     &  313495 &  0.5112 &  -6765 \\
3d6\phantom{()}          & 5D  &  2 &     1489.9 &    1790 &  1.4995 &   1382 \\
3d6\phantom{()}          & 3P2 &  2 &    26153.0 &   30208 &  1.4965 &   -425 \\
3d6\phantom{()}          & 3F2 &  2 &    29899.2 &   33189 &  0.6701 &    933 \\
3d6\phantom{()}          & 3D  &  2 &    41626.9 &   45802 &  1.1653 &    433 \\
3d6\phantom{()}          & 1D2 &  2 &    48607.0 &   56120 &  1.0043 &   1085 \\
3d6\phantom{()}          & 3F1 &  2 &    68632.1 &   74836 &  0.6698 &    542 \\
3d6\phantom{()}          & 3P1 &  2 &    69156.1 &   76951 &  1.4946 &   1648 \\
3d6\phantom{()}          & 1D1 &  2 &   104420.5 &  114888 &  0.9999 &    460 \\
3d5.(6S).4s  & 5S  &  2 &   178019.8 &  165585 &  1.9991 &  -6212 \\
3d5.(4G).4s  & 5G  &  2 &   208151.5 &  200680 &  0.3353 &  -7018 \\
3d5.(4P).4s  & 5P  &  2 &   212253.4 &  206958 &  1.8033 &  -7509 \\
3d5.(4D).4s  & 5D  &  2 &   216590.5 &  211742 &  1.5257 &  -6305 \\
3d5.(4P).4s  & 3P  &  2 &   221087.6 &  215453 &  1.4642 &  -7412 \\
3d5.(4D).4s  & 3D  &  2 &   225616.5 &  220456 &  1.1951 &  -5872 \\
3d5.(2D3).4s & 3D  &  2 &   232655.6 &  229615 &  1.0277 &  -8247 \\
3d5.(4F).4s  & 5F  &  2 &   234412.7 &  231112 &  0.9074 &  -7170 \\
3d5.(2F1).4s & 3F  &  2 &   235736.5 &  232870 &  0.9695 &  -6243 \\
3d5.(2D3).4s & 1D  &  2 &   239107.7 &  236370 &  0.9117 &  -4938 \\
3d5.(4F).4s  & 3F  &  2 &   243266.2 &  239538 &  0.6925 &  -6121 \\
3d5.(2F2).4s & 3F  &  2 &   247165.0 &  245934 &  0.6697 &  -6562 \\
3d5.(2D2).4s & 3D  &  2 &   263735.7 &  265044 &  1.1658 &  -6783  \\
3d5.(2D2).4s & 1D  &  2 &   268273.9 &  269325 &  0.9996 &  -6576  \\
3d5.(2D1).4s & 3D  &  2 &   307025.2 &  302060 &  1.4971 &  -7180  \\
3d5.(2D1).4s & 1D  &  2 &   311470.3 &  313629 &  1.1682 &  -6917  \\
\phantom{()} 	     &     &  2 & 	     &  317855 &  1.0012 &  -6809  \\

3d6\phantom{()}           & 5D  &  3 &      889.7 &     1250 &  1.4992 &    896 \\
3d6\phantom{()}           & 3F2 &  3 &    29570.8 &    33285 &  1.0622 &    505 \\
3d6\phantom{()}           & 3G  &  3 &    34416.4 &    37773 &  0.7727 &   1769 \\
3d6\phantom{()}           & 3D  &  3 &    41920.2 &    46657 &  1.3319 &    848 \\
3d6\phantom{()}           & 1F  &  3 &    57924.1 &    64500 &  1.0033 &    293 \\
3d6\phantom{()}           & 3F1 &  3 &    68854.7 &    75646 &  1.0806 &   1183 \\
3d5.(6S).4s  & 7S  &  3 &   164525.9 &   152772 &  1.9995 &  -6678 \\
3d5.(4G).4s  & 5G  &  3 &   208164.6 &   200761 &  0.9175 &  -6922 \\
3d5.(4P).4s  & 5P  &  3 &   212095.8 &   206908 &  1.6488 &  -7762 \\
3d5.(4D).4s  & 5D  &  3 &   216596.0 &   209293 &  0.7523 &  -6702 \\
3d5.(4G).4s  & 3G  &  3 &   217101.0 &   211921 &  1.5126 &  -6230 \\
3d5.(4D).4s  & 3D  &  3 &   225200.7 &   219960 &  1.3318 &  -6610 \\
3d5.(2D3).4s & 3D  &  3 &   232545.9 &   229644 &  1.2580 &  -8398 \\
3d5.(4F).4s  & 5F  &  3 &   234275.2 &   230996 &  1.2337 &  -6944 \\
3d5.(2F1).4s & 3F  &  3 &   236454.1 &   233743 &  1.1427 &  -5430 \\
3d5.(2F1).4s & 1F  &  3 &   240193.8 &   237218 &  1.0167 &  -5882 \\
3d5.(2G2).4s & 3G  &  3 &   242290.4 &   239187 &  0.9910 &  -6611 \\
3d5.(4F).4s  & 3F  &  3 &   243370.5 &   240142 &  0.8601 &  -6183 \\
3d5.(2F2).4s & 3F  &  3 &   247104.9 &   245961 &  1.0846 &  -6647 \\
3d5.(2F2).4s & 1F  &  3 &   251654.9 &   250168 &  1.0023 &  -6475 \\
\phantom{()} &&&&&& \\
3d5.(2D2).4s & 3D  &  3 &   263805.8 &   265275 &  1.3313 &  -6590 \\
3d5.(2G1).4s & 3G  &  3 &   274773.5 &   276114 &  0.7506 &  -6784 \\
3d5.(2D1).4s & 3D  &  3 &   306962.9 &   313547 &  1.3333 &  -7018 \\

3d6\phantom{()}          &  5D  &  4 &        0.0 &      -0 &  1.4988  &     0  \\
3d6\phantom{()}          &  3H  &  4 &    27858.8 &   30010 &  0.8318  &   387  \\
3d6\phantom{()}          &  3F2 &  4 &    29123.7 &   32475 &  1.2046  &   183  \\
3d6\phantom{()}          &  3G  &  4 &    34061.7 &   36822 &  1.0612  &  1600  \\
3d6\phantom{()}          &  1G2 &  4 &    42208.1 &   45917 &  1.0037  &  1454  \\
3d6\phantom{()}          &  3F1 &  4 &    68718.7 &   74766 &  1.2476  &   636  \\
3d6\phantom{()}          &  1G1 &  4 &    77899.5 &   85539 &  1.0023  &   773 \\
3d5.(4G).4s &  5G  &  4 &   208163.7 &  200835 &  1.1503  & -6829 \\
3d5.(4D).4s &  5D  &  4 &   216189.9 &  209318 &  1.0509  & -6518 \\
3d5.(4G).4s &  3G  &  4 &   217129.1 &  211531 &  1.4982  & -6927 \\
3d5.(4F).4s &  5F  &  4 &   234125.4 &  230800 &  1.3452  & -7025 \\
3d5.(2F1).4s&  3F  &  4 &   235420.6 &  232594 &  1.2505  & -6581 \\
3d5.(2H).4s &  3H  &  4 &   240959.6 &  236976 &  0.8335  & -7430 \\
3d5.(2G2).4s&  3G  &  4 &   242504.3 &  238958 &  1.1901  & -6927 \\
3d5.(4F).4s &  3F  &  4 &   243331.5 &  240278 &  1.0715  & -6057 \\
3d5.(2G2).4s&  1G  &  4 &   247049.1 &  244341 &  1.0094  & -5918   \\
3d5.(2F2).4s&  3F  &  4 &   247281.8 &  245917 &  1.2503  & -6592   \\
3d5.(2G1).4s&  3G  &  4 &   274738.6 &  276093 &  1.0501  & -6822   \\
3d5.(2G1).4s&  1G  &  4 &   279199.5 &  280311 &  1.0002  & -6684   \\
3d6 \phantom{()}         &  3H  &  5 &    27578.2 &    29506&   1.0404 &    503  \\
3d6 \phantom{()}         &  3G  &  5 &    33256.5 &    36405&   1.1929 &    768  \\
3d5.(4G).4s &  5G  &  5 &   208131.0 &   200830&   1.2666 &  -6770  \\
3d5.(4G).4s &  3G  &  5 &   217048.7 &   209359&   1.1996 &  -6483  \\
3d5.(2I).4s &  3I  &  5 &   229413.0 &   222780&   0.8355 &  -7026 \\
3d5.(4F).4s &  5F  &  5 &   234082.1 &   230750&   1.3953 &  -7077 \\
3d5.(2H).4s &  3H  &  5 &   241082.2 &   237200&   1.0521 &  -7095 \\
3d5.(2G2).4s&  3G  &  5 &   242862.6 &   239964&   1.1575 &  -6323 \\
3d5.(2H).4s &  1H  &  5 &   246240.9 &   242144&   1.0268 &  -5814 \\
3d5.(2G1).4s&  3G  &  5 &   274695.4 &   276128&   1.2000 &  -6866 \\
3d6\phantom{()}         &  3H  &  6 &    27111.2 &    28827&   1.1653 &    279 \\
3d6\phantom{()}         &  1I  &  6 &    41252.2 &    43790&   1.0014 &    945 \\
3d5.(4G).4s &  5G  &  6 &   208046.4 &   200881&   1.3330 &  -6792 \\
3d5.(2I).4s &  3I  &  6 &   229408.8 &   222883&   1.0250 &  -6908 \\
3d5.(2I).4s &  1I  &  6 &   233839.2 &   227172&   1.0021 &  -6762 \\
3d5.(2H).4s &  3H  &  6 &   241773.6 &   237776&   1.1638 &  -6164 \\
\end{longtable}
 
\begin{longtable}{|llcr | rrr|}
\caption{\label{t:NiVodd} \raggedright Odd energy levels.
} \\
\hline
 \multicolumn{4}{|c|}{Experimental data} & \multicolumn{3}{c|}{Theoretical data} \\
Configuration & Term & $J$ & Energy      & Energy      & $g$~~~ & $q$~~~ \\
           &         &        & ~[cm$^{-1}$]~ & ~[cm$^{-1}$]~  &        & ~[cm$^{-1}$]~ \\
\hline
\endfirsthead
 \multicolumn{7}{c}%
 {\tablename\ \thetable\ -- \textit{Continued from previous page}} \\
\hline

\multicolumn{4}{|c|}{Experimental data \cite{NIST_ASD}} & \multicolumn{3}{c|}{Theoretical data} \\
Configuration & Term & $J$ & Energy      & Energy      & $g$~~~ & $q$~~~ \\
\phantom{()}            &         &        & ~[cm$^{-1}$]~ & ~[cm$^{-1}$]~  &        & ~[cm$^{-1}$]~ \\
\hline
\endhead
 \hline \multicolumn{7}{r}{\textit{Continued on next page}} \\
 \endfoot
 \hline
 \endlastfoot

\phantom{()} &&&&&& \\
3d5.(4P).4p  &   5D* &  0 &  290262.0 & 281464 &  0.0000 &  -5470 \\
3d5.(4P).4p  &   3P* &  0 &  293867.0 & 287790 &  0.0000 &  -3601 \\
3d5.(4D).4p  &   5D* &  0 &  298060.0 & 291969 &  0.0000 &  -2663 \\
3d5.(4D).4p  &   3P* &  0 &  305386.9 & 299362 &  0.0000 &  -4301 \\
3d5.(2D3).4p &   3P* &  0 &  313577.3 & 309703 &  0.0000 &  -3418 \\
3d5.(4F).4p  &   5D* &  0 &  317462.3 & 313175 &  0.0000 &  -2326 \\
3d5.(2S).4p  &   3P* &  0 &  329618.5 & 329372 &  0.0000 &  -5126 \\
3d5.(2D2).4p &   3P* &  0 &  346920.2 & 347460 &  0.0000 &  -3490 \\
3d5.(2P).4p  &   3P* &  0 &  368440.5 & 373470 &  0.0000 &  -4946 \\
\phantom{()}              &       &    &           & 378828 &  0.0000 &  -3974 \\
\phantom{()}              &       &    &           & 399017 &  0.0000 &  -2695 \\
                               
3d5.(6S).4p   &  5P* &  1 &  254885.0  &  241377 &  2.4987 &  -2708 \\
3d5.(4P).4p   &  5D* &  1 &  287755.5  &  280878 &  0.2950 &  -3976 \\
3d5.(4G).4p   &  5F* &  1 &  289163.0  &  281723 &  1.1991 &  -4762 \\
3d5.(4P).4p   &  5P* &  1 &  291541.7  &  285184 &  2.3645 &  -3997 \\
3d5.(4P).4p   &  3P* &  1 &  293420.0  &  287310 &  1.4579 &  -3978 \\
3d5.(4D).4p   &  5F* &  1 &  293833.8  &  287628 &  0.1650 &  -4603 \\
3d5.(4D).4p   &  5D* &  1 &  297417.9  &  291341 &  1.3386 &  -3715 \\
3d5.(4D).4p   &  5P* &  1 &  297982.8  &  291883 &  1.1669 &  -3649 \\
3d5.(4P).4p   &    * &  1 &  298600.6  &  292692 &  1.9791 &  -3134 \\
3d5.(4D).4p   &  3D* &  1 &  300563.3  &  294473 &  0.5265 &  -2497 \\
3d5.(4P).4p   &  3S* &  1 &  303249.5  &  296357 &  1.9601 &  -3800 \\
3d5.(4D).4p    & 3P* &  1 &  305838.1  &  299924 &  1.5284 &  -3576 \\
3d5.(2D3).4p   &   * &  1 &  312291.0  &  308444 &  0.9373 &  -5157 \\
3d5.(2D3).4p   &   * &  1 &  313679.0  &  309782 &  0.8141 &  -3762 \\
3d5.(4F).4p    & 5F* &  1 &  315152.8  &  310866 &  0.3252 &  -3727 \\
3d5.(2F1).4p   & 3D* &  1 &  315300.7  &  311524 &  0.7944 &  -4032 \\
3d5.(4F).4p    & 5D* &  1 &  317477.9  &  313240 &  1.2865 &  -2263 \\
3d5.(2D3).4p   & 1P* &  1 &  319073.4  &  315355 &  0.8106 &  -1346 \\
3d5.(4F).4p    & 3D* &  1 &  322984.5  &  318413 &  0.5500 &  -2697 \\
3d5.(2F2).4p   & 3D* &  1 &  329462.3  &  326703 &  0.5249 &  -3574 \\
3d5.(2S).4p    & 3P* &  1 &  330370.7  &  329930 &  1.4656 &  -4581 \\
\phantom{()} &&&&&& \\
3d5.(2S).4p    & 1P* &  1 &  334477.2  &  334712 &  1.0120 &  -3450 \\
3d5.(2D2).4p   & 3D* &  1 &  343478.2  &  343245 &  0.5459 &  -4549 \\
3d5.(2D2).4p   & 3P* &  1 &  346959.5  &  347592 &  1.4304 &  -3410 \\
3d5.(2D2).4p   & 1P* &  1 &  348477.9  &  348816 &  1.0233 &  -3251 \\
3d5.(2P).4p    & 3P* &  1 &  368749.7  &  373807 &  1.5033 &  -4797 \\
3d5.(2P).4p    & 3D* &  1 &  374828.1  &  379609 &  0.5133 &  -5064 \\
3d5.(2P).4p    & 3S* &  1 &  378555.0  &  383711 &  1.6646 &  -3986 \\
3d5.(2P).4p    & 1P* &  1 &  380165.6  &  384557 &  1.3169 &  -3363 \\
3d5.(2D1).4p   & 3D* &  1 &  388746.1  &  394103 &  0.5120 &  -4887 \\
\phantom{()}                &     &    &            &  398508 &  1.4944 &  -3300 \\
\phantom{()}                &     &    &            &  404010 &  0.9956 &  -3522 \\
                                                                    
3d5.(6S).4p   & 5P*  & 2 &  254495.6  &  229894 &  2.3277 &  -4828 \\
3d5.(4G).4p   & 5G*  & 2 &  284215.5  &  241014 &  1.8375 &  -3014 \\
3d5.(4P).4p   & 5D*  & 2 &  287782.1  &  276075 &  0.3521 &  -4786 \\
3d5.(4P).4p   &   *  & 2 &  288877.9  &  280657 &  1.0598 &  -4086 \\
3d5.(4G).4p   & 5F*  & 2 &  289247.1  &  282019 &  1.4292 &  -4927 \\
3d5.(4G).4p   & 3F*  & 2 &  291097.7  &  282941 &  0.9106 &  -4043 \\
3d5.(4P).4p   & 5P*  & 2 &  291390.0  &  283078 &  1.6798 &  -4724 \\
3d5.(4P).4p   & 3P*  & 2 &  292983.0  &  285114 &  1.7959 &  -3535 \\
3d5.(4D).4p   & 5F*  & 2 &  294086.0  &  286909 &  1.5530 &  -4015 \\
3d5.(4D).4p   & 5D*  & 2 &  297013.9  &  287948 &  1.0263 &  -4325 \\
3d5.(4P).4p   & 3D*  & 2 &  297842.5  &  290869 &  1.4007 &  -4204 \\
3d5.(4D).4p   & 5P*  & 2 &  299045.6  &  291690 &  1.3451 &  -3841 \\
3d5.(4D).4p   & 3D*  & 2 &  300224.9  &  293244 &  1.6688 &  -2780 \\
3d5.(4D).4p   & 3F*  & 2 &  301553.0  &  294134 &  1.2039 &  -2872 \\
3d5.(4D).4p   & 3P*  & 2 &  306377.8  &  295583 &  0.7273 &  -2664 \\
3d5.(2D3).4p  & 3F*  & 2 &  307731.1  &  300678 &  1.4975 &  -3224 \\
3d5.(2D3).4p  &   *  & 2 &  308943.0  &  303830 &  0.7770 &  -5966 \\
3d5.(2D3).4p  & 3P*  & 2 &  311966.5  &  305575 &  0.9302 &  -5128 \\
3d5.(4F).4p   & 5G*  & 2 &  312778.2  &  307992 &  0.7962 &  -5100 \\
3d5.(2D3).4p  & 3D*  & 2 &  313686.6  &  308407 &  0.9480 &  -4171 \\
3d5.(4F).4p   & 5F*  & 2 &  314834.7  &  309715 &  1.0745 &  -4288 \\
3d5.(2F1).4p  &   *  & 2 &  315366.1  &  310696 &  1.1579 &  -3761 \\
3d5.(2F1).4p  &   *  & 2 &  316165.4  &  311537 &  0.7948 &  -2882 \\
3d5.(4F).4p   & 5D*  & 2 &  317517.5  &  312447 &  1.2081 &  -2480 \\
3d5.(2F1).4p  & 1D*  & 2 &  319926.5  &  313434 &  1.3131 &  -1661 \\
3d5.(2G2).4p  & 3F*  & 2 &  321018.3  &  316167 &  0.9754 &  -3391 \\
3d5.(4F).4p   & 3D*  & 2 &  322436.4  &  317360 &  0.7783 &  -3505 \\
3d5.(4F).4p   & 3F*  & 2 &  323853.1  &  318193 &  1.0621 &  -3368 \\
3d5.(2F2).4p  & 3F*  & 2 &  325982.2  &  319793 &  0.6943 &  -2601 \\
3d5.(2F2).4p  & 1D*  & 2 &  327122.7  &  323571 &  0.7952 &  -4036 \\
3d5.(2F2).4p  & 3D*  & 2 &  329776.3  &  325064 &  0.8932 &  -2819 \\
3d5.(2S).4p   & 3P*  & 2 &  331678.2  &  326810 &  1.1605 &  -3332 \\
3d5.(2D2).4p  & 3F*  & 2 &  342894.6  &  331480 &  1.4933 &  -2963 \\
3d5.(2D2).4p  & 3D*  & 2 &  343905.7  &  342876 &  0.7859 &  -5049 \\
3d5.(2D2).4p  & 3P*  & 2 &  346912.4  &  343840 &  1.0749 &  -4135 \\
3d5.(2D2).4p  & 1D*  & 2 &  349546.0  &  347523 &  1.4455 &  -3437 \\
3d5.(2G1).4p  & 3F*  & 2 &  355150.0  &  349161 &  1.0256 &  -3122 \\
3d5.(2P).4p   & 3P*  & 2 &  369649.1  &  355824 &  0.6675 &  -2843 \\
3d5.(2P).4p   & 3D*  & 2 &  374803.7  &  374877 &  1.4917 &  -3844 \\
3d5.(2P).4p   & 1D*  & 2 &  377059.1  &  379589 &  1.1140 &  -5230 \\
3d5.(2D1).4p  & 3F*  & 2 &  386968.8  &  382070 &  1.0570 &  -3051 \\
3d5.(2D1).4p  & 3D*  & 2 &  389571.8  &  392115 &  0.7063 &  -5044 \\
3d5.(2D1).4p  & 1D*  & 2 &  390675.1  &  395085 &  1.1438 &  -3946 \\
3d5.(2D1).4p  & 3P*  & 2 &  392413.5  &  396028 &  1.0955 &  -4338 \\
3d5.(6S).5p   & 5P*  & 2 &  423782    &  397840 &  1.3919 &  -3591 \\

3d5.(6S).4p   & 7P*  & 3 &  243608.5  &  230698 &  1.9118 &  -4119 \\
3d5.(6S).4p   & 5P*  & 3 &  253862.7  &  240404 &  1.6705 &  -3569 \\
3d5.(4G).4p   & 5G*  & 3 &  284249.0  &  276109 &  0.8765 &  -4826 \\
3d5.(4G).4p   & 5H*  & 3 &  286293.6  &  277563 &  0.5556 &  -4806 \\
3d5.(4G).4p   &   *  & 3 &  287960.0  &  280438 &  1.2565 &  -4072 \\
3d5.(4G).4p   & 5F*  & 3 &  289298.0  &  282481 &  1.4906 &  -4859 \\
3d5.(4P).4p   & 5P*  & 3 &  290757.0  &  283221 &  1.0889 &  -3425 \\
3d5.(4G).4p   & 3F*  & 3 &  291328.5  &  284588 &  1.6263 &  -3939 \\
3d5.(4D).4p   & 5F*  & 3 &  294443.3  &  288126 &  0.7528 &  -3554 \\
3d5.(4D).4p   & 5D*  & 3 &  296574.0  &  288491 &  1.2649 &  -3921 \\
3d5.(4G).4p   & 3G*  & 3 &  296847.1  &  290332 &  1.4271 &  -4843 \\
3d5.(4P).4p   & 3D*  & 3 &  297418.1  &  291221 &  1.4131 &  -4357 \\
3d5.(4D).4p   & 3D*  & 3 &  298972.3  &  293092 &  1.4568 &  -3460 \\
3d5.(4D).4p   &   *  & 3 &  300201.0  &  294296 &  1.4822 &  -2654 \\
3d5.(4D).4p   & 3F*  & 3 &  301470.2  &  295588 &  1.1320 &  -2604 \\
3d5.(2D3).4p  & 3F*  & 3 &  308592.0  &  304514 &  1.0407 &  -5969 \\
3d5.(2F1).4p  & 3G*  & 3 &  312463.3  &  308099 &  0.9539 &  -5016 \\
3d5.(4F).4p   & 5G*  & 3 &  312889.4  &  308350 &  0.8926 &  -4790 \\
3d5.(2F1).4p  &   *  & 3 &  312953.6  &  309183 &  1.2718 &  -4865 \\
3d5.(2F1).4p  &   *  & 3 &  313919.8  &  310109 &  1.3019 &  -4626 \\
3d5.(4F).4p   & 5F*  & 3 &  314562.8  &  310586 &  1.2231 &  -4363 \\
3d5.(2F1).4p  &   *  & 3 &  315326.2  &  311337 &  0.9816 &  -2939 \\
\phantom{()} &&&&&& \\
3d5.(2F1).4p  &   *  & 3 &  316280.3  &  312453 &  1.0733 &  -2638 \\
3d5.(4F).4p   & 5D*  & 3 &  317232.0  &  312947 &  1.1158 &  -2667 \\
3d5.(4F).4p   &   *  & 3 &  317376.8  &  313303 &  1.1083 &  -1828 \\
3d5.(4F).4p   & 3G*  & 3 &  319620.2  &  315318 &  0.7914 &  -3194 \\
3d5.(2G2).4p  & 3F*  & 3 &  320513.8  &  316580 &  1.0024 &  -3333 \\
3d5.(2F1).4p  & 1F*  & 3 &  321081.9  &  317033 &  1.1152 &  -3815 \\
3d5.(4F).4p   & 3D*  & 3 &  322617.6  &  318370 &  1.2656 &  -3181 \\
3d5.(4F).4p   & 3F*  & 3 &  323532.2  &  319057 &  1.0834 &  -3476 \\
3d5.(2G2).4p  & 3G*  & 3 &  325211.9  &  321602 &  0.7825 &  -2976 \\
3d5.(2F2).4p  & 3F*  & 3 &  326029.9  &  323176 &  1.0058 &  -2709 \\
3d5.(2G2).4p  & 1F*  & 3 &  326739.0  &  323938 &  1.0686 &  -3747 \\
3d5.(2F2).4p  & 3G*  & 3 &  329614.3  &  326076 &  0.8248 &  -4116 \\
3d5.(2F2).4p  & 3D*  & 3 &  329872.9  &  326723 &  1.2729 &  -3339 \\
3d5.(2F2).4p  & 1F*  & 3 &  334727.6  &  332526 &  1.0048 &  -3182 \\
3d5.(2D2).4p  & 3F*  & 3 &  343281.0  &  343388 &  1.1491 &  -4646 \\
3d5.(2D2).4p  & 3D*  & 3 &  344805.3  &  344753 &  1.2438 &  -2962 \\
3d5.(2D2).4p  & 1F*  & 3 &  345936.1  &  346185 &  1.0196 &  -3164 \\
3d5.(2G1).4p  & 3F*  & 3 &  353944.1  &  354411 &  0.9271 &  -4699 \\
3d5.(2G1).4p  & 3G*  & 3 &  355398.0  &  355868 &  0.9137 &  -3176 \\
3d5.(2G1).4p  & 1F*  & 3 &  360059.7  &  360180 &  0.9941 &  -3387 \\
3d5.(2P).4p   & 3D*  & 3 &  376471.6  &  381384 &  1.3320 &  -3382 \\
3d5.(2D1).4p  & 3F*  & 3 &  387333.4  &  392545 &  1.1028 &  -4947 \\
3d5.(2D1).4p  & 3D*  & 3 &  390478.2  &  396043 &  1.3067 &  -2888 \\
3d5.(2D1).4p  & 1F*  & 3 &  392957.1  &  397826 &  1.0084 &  -3427 \\

3d5.(6S).4p   & 7P*  & 4 &  244900.5  &  231980  & 1.7496 &  -2661 \\
3d5.(4G).4p   & 5G*  & 4 &  284308.9  &  276164  & 1.1113 &  -4773 \\
3d5.(4G).4p   & 5H*  & 4 &  286706.6  &  278070  & 0.9487 &  -4203 \\
3d5.(4G).4p   & 5F*  & 4 &  288161.6  &  280235  & 1.3385 &  -3939 \\
3d5.(4G).4p   & 3F*  & 4 &  291554.6  &  283505  & 1.2629 &  -3200 \\
3d5.(4G).4p   & 3H*  & 4 &  292631.0  &  283699  & 0.8043 &  -3264 \\
3d5.(4D).4p   & 5F*  & 4 &  294939.6  &  284038  & 1.4715 &  -3334 \\
3d5.(4G).4p   & 3G*  & 4 &  296897.0  &  288262  & 1.0531 &  -3357 \\
3d5.(4D).4p   & 5D*  & 4 &  296919.3  &  289118  & 1.3559 &  -3330 \\
3d5.(4D).4p   & 3F*  & 4 &  300918.1  &  290810  & 1.4875 &  -4232 \\
3d5.(2I).4p   & 3H*  & 4 &  309952.5  &  295022  & 1.2618 &  -3416 \\
3d5.(2D3).4p  & 3F*  & 4 &  310212.6  &  302913  & 0.8060 &  -3389 \\
3d5.(2F1).4p  & 1G*  & 4 &  312008.3  &  306255  & 1.2059 &  -4108 \\
3d5.(4F).4p   & 5G*  & 4 &  313281.3  &  308059  & 1.0640 &  -4996 \\
3d5.(2F1).4p  &   *  & 4 &  314208.8  &  308751  & 1.1121 &  -4469 \\
3d5.(4F).4p   & 5F*  & 4 &  314599.2  &  309768  & 1.2790 &  -4390 \\
3d5.(2G2).4p  & 3H*  & 4 &  315370.1  &  310292  & 1.2267 &  -3672 \\
3d5.(2D3).4p  &   *  & 4 &  316068.8  &  310978  & 0.8805 &  -4816 \\
3d5.(4F).4p   & 5D*  & 4 &  316744.0  &  311935  & 1.0965 &  -2639 \\
3d5.(2H).4p   &   *  & 4 &  316887.8  &  312339  & 1.4220 &  -3314 \\
3d5.(2G2).4p  &   *  & 4 &  319138.7  &  312830  & 1.0755 &  -2983 \\
3d5.(4F).4p   & 3G*  & 4 &  319899.1  &  315218  & 1.0526 &  -3272 \\
3d5.(2G2).4p  & 3F*  & 4 &  321056.4  &  315252  & 1.0822 &  -3461 \\
3d5.(4F).4p   & 3F*  & 4 &  322820.8  &  317653  & 1.1721 &  -2969 \\
3d5.(2H).4p   & 3H*  & 4 &  323926.3  &  318562  & 1.2436 &  -3761 \\
3d5.(2G2).4p  & 3G*  & 4 &  325222.9  &  319411  & 0.8302 &  -3839 \\
3d5.(2F2).4p  &   *  & 4 &  325558.6  &  321537  & 1.0597 &  -2853 \\
3d5.(2F2).4p  & 3F*  & 4 &  326876.3  &  321989  & 1.0144 &  -3604 \\
3d5.(2F2).4p  & 3G*  & 4 &  330297.6  &  324418  & 1.2174 &  -3042 \\
3d5.(2H).4p   & 1G*  & 4 &  332995.6  &  326849  & 1.0624 &  -3158 \\
3d5.(2D2).4p  & 3F*  & 4 &  344911.2  &  329270  & 1.0025 &  -3365 \\
3d5.(2G1).4p  & 3H*  & 4 &  353071.6  &  345121  & 1.2487 &  -2780 \\
3d5.(2G1).4p  & 3F*  & 4 &  353347.1  &  353117  & 0.8394 &  -5094 \\
3d5.(2G1).4p  & 3G*  & 4 &  355765.2  &  353776  & 1.1998 &  -4993 \\
3d5.(2G1).4p  & 1G*  & 4 &  358760.0  &  356094  & 1.0564 &  -3153 \\
3d5.(2D1).4p  & 3F*  & 4 &  388698.9  &  358898  & 1.0052 &  -3293 \\
\phantom{()}               &      & 4 &            &  393842  & 1.2500 &  -3205 \\

3d5.(4G).4p   & 5G*  & 5 &  284402.5  &  276314  & 1.2360 &  -4635 \\
3d5.(4G).4p   & 5H*  & 5 &  287127.2  &  278555  & 1.1406 &  -3702 \\
3d5.(4G).4p   & 5F*  & 5 &  287906.9  &  279860  & 1.3839 &  -4154 \\
3d5.(4G).4p   & 3H*  & 5 &  292353.4  &  283513  & 1.0367 &  -3287 \\
3d5.(4D).4p   & 5F*  & 5 &  295444.3  &  288318  & 1.2024 &  -3264 \\
3d5.(4G).4p   & 3G*  & 5 &  296932.9  &  289653  & 1.3986 &  -2900 \\
3d5.(2I).4p   & 3I*  & 5 &  306049.0  &  298775  & 0.8810 &  -5436 \\
3d5.(2I).4p   & 1H*  & 5 &  308804.1  &  301829  & 0.9656 &  -3298 \\
3d5.(2I).4p   & 3H*  & 5 &  309919.5  &  302843  & 1.0266 &  -3045 \\
3d5.(4F).4p   & 5G*  & 5 &  313464.7  &  308902  & 1.2535 &  -4880 \\
3d5.(2F1).4p  & 3G*  & 5 &  314702.2  &  310339  & 1.2514 &  -3228 \\
3d5.(4F).4p   & 5F*  & 5 &  315168.2  &  310829  & 1.3139 &  -3418 \\
3d5.(2H).4p   & 3H*  & 5 &  315990.5  &  311510  & 1.0211 &  -4600 \\
3d5.(2H).4p   & 3G*  & 5 &  316726.6  &  312377  & 1.1978 &  -3475 \\
3d5.(2H).4p   & 3I*  & 5 &  319076.2  &  313799  & 0.8816 &  -3940 \\
3d5.(4F).4p   & 3G*  & 5 &  319652.7  &  314973  & 1.2030 &  -3334 \\
3d5.(2G2).4p  & 3H*  & 5 &  323908.6  &  319610  & 1.0479 &  -3862 \\
\phantom{()} &&&&&& \\
3d5.(2G2).4p  & 3G*  & 5 &  324980.2  &  320884  & 1.0781 &  -3251 \\
3d5.(2G2).4p  & 1H*  & 5 &  326337.1  &  321315  & 1.1031 &  -3224 \\
3d5.(2H).4p   & 1H*  & 5 &  327356.6  &  322781  & 1.0108 &  -2583 \\
3d5.(2F2).4p  & 3G*  & 5 &  330718.1  &  327319  & 1.1995 &  -2853 \\
3d5.(2G1).4p  & 3H*  & 5 &  353548.7  &  353578  & 1.0513 &  -4887 \\
3d5.(2G1).4p  & 3G*  & 5 &  356036.3  &  356255  & 1.1714 &  -2970 \\
3d5.(2G1).4p  & 1H*  & 5 &  358475.6  &  358251  & 1.0110 &  -3151 \\

3d5.(4G).4p   & 5G*  & 6 &  284579.5  &  276594  & 1.3127 &  -4347 \\
3d5.(4G).4p   & 5H*  & 6 &  287645.9  &  279047  & 1.2302 &  -3005 \\
3d5.(4G).4p   & 3H*  & 6 &  291891.4  &  283182  & 1.1705 &  -3539 \\
3d5.(2I).4p   & 3K*  & 6 &  305590.8  &  297831  & 0.8946 &  -5614 \\
3d5.(2I).4p   & 3I*  & 6 &  307399.7  &  299924  & 1.0135 &  -4077 \\
3d5.(2I).4p   & 3H*  & 6 &  309264.0  &  302007  & 1.1396 &  -3803 \\
3d5.(2I).4p   & 1I*  & 6 &  314392.0  &  306937  & 1.0078 &  -3467 \\
3d5.(4F).4p   & 5G*  & 6 &  314756.4  &  309862  & 1.3209 &  -3333 \\
3d5.(2H).4p   & 3H*  & 6 &  317327.3  &  312617  & 1.1383 &  -3348 \\
3d5.(2H).4p   & 3I*  & 6 &  319860.4  &  314709  & 1.0536 &  -2904 \\
3d5.(2H).4p   & 1I*  & 6 &  322324.2  &  316934  & 1.0174 &  -3629 \\
3d5.(2G2).4p  & 3H*  & 6 &  325148.4  &  320798  & 1.1533 &  -2232 \\
3d5.(2G1).4p  & 3H*  & 6 &  354989.6  &  354924  & 1.1667 &  -3051 \\

3d5.(4G).4p  & 5H*  & 7 &  288021.6  &  279472  & 1.2854  & -2838 \\
3d5.(2I).4p  & 3K*  & 7 &  305996.3  &  298337  & 1.0514  & -5106 \\
3d5.(2I).4p  & 3I*  & 7 &  308317.3  &  300798  & 1.1022  & -2731 \\
3d5.(2I).4p  & 1K*  & 7 &  309743.6  &  301788  & 1.0093  & -3396 \\
3d5.(2H).4p  & 3I*  & 7 &  320783.1  &  315488  & 1.1411  & -2385 \\
\end{longtable}
\endgroup

%% file: SMtables/G191_0.5.txt
\begin{lstlisting}
lam_K       error    lam_obs     error    q      lam_rest      match
1159.018	4	1159.115	2	3275	1159.022987	1.115169125
1159.036	5	1159.115	2	3368	1159.022987	2.416418536
1167.102	4	1167.194	2	2415	1167.101346	0.146269895
1171.114	11	1171.207	2	2842	1171.114027	0.002441909
1174.186	2.5	1174.276	2	3152	1174.182784	1.004610164
1179.2264	2.5	1179.323	1	4405	1179.229383	1.107872292
1182.606	3	1182.704	3	2791	1182.610115	0.969831612
1185.944	3	1186.045	2	3588	1185.950849	1.899690625
1194.2807	2.1	1194.368	5.6	2121	1194.273189	1.255895657
1195.1891	1.6	1195.283	2	3065	1195.188116	0.384147835
1198.519	1.6	1198.619	2.1	3739	1198.523851	1.837554574
1202.034	3	1202.138	2	3605	1202.042572	2.377427898
1202.427	3	1202.516	1	2890	1202.420542	2.042220476
1203.265	3	1203.361	2	2455	1203.265475	0.131700821
1204.7271	1.7	1204.822	1.5	2833	1204.726359	0.326895329
1229.402	5	1229.504	1.9	3317	1229.4064	0.822529472
1229.410 	5	1229.504	1.9	3896	1229.4064	0.673124257
1230.069	4	1230.155	6	3852	1230.057348	1.615856368
1230.440 	4	1230.546	2.1	3139	1230.448317	1.840931644
1231.0942	2.4	1231.202	3	2161	1231.104265	2.619757883
1232.8074	1.7	1232.910 	1	3231	1232.812129	2.397797931
1232.970 	5	1233.073	1	3225	1232.975116	1.003380644
1233.115	3	1233.211	1	3272	1233.113105	0.599155871
1233.324	5	1233.429	1	1995	1233.331088	1.390070677
1234.3974	1.8	1234.496	1.5	2256	1234.398003	0.257480871
1234.5839	2	1234.679	2	3628	1234.580989	1.029275213
1235.8376	1.8	1235.939	1	3429	1235.840889	1.597157611
1236.277	4	1236.377	1.6	2067	1236.278854	0.430344191
1236.292	5	1236.377	1.6	4227	1236.278854	2.50411747
1236.703	5	1236.803	2	3297	1236.70482	0.337995747
1237.2054	1.8	1237.305	3.5	2590	1237.20678	0.35071302
1241.0412	1.8	1241.139	2	2195	1241.040476	0.26908636
1241.333	4	1241.436	2	2656	1241.337452	0.995583824
1241.429	5	1241.533	1.8	3295	1241.434445	1.024567335
1241.440 	4	1241.533	1.8	2535	1241.434445	1.266502539
1241.6405	1.9	1241.741	1.1	3610	1241.642428	0.878259518
1241.9881	1.5	1242.086	2	2687	1241.987401	0.279684822
1243.5042	1.7	1243.603	2	2377	1243.50428	0.030616826
1244.026	3	1244.127	2	2597	1244.028239	0.620922907
1244.183	5	1244.288	0.5	3662	1244.189226	1.239018097
1244.1847	2.2	1244.288	0.5	4017	1244.189226	2.006109303
1244.788	3	1244.877	2	3512	1244.778179	2.723790715
1244.9596	2.5	1245.065	2	2554	1244.966164	2.050345695
1244.965	3	1245.065	2	3372	1244.966164	0.322921245
1245.0666	1.6	1245.170 	2	3958	1245.071156	1.778808998
1245.184	6	1245.285	1	3610	1245.186147	0.352939157
1245.448	17	1245.556	1.2	3815	1245.457125	0.535451928
1245.5389	2.5	1245.643	2.1	2581	1245.544118	1.598309774
1246.549	5	1246.647	3.1	2296	1246.548039	0.163397726
1246.816	6	1246.900 	2	1829	1246.801019	2.368760492
1246.836	14	1246.900  	2	3325	1246.801019	2.473555459
1246.836	14	1246.945	2	3325	1246.846015	0.708172464
1247.612	5	1247.718	2	2707	1247.618954	1.291271305
1248.4946	2.5	1248.592	2.4	2590	1248.492884	0.495065334
1249.5321	1.9	1249.635	0.9	2831	1249.535802	1.76063998
1250.045	3	1250.151	1	2890	1250.051761	2.137880738
1250.065	5	1250.151	1	2983	1250.051761	2.596465353
1250.342	4	1250.444	1	2781	1250.344737	0.663896061
1250.396	22	1250.444	1	3954	1250.344737	2.327718683
1250.396	22	1250.501	-	3954	1250.401733	0.26058131
1250.405	3	1250.501	-	3634	1250.401733	1.089070396
1251.8353	1.9	1251.939	1	3583	1251.839619	2.011390041
1252.075	5	1252.160 	2	2727	1252.060601	2.67380969
1252.075	5	1252.188	2	2727	1252.088599	2.525247034
1252.146	16	1252.270 	1	4012	1252.170592	1.534029403
1252.173	11	1252.270 	1	2638	1252.170592	0.21797729
1252.281	3	1252.382	1.3	3105	1252.282583	0.484307685
1253.015	5	1253.119	3	3231	1253.019525	0.776025415
1253.017	3	1253.119	3	2159	1253.019525	0.595140397
1253.222	5	1253.312	1.4	2063	1253.21251	1.827773859
1253.301	3	1253.410 	1.1	3597	1253.310502	2.973692538
1253.491	11	1253.604	2.4	2329	1253.504486	1.197862768
1253.500  	11	1253.604	2.4	2616	1253.504486	0.398486236
1253.663	3	1253.772	1	3570	1253.672473	2.995666878
1253.983	8	1254.100  	0.8	1387	1254.000447	2.17006332
1254.016	11	1254.100  	0.8	2675	1254.000447	1.410176148
1254.1965	1.7	1254.302	1.4	4221	1254.202431	2.693154711
1254.420 	6	1254.525	1	3424	1254.425413	0.889950217
1254.429	5	1254.525	1	1323	1254.425413	0.703398792
1254.442	6	1254.525	1	2084	1254.425413	2.726827503
1254.8657	2.4	1254.956	2	2171	1254.856379	2.983533873
1255.747	3	1255.849	1	2900	1255.749308	0.729934047
1255.818	5	1255.909	2	2232	1255.809303	1.61490114
1256.024	6	1256.135	2	2136	1256.035286	1.784402273
1256.222	4	1256.326	3.1	3042	1256.22627	0.843844924
1256.371	3	1256.472	2.8	2988	1256.372259	0.306750589
1256.9159	1.9	1257.018	1.8	3531	1256.918215	0.884690502
1257.637	4	1257.738	0.6	3765	1257.638158	0.286371645
1258.0243	1.6	1258.124	2	3088	1258.024128	0.067287392
1258.537	1.7	1258.646	2.9	2845	1258.546086	2.702988267
1259.130 	4	1259.232	2	3270	1259.13204	0.4560919
1259.592	3	1259.687	1	3728	1259.587004	1.580004775
1259.727	3	1259.829	2	2197	1259.728992	0.552568472
1260.407	3	1260.507	-	2145	1260.406938	0.020502335
1261.118	3	1261.229	3	3038	1261.128881	2.564718539
1261.219	10	1261.334	2	3512	1261.233873	1.458402354
1261.341	5	1261.451	1	3353	1261.350864	1.934402566
1261.437	5	1261.544	2	3533	1261.443856	1.273159533
1261.758	3	1261.862	1	3254	1261.761831	1.211446578
1262.552	5	1262.654	2.1	2123	1262.553768	0.326023926
1262.558	8	1262.654	2.1	2971	1262.553768	0.511657933
1263.3452	1.8	1263.452	3	2029	1263.351705	1.859248502
1264.436	10	1264.534	2	3119	1264.433619	0.233493726
1264.517	8	1264.637	-	2559	1264.536611	2.451330691
1264.755	5	1264.854	3	2678	1264.753593	0.24122654
1265.667	5	1265.770 	1	3779	1265.669521	0.494351075
1265.733	3	1265.832	-	1948	1265.731516	0.494738634
1265.8763	1.9	1265.977	2	3068	1265.876504	0.074049164
1266.403	3	1266.510 	-	3127	1266.409462	2.153987711
1266.8645	1.9	1266.971	1	2392	1266.870425	2.759719034
1267.305	5	1267.398	1	4177	1267.297391	1.492155113
1268.1818	1.8	1268.281	3	1979	1268.180321	0.422636055
1268.8826	1.8	1268.989	1.6	1935	1268.888265	2.352335995
1269.384	2.4	1269.494	2	4191	1269.393225	2.952878515
1270.205	1.9	1270.312	2.7	3220	1270.21116	1.865856528
1271.259	3	1271.369	2.1	4337	1271.268076	2.478516875
1271.448	3	1271.541	2	1772	1271.440063	2.201440741
1272.626	4	1272.736	2	3112	1272.634968	2.005245633
1272.656	10	1272.736	2	2567	1272.634968	2.062383645
1272.656	10	1272.772	2	2567	1272.670965	1.467426562
1272.731	6	1272.834	3	1440	1272.73296	0.292172344
1273.188	10	1273.315	1	2626	1273.213922	2.579312414
1273.209	8	1273.315	1	2719	1273.213922	0.610470305
1273.817	3	1273.922	1	4084	1273.820874	1.224934846
1274.2628	1.6	1274.365	2.1	2380	1274.263838	0.393328825
1276.4314	2	1276.535	0.7	3368	1276.433666	1.069466582
1276.6323	2.4	1276.725	3	4042	1276.623651	2.251224869
1278.872	3	1278.979	2.3	2862	1278.877472	1.447577568
1279.7132	2	1279.820 	1	2212	1279.718405	2.327920884
1280.108	5	1280.214	1.4	2808	1280.112374	0.842422659
1280.120 	4	1280.214	1.4	3490	1280.112374	1.799439327
1280.283	3	1280.376	4	1067	1280.274361	1.727749404
1281.285	5	1281.388	3	3586	1281.286281	0.2196757
1282.212	7	1282.308	2	2831	1282.206208	0.795607902
1282.281	7	1282.385	1	2949	1282.283202	0.311377955
1282.732	3	1282.843	1.9	3571	1282.741165	2.581039903
1283.1857	2.3	1283.284	2	2667	1283.18213	1.171144404
1284.475	5	1284.58n	1	2751	1284.478028	0.593747728
1284.566	5	1284.671	1.6	3148	1284.56902	0.575322757
1285.806	10	1285.894	1.8	3493	1285.791923	1.385412882
1287.5713	2	1287.675	1	5009	1287.572782	0.66270083
1287.807	3	1287.918	2	2642	1287.815763	2.430295292
1289.852	8	1289.955	2	2368	1289.852601	0.07286416
1290.3967	1.8	1290.502	2	2290	1290.399557	1.061955989
1292.002	7	1292.110 	3.7	2054	1292.00743	0.685777939
1292.071	7	1292.179	2	3655	1292.076424	0.745085977
1296.206	2.4	1296.305	3	3593	1296.202097	1.015968143
1298.7322	1.9	1298.835	1	3419	1298.731896	0.141614155
1300.226	4	1300.340 	3	2747	1300.236776	2.155294335
1300.9807	2	1301.089	0.5	3504	1300.985717	2.43360954
1302.252	3	1302.352	2	2989	1302.248617	0.938343281
1302.838	3	1302.950 	1	3208	1302.846569	2.709845782
1303.320 	4	1303.428	2	1227	1303.324531	1.013238473
1303.976	5	1304.082	2	3194	1303.978479	0.460417554
1304.813	3	1304.911	1	1472	1304.807414	1.766569489
1304.884	3	1304.986	1	1590	1304.882408	0.503541133
1305.704	3	1305.808	1	1802	1305.704342	0.108279915
1305.941	4	1306.038	1	2987	1305.934324	1.61913065
1306.620 	15	1306.730 	-	2445	1306.626269	0.417948066
1307.5993	1.9	1307.706	1	1384	1307.602192	1.346819551
1308.594	5	1308.686	3	3171	1308.582114	2.038440775
1308.660 	4	1308.759	1	2554	1308.655108	1.186446681
1309.137	5	1309.234	4	3167	1309.13007	1.082214112
1309.660 	3	1309.756	1	2157	1309.652029	2.520647979
1310.0784	1.9	1310.182	1.1	3055	1310.077995	0.184383854
1310.2561	1.9	1310.360 	2	2723	1310.255981	0.043114082
1311.111	1.9	1311.219	1	2135	1311.114913	1.822407765
1311.5569	1.9	1311.662	2	3747	1311.557878	0.354419311
1312.656	12	1312.758	1	1294	1312.653791	0.183471832
1312.728	12	1312.839	1	1412	1312.734784	0.56340351
1312.740 	4	1312.839	1	1997	1312.734784	1.264998723
1313.3056	2.1	1313.412	1	2056	1313.307739	0.919537982
1314.304	14	1314.439	1	2096	1314.334657	2.184239729
1314.690 	10	1314.789	1	4004	1314.684629	0.53438653
1314.916	10	1315.024	2	3908	1314.919611	0.354070726
1314.995	7	1315.101	1.8	2429	1314.996605	0.222022117
1315.6679	1.9	1315.774	1.6	3216	1315.669551	0.664784571
1316.897	7	1317.007	1.8	2278	1316.902453	0.75451309
1316.912	5	1317.007	1.8	3000	1316.902453	1.796452769
1317.4449	1.9	1317.551	0.7	3366	1317.44641	0.745848885
1318.151	1.9	1318.257	2	2903	1318.152354	0.490891762
1318.354	4	1318.456	0.8	3268	1318.351338	0.652481269
1318.5112	1.9	1318.618	1	2643	1318.513326	0.989957244
1319.752	2.4	1319.849	3	2949	1319.744228	2.023020469
1320.7152	1.9	1320.823	0.7	2378	1320.71815	1.457143711
1320.901	3	1321.005	3	2650	1320.900136	0.203636471
1321.251	4	1321.363	2	2976	1321.258108	1.58931324
1323.547	3	1323.655	1.1	2816	1323.549926	0.915617673
1323.557	4	1323.655	1.1	1322	1323.549926	1.705274024
1323.975	9	1324.094	1	2966	1323.988891	1.533985861
1324.464	5	1324.576	3	3533	1324.470853	1.175206175
1324.487	9	1324.576	3	3663	1324.470853	1.702088501
1324.733	4	1324.838	3	3491	1324.732832	0.03364548
1325.094	5	1325.201	3	2439	1325.095803	0.309204566
1331.436	11	1331.522	3	2660	1331.416301	1.727700487
1332.802	3	1332.908	3.5	2830	1332.802191	0.041468478
1332.928	4	1333.040 	1.2	3830	1332.934181	1.480004878
1333.9644	1.8	1334.069	1.2	3464	1333.963099	0.601388483
1334.180 	5	1334.289	2.7	2689	1334.183082	0.542291638
1334.205	11	1334.289	2.7	2308	1334.183082	1.935146096
1337.007	3	1337.114	2	1978	1337.007857	0.237766622
1337.074	3	1337.183	3	3192	1337.076852	0.67217627
1339.065	5	1339.161	2	3308	1339.054695	1.913630405
1340.3529	2	1340.455	3	3092	1340.348592	1.194806294
1341.0684	1.8	1341.179	1.8	3093	1341.072535	1.624221177
1342.1803	1.8	1342.292	1.1	3283	1342.185446	2.439551723
1343.8962	1.8	1343.998	2	3758	1343.891311	1.817051524
1347.740 	4	1347.844	2	2988	1347.737006	0.669588008
1351.4142	2	1351.519	2	2187	1351.411714	0.879010549
1355.553	4	1355.666	3	3150	1355.558385	1.076917152
1366.671	7	1366.799	2.9	612	1366.690501	2.573707834
1384.8712	1.9	1384.989	2.8	1326	1384.879057	2.321917345
1390.917	6	1391.009	3.5	3877	1390.898579	2.651946714
1392.494	4	1392.611	3	1774	1392.500452	1.290363904
\end{lstlisting}

%% file: SMtables/G191_0.75.txt
\begin{lstlisting}
lam_K       error    lam_obs     error    q      lam_rest      match
1159.018	4	1159.115	3	3275	1159.022987	0.997437588
1159.036	5	1159.115	3	3368	1159.022987	2.231678857
1167.102	4	1167.194	2	2415	1167.101346	0.146269895
1171.114	11	1171.207	1	2842	1171.114027	0.00247175
1174.186	2.5	1174.276	2	3152	1174.182784	1.004610164
1179.2264	2.5	1179.322	1	4405	1179.228383	0.736511098
1182.606	3	1182.704	3	2791	1182.610115	0.969831612
1185.944	3	1186.045	2	3588	1185.950849	1.899690625
1195.1891	1.6	1195.283	2	3065	1195.188116	0.384147835
1198.519	1.6	1198.619	2.2	3739	1198.523851	1.783367202
1202.034	3	1202.139	2	3605	1202.043572	2.654755979
1202.427	3	1202.516	1	2890	1202.420542	2.042220476
1203.265	3	1203.361	2	2455	1203.265475	0.131700821
1204.7271	1.7	1204.821	1.6	2833	1204.725359	0.745781135
1229.402	5	1229.504	1.8	3317	1229.4064	0.827900321
1229.410      5      1229.504      1.8    3896   1229.4064     0.677519538
1230.069	4	1230.164	5	3852	1230.066347	0.414300944
1230.440      4      1230.546      2      3139   1230.448317   1.859705466
1231.0942	2.4	1231.202	2.8	2161	1231.104265	2.72919697
1232.8074     1.7    1232.910      1      3231   1232.812129   2.397797931
1232.970      5      1233.074      1      3225   1232.976116   1.199481211
1233.115	3	1233.211	1	3272	1233.113105	0.599155871
1233.324	5	1233.428	1	1995	1233.330088	1.19397011
1234.3974	1.8	1234.496	1.4	2256	1234.398003	0.264563221
1234.5839	2	1234.678	2	3628	1234.579989	1.382800538
1235.8376	1.8	1235.939	1	3429	1235.840889	1.597157611
1236.277	4	1236.377	1.6	2067	1236.278854	0.430344191
1236.292	5	1236.377	1.6	4227	1236.278854	2.50411747
1236.703	5	1236.803	2	3297	1236.70482	0.337995747
1237.2054	1.8	1237.309	2	2590	1237.21078	1.999459593
1241.0412	1.8	1241.141	2	2195	1241.042476	0.474148782
1241.333	4	1241.436	1.8	2656	1241.337452	1.015056685
1241.429	5	1241.534	2.1	3295	1241.435445	1.188362426
1241.440      4      1241.534      2.1    2535   1241.435445   1.0083339
1241.6405	1.9	1241.742	1.1	3610	1241.643428	1.333710879
1241.9881	1.5	1242.086	2	2687	1241.987401	0.279684822
1243.5042	1.7	1243.603	2	2377	1243.50428	0.030616826
1244.026	3	1244.127	2	2597	1244.028239	0.620922907
1244.183	5	1244.288	0.6	3662	1244.189226	1.236328032
1244.1847	2.2	1244.288	0.6	4017	1244.189226	1.984777424
1244.788	3	1244.877	2	3512	1244.778179	2.723790715
1244.9596	2.5	1245.065	1.5	2554	1244.966164	2.251539449
1244.965	3	1245.065	1.5	3372	1244.966164	0.347129908
1245.0666     1.6    1245.170      2      3958   1245.071156   1.778808998
1245.184	6	1245.286	1	3610	1245.187147	0.517325094
1245.448	17	1245.556	1.2	3815	1245.457125	0.535451928
1245.5389	2.5	1245.643	2.1	2581	1245.544118	1.598309774
1246.365	3	1246.458	3	3457	1246.359054	1.401549269
1246.549	5	1246.646	3.3	2296	1246.547039	0.327366152
1246.816	6	1246.901	3	1829	1246.802019	2.084229445
1246.836	14	1246.901	3	3325	1246.802019	2.373366451
1246.836	14	1246.944	2	3325	1246.845015	0.637467399
1247.612	5	1247.718	2	2707	1247.618954	1.291271305
1248.4946	2.5	1248.592	2.4	2590	1248.492884	0.495065334
1249.5321	1.9	1249.635	0.9	2831	1249.535802	1.76063998
1250.045	3	1250.151	1.3	2890	1250.051761	2.067733775
1250.065	5	1250.151	1.3	2983	1250.051761	2.562683183
1250.342	4	1250.444	1	2781	1250.344737	0.663896061
1250.396	22	1250.444	1	3954	1250.344737	2.327718683
1250.396	22	1250.501	-	3954	1250.401733	0.26058131
1250.405	3	1250.501	-	3634	1250.401733	1.089070396
1251.8353	1.9	1251.939	1	3583	1251.839619	2.011390041
1252.075      5      1252.160      2      2727   1252.060601   2.67380969
1252.075	5	1252.189	2	2727	1252.089599	2.710927632
1252.146      16     1252.270      1      4012   1252.170592   1.534029403
1252.173      11     1252.270      1      2638   1252.170592   0.21797729
1252.281	3	1252.382	1.2	3105	1252.282583	0.490072083
1253.015      5      1253.120      3      3231   1253.020525   0.947510386
1253.017      3      1253.120      3      2159   1253.020525   0.830823946
1253.222	5	1253.312	1.6	2063	1253.21251	1.807768435
1253.301      3      1253.410      2      3597   1253.310502   2.635343665
1253.491	11	1253.604	2.3	2329	1253.504486	1.200089677
1253.500      11     1253.604      2.3    2616   1253.504486   0.399227049
1253.663	3	1253.772	1	3570	1253.672473	2.995666878
1253.983      8      1254.100      1      1387   1254.000447   2.16404557
1254.016      11     1254.100      1      2675   1254.000447   1.408094023
1254.1965	1.7	1254.302	1.5	4221	1254.202431	2.616077541
1254.420      6      1254.525      1      3424   1254.425413   0.889950217
1254.429	5	1254.525	1	1323	1254.425413	0.703398792
1254.442	6	1254.525	1	2084	1254.425413	2.726827503
1254.8657	2.4	1254.957	2	2171	1254.857379	2.663467083
1255.747	3	1255.849	2	2900	1255.749308	0.64019451
1255.818	5	1255.908	3	2232	1255.808304	1.662923924
1256.024	6	1256.136	2.8	2136	1256.036285	1.855481192
1256.222	4	1256.326	3	3042	1256.22627	0.854077788
1256.371	3	1256.473	2.7	2988	1256.373259	0.559631066
1256.9159	1.9	1257.019	1.7	3531	1256.919215	1.300397944
1257.637	4	1257.738	0.6	3765	1257.638158	0.286371645
1258.0243	1.6	1258.124	2	3088	1258.024128	0.067287392
1259.130      4      1259.232      2      3270   1259.13204    0.4560919
1259.592	3	1259.687	2	3728	1259.587004	1.385755859
1259.727	3	1259.829	2	2197	1259.728992	0.552568472
1260.407	3	1260.507	-	2145	1260.406938	0.020502335
1261.118	3	1261.229	3	3038	1261.128881	2.564718539
1261.219	10	1261.334	2	3512	1261.233873	1.458402354
1261.341	5	1261.451	1	3353	1261.350864	1.934402566
1261.437	5	1261.544	2	3533	1261.443856	1.273159533
1261.758	3	1261.862	1	3254	1261.761831	1.211446578
1262.552	5	1262.654	2.1	2123	1262.553768	0.326023926
1262.558	8	1262.654	2.1	2971	1262.553768	0.511657933
1263.3452	1.8	1263.452	3	2029	1263.351705	1.859248502
1264.436	10	1264.534	2	3119	1264.433619	0.233493726
1264.517	8	1264.637	-	2559	1264.536611	2.451330691
1264.755	5	1264.854	3	2678	1264.753593	0.24122654
1265.667      5      1265.770      1      3779   1265.669521   0.494351075
1265.733	3	1265.832	-	1948	1265.731516	0.494738634
1265.8763	1.9	1265.977	2	3068	1265.876504	0.074049164
1266.403      3      1266.510      -      3127   1266.409462   2.153987711
1266.8645	1.9	1266.971	1	2392	1266.870425	2.759719034
1267.160      3      1267.255      3.1    1758   1267.154403   1.297465472
1267.305	5	1267.398	1	4177	1267.297391	1.492155113
1268.1818	1.8	1268.281	3	1979	1268.180321	0.422636055
1268.8826	1.8	1268.990      1.5	1935	1268.889265	2.844593608
1269.384	2.4	1269.488	1	4191	1269.387226	1.240601415
1270.205	1.9	1270.309	1.7	3220	1270.20816	1.239607374
1271.259	3	1271.369	2	4337	1271.268076	2.517297781
1271.448	3	1271.542	2	1772	1271.441063	1.924112659
1272.626	4	1272.736	2	3112	1272.634968	2.005245633
1272.656	10	1272.736	2	2567	1272.634968	2.062383645
1272.656	10	1272.771	2	2567	1272.669965	1.369376278
1272.731	6	1272.835	4	1440	1272.73396	0.410460434
1273.188	10	1273.315	1	2626	1273.213922	2.579312414
1273.209	8	1273.315	1	2719	1273.213922	0.610470305
1273.817	3	1273.922	1	4084	1273.820874	1.224934846
1274.2628	1.6	1274.366	2.1	2380	1274.264838	0.772075766
1274.514	6	1274.588	7.1	2193	1274.486821	2.923856227
1276.4314	2	1276.535	0.7	3368	1276.433666	1.069466582
1276.6323	2.4	1276.724	4	4042	1276.622651	2.068453896
1278.872	3	1278.979	2	2862	1278.877472	1.517701219
1279.7132     2      1279.820      1      2212   1279.718405   2.327920884
1280.108	5	1280.214	1.5	2808	1280.112374	0.837928053
1280.120      4      1280.214      1.5    3490   1280.112374   1.78508516
1280.283	3	1280.384	2	1067	1280.282361	0.177332682
1280.298	5	1280.384	2	3671	1280.282361	2.904160344
1281.285	5	1281.388	3	3586	1281.286281	0.2196757
1282.212	7	1282.307	2	2831	1282.205208	0.932957562
1282.281	7	1282.385	1	2949	1282.283202	0.311377955
1282.732	3	1282.843	1.8	3571	1282.741165	2.61976028
1283.1857	2.3	1283.284	2	2667	1283.18213	1.171144404
1284.475	5	1284.580      0.9	2751	1284.478028	0.595929154
1284.566	5	1284.671	1.5	3148	1284.56902	0.578585981
1285.806	10	1285.894	2	3493	1285.791923	1.380341515
1287.5713	2	1287.675	1	5009	1287.572782	0.66270083
1287.807	3	1287.917	2	2642	1287.814763	2.152967211
1289.852	8	1289.955	2	2368	1289.852601	0.07286416
1290.3967	1.8	1290.502	2	2290	1290.399557	1.061955989
1292.002	7	1292.109	3.2	2054	1292.00643	0.575549817
1292.071	7	1292.179	2	3655	1292.076424	0.745085977
1296.206	2.4	1296.305	2.6	3593	1296.202097	1.103115799
1298.7322	1.9	1298.835	1	3419	1298.731896	0.141614155
1300.226      4      1300.340      2      2747   1300.236776   2.409692322
1300.9807	2	1301.089	0.5	3504	1300.985717	2.43360954
1302.252	3	1302.352	2	2989	1302.248617	0.938343281
1302.838	3	1302.950      1	3208	1302.846569	2.709845782
1303.320      4      1303.428      2      1227   1303.324531   1.013238473
1303.976	5	1304.082	2	3194	1303.978479	0.460417554
1304.813	3	1304.912	1	1472	1304.808414	1.450366826
1304.884	3	1304.986	1	1590	1304.882408	0.503541133
1305.704	3	1305.808	1	1802	1305.704342	0.108279915
1305.941	4	1306.038	1	2987	1305.934324	1.61913065
1306.620      15	1306.730      -	2445	1306.626269	0.417948066
1307.5993	1.9	1307.706	1	1384	1307.602192	1.346819551
1307.779	9	1307.881	2	2244	1307.777178	0.19763966
1308.660      4	1308.759	1	2554	1308.655108	1.186446681
1309.137	5	1309.232	3	3167	1309.128071	1.531378205
1309.660      3	1309.757	1	2157	1309.653029	2.204445315
1310.0784	1.9	1310.182	1.1	3055	1310.077995	0.184383854
1310.2561	1.9	1310.360      2	2723	1310.255981	0.043114082
1311.111	1.9	1311.218	1	2135	1311.113913	1.356698304
1311.5569	1.9	1311.665	2	3747	1311.560877	1.441832133
1312.656	12	1312.758	1	1294	1312.653791	0.183471832
1312.728	12	1312.839	1	1412	1312.734784	0.56340351
1312.740      4	1312.839	1	1997	1312.734784	1.264998723
1313.3056	2.1	1313.412	1	2056	1313.307739	0.919537982
1314.304	14	1314.439	1	2096	1314.334657	2.184239729
1314.690      10	1314.789	1	4004	1314.684629	0.53438653
1314.916	10	1315.024	2	3908	1314.919611	0.354070726
1314.995	7	1315.101	1.8	2429	1314.996605	0.222022117
1315.6679	1.9	1315.774	1.7	3216	1315.669551	0.647689468
1316.897	7	1317.006	1.7	2278	1316.901453	0.618242485
1316.912	5	1317.006	1.7	3000	1316.901453	1.997029256
1317.4449	1.9	1317.551	0.7	3366	1317.44641	0.745848885
1318.151	1.9	1318.257	2	2903	1318.152354	0.490891762
1318.354	4	1318.456	0.8	3268	1318.351338	0.652481269
1318.5112	1.9	1318.618	1	2643	1318.513326	0.989957244
1319.752	2.4	1319.849	3	2949	1319.744228	2.023020469
1320.7152	1.9	1320.823	1	2378	1320.71815	1.374180723
1320.901	3	1321.004	2.8	2650	1320.899136	0.454199032
1321.251	4	1321.363	2	2976	1321.258108	1.58931324
1323.547	3	1323.655	1.2	2816	1323.549926	0.905475677
1323.557	4	1323.655	1.2	1322	1323.549926	1.693992065
1323.975	9	1324.094	1	2966	1323.988891	1.533985861
1324.464	5	1324.576	3	3533	1324.470853	1.175206175
1324.487	9	1324.576	3	3663	1324.470853	1.702088501
1324.733	4	1324.838	3	3491	1324.732832	0.03364548
1325.094	5	1325.201	2	2439	1325.095803	0.3348007
1331.436	11	1331.528	5.8	2660	1331.422301	1.101633726
1332.928	4	1333.040      1.2	3830	1332.934181	1.480004878
1333.9644	1.8	1334.069	1.2	3464	1333.963099	0.601388483
1334.180      5	1334.289	2.7	2689	1334.183082	0.542291638
1334.205	11	1334.289	2.7	2308	1334.183082	1.935146096
1336.1601	1.8	1336.265	1	2944	1336.158925	0.570788266
1336.548	5	1336.664	3.7	4603	1336.557893	1.590481843
1337.007	3	1337.114	2	1978	1337.007857	0.237766622
1337.074	3	1337.183	3	3192	1337.076852	0.67217627
1339.065	5	1339.160      2	3308	1339.053695	2.099311002
1340.3529	2	1340.455	3	3092	1340.348592	1.194806294
1341.0684	1.8	1341.179	1.8	3093	1341.072535	1.624221177
1342.1803	1.8	1342.292	1.2	3283	1342.185446	2.37885028
1343.8962	1.8	1343.999	2	3758	1343.892311	1.445433952
1347.740      4	1347.844	2	2988	1347.737006	0.669588008
1351.4142	2	1351.519	2	2187	1351.411714	0.879010549
1355.553	4	1355.666	3	3150	1355.558385	1.076917152
1366.671	7	1366.799	3	612	1366.690501	2.560583963
1390.917	6	1391.009	3.4	3877	1390.898579	2.67111465
1392.494	4	1392.611	3.2	1774	1392.500452	1.259506157
\end{lstlisting}

%% file: SMtables/G191_1.0.txt
\begin{lstlisting}
lam_K       error    lam_obs     error    q      lam_rest      match
1159.018	4	1159.115	2	3275	1159.022987	1.115169125
1159.036	5	1159.115	2	3368	1159.022987	2.416418536
1167.102	4	1167.194	2	2415	1167.101346	0.146269895
1171.114	11	1171.207	2	2842	1171.114027	0.002441909
1174.186	2.5	1174.276	2	3152	1174.182784	1.004610164
1182.606	3	1182.704	3	2791	1182.610115	0.969831612
1185.944	3	1186.045	1	3588	1185.950849	2.165980565
1195.1891	1.6	1195.283	2	3065	1195.188116	0.384147835
1197.994	4	1198.087	3.1	2374	1197.991894	0.416249465
1197.9969	1.7	1198.087	3.1	3769	1197.991894	1.416047944
1198.519	1.6	1198.619	2	3739	1198.523851	1.8941079
1202.034	3	1202.139	2	3605	1202.043572	2.654755979
1202.427	3	1202.516	1	2890	1202.420542	2.042220476
1203.265	3	1203.361	2	2455	1203.265475	0.131700821
1204.7271	1.7	1204.821	2	2833	1204.725359	0.663284782
1229.402	5	1229.503	2	3317	1229.4054	0.631299352
1229.410      5	1229.503	2	3896	1229.4054	0.854263353
1230.440      4	1230.546	1.7	3139	1230.448317	1.913564856
1231.0942	2.4	1231.202	3	2161	1231.104265	2.619757883
1232.8074	1.7	1232.910      1	3231	1232.812129	2.397797931
1232.970      5	1233.073	1	3225	1232.975116	1.003380644
1233.115	3	1233.211	1	3272	1233.113105	0.599155871
1233.324	5	1233.430	1	1995	1233.332088	1.586171244
1234.3974	1.8	1234.496	1.4	2256	1234.398003	0.264563221
1234.5839	2	1234.678	2	3628	1234.579989	1.382800538
1235.8376	1.8	1235.939	1	3429	1235.840889	1.597157611
1236.277	4	1236.377	1.6	2067	1236.278854	0.430344191
1236.292	5	1236.377	1.6	4227	1236.278854	2.50411747
1236.703	5	1236.803	1	3297	1236.70482	0.356963294
1237.2054	1.8	1237.308	2.1	2590	1237.20978	1.583619428
1241.0412	1.8	1241.142	3	2195	1241.043476	0.650472556
1241.059	5	1241.142	3	3797	1241.043476	2.662391282
1241.059	5	1241.173	2	3797	1241.074473	2.87331293
1241.333	4	1241.436	2	2656	1241.337452	0.995583824
1241.429	5	1241.534	1.6	3295	1241.435445	1.227599824
1241.440      4	1241.534	1.6	2535	1241.435445	1.057394105
1241.6405	1.9	1241.742	1.1	3610	1241.643428	1.333710879
1241.9881	1.5	1242.086	2	2687	1241.987401	0.279684822
1243.5042	1.7	1243.603	2	2377	1243.50428	0.030616826
1244.026	3	1244.127	2	2597	1244.028239	0.620922907
1244.183	5	1244.288	0.6	3662	1244.189226	1.236328032
1244.1847	2.2	1244.288	0.6	4017	1244.189226	1.984777424
1244.788	3	1244.877	2	3512	1244.778179	2.723790715
1244.9596	2.5	1245.065	1.5	2554	1244.966164	2.251539449
1244.965	3	1245.065	1.5	3372	1244.966164	0.347129908
1245.0666	1.6	1245.170      3	3958	1245.071156	1.339992354
1245.184	6	1245.286	1	3610	1245.187147	0.517325094
1245.448	17	1245.557	1.6	3815	1245.458125	0.59298256
1245.5389	2.5	1245.643	1.9	2581	1245.544118	1.661885506
1246.365	3	1246.462	5.5	3457	1246.363053	0.310709187
1246.549	5	1246.647	2.8	2296	1246.548039	0.16774332
1246.816	6	1246.902	3	1829	1246.803018	1.93517008
1246.836	14	1246.902	3	3325	1246.803018	2.303528965
1246.836	14	1246.945	2	3325	1246.846015	0.708172464
1247.612	5	1247.718	2	2707	1247.618954	1.291271305
1248.4946	2.5	1248.592	2.3	2590	1248.492884	0.505046209
1249.5321	1.9	1249.634	0.9	2831	1249.534802	1.285026244
1250.045	3	1250.151	1.3	2890	1250.051761	2.067733775
1250.065	5	1250.151	1.3	2983	1250.051761	2.562683183
1250.342	4	1250.444	1	2781	1250.344737	0.663896061
1250.396	22	1250.444	1	3954	1250.344737	2.327718683
1250.396	22	1250.501	-	3954	1250.401733	0.26058131
1250.405	3	1250.501	-	3634	1250.401733	1.089070396
1251.8353	1.9	1251.939	1	3583	1251.839619	2.011390041
1252.075	5	1252.160      2	2727	1252.060601	2.67380969
1252.075	5	1252.189	2	2727	1252.089599	2.710927632
1252.146	16	1252.270      1	4012	1252.170592	1.534029403
1252.173	11	1252.270      1	2638	1252.170592	0.21797729
1252.281	3	1252.382	1.2	3105	1252.282583	0.490072083
1253.015	5	1253.120      3	3231	1253.020525	0.947510386
1253.017	3	1253.120      3	2159	1253.020525	0.830823946
1253.222	5	1253.312	1.6	2063	1253.21251	1.807768435
1253.301	3	1253.410      2	3597	1253.310502	2.635343665
1253.491	11	1253.603	2.2	2329	1253.503487	1.113096883
1253.500      11	1253.603	2.2	2616	1253.503487	0.310803603
1253.663	3	1253.772	1	3570	1253.672473	2.995666878
1253.983	8	1254.100      1	1387	1254.000447	2.16404557
1254.016	11	1254.100      1	2675	1254.000447	1.408094023
1254.1965	1.7	1254.302	1.5	4221	1254.202431	2.616077541
1254.420      6	1254.525	1	3424	1254.425413	0.889950217
1254.429	5	1254.525	1	1323	1254.425413	0.703398792
1254.442	6	1254.525	1	2084	1254.425413	2.726827503
1254.8657	2.4	1254.957	2	2171	1254.857379	2.663467083
1255.747	3	1255.849	2	2900	1255.749308	0.64019451
1255.818	5	1255.908	3	2232	1255.808304	1.662923924
1256.024	6	1256.136	2.5	2136	1256.036285	1.89007254
1256.222	4	1256.326	2.9	3042	1256.22627	0.864337871
1256.371	3	1256.473	2.7	2988	1256.373259	0.559631066
1256.9159	1.9	1257.019	1.7	3531	1256.919215	1.300397944
1257.637	4	1257.738	0.7	3765	1257.638158	0.285240597
1258.0243	1.6	1258.124	2	3088	1258.024128	0.067287392
1259.130      4	1259.232	2	3270	1259.13204	0.4560919
1259.592	3	1259.687	1	3728	1259.587004	1.580004775
1259.727	3	1259.829	1.9	2197	1259.728992	0.561048279
1260.407	3	1260.507	-	2145	1260.406938	0.020502335
1261.118	3	1261.229	3	3038	1261.128881	2.564718539
1261.219	10	1261.334	2	3512	1261.233873	1.458402354
1261.341	5	1261.451	1	3353	1261.350864	1.934402566
1261.437	5	1261.544	2	3533	1261.443856	1.273159533
1261.758	3	1261.863	1	3254	1261.762831	1.527649241
1262.552	5	1262.654	2	2123	1262.553768	0.328320489
1262.558	8	1262.654	2	2971	1262.553768	0.513198113
1262.739	5	1262.852	3.3	3915	1262.751752	2.128644971
1263.3452	1.8	1263.452	3	2029	1263.351705	1.859248502
1264.264	4	1264.362	4	1955	1264.261632	0.418523145
1264.436	10	1264.532	3	3119	1264.431619	0.41962555
1264.436	10	1264.564	2	3119	1264.463616	2.70801478
1264.517	8	1264.637	-	2559	1264.536611	2.451330691
1264.755	5	1264.853	3	2678	1264.752593	0.412711512
1265.667	5	1265.77	1	3779	1265.669521	0.494351075
1265.733	3	1265.832	-	1948	1265.731516	0.494738634
1265.8763	1.9	1265.977	2	3068	1265.876504	0.074049164
1266.403	3	1266.51	-	3127	1266.409462	2.153987711
1267.305	5	1267.398	1	4177	1267.297391	1.492155113
1268.1818	1.8	1268.281	2	1979	1268.180321	0.549525652
1268.8826	1.8	1268.99	1.7	1935	1268.889265	2.692006875
1269.384	2.4	1269.488	2	4191	1269.387226	1.032477774
1270.205	1.9	1270.309	1.7	3220	1270.20816	1.239607374
1271.259	3	1271.369	2.2	4337	1271.268076	2.4397101
1271.448	3	1271.543	2	1772	1271.442062	1.646784578
1272.626	4	1272.736	2	3112	1272.634968	2.005245633
1272.656	10	1272.736	2	2567	1272.634968	2.062383645
1272.656	10	1272.771	2	2567	1272.669965	1.369376278
1272.731	6	1272.835	4	1440	1272.73396	0.410460434
1273.188	10	1273.315	1	2626	1273.213922	2.579312414
1273.209	8	1273.315	1	2719	1273.213922	0.610470305
1273.817	3	1273.922	1	4084	1273.820874	1.224934846
1274.2628	1.6	1274.367	2.4	2380	1274.265838	1.053326835
1276.4314	2	1276.535	0.7	3368	1276.433666	1.069466582
1276.6323	2.4	1276.724	4	4042	1276.622651	2.068453896
1278.872	3	1278.979	2	2862	1278.877472	1.517701219
1279.7132	2	1279.820      1	2212	1279.718405	2.327920884
1280.108	5	1280.214	1.5	2808	1280.112374	0.837928053
1280.120      4	1280.214	1.5	3490	1280.112374	1.78508516
1280.283	3	1280.384	2	1067	1280.282361	0.177332682
1280.298	5	1280.384	2	3671	1280.282361	2.904160344
1281.285	5	1281.388	3	3586	1281.286281	0.2196757
1282.212	7	1282.307	2	2831	1282.205208	0.932957562
1282.281	7	1282.385	1	2949	1282.283202	0.311377955
1282.732	3	1282.843	1.9	3571	1282.741165	2.581039903
1283.1857	2.3	1283.284	1.7	2667	1283.18213	1.248077784
1284.475	5	1284.581	0.9	2751	1284.479027	0.792750193
1284.566	5	1284.670      1.5	3148	1284.56802	0.387035931
1285.806	10	1285.894	2	3493	1285.791923	1.380341515
1287.5713	2	1287.675	1	5009	1287.572782	0.66270083
1287.628	7	1287.731	3	511	1287.628777	0.102077452
1287.635	8	1287.731	3	3412	1287.628777	0.728300393
1287.807	3	1287.917	2	2642	1287.814763	2.152967211
1289.852	8	1289.955	2	2368	1289.852601	0.07286416
1290.3967	1.8	1290.502	2	2290	1290.399557	1.061955989
1292.002	7	1292.109	3.1	2054	1292.00643	0.578634847
1292.071	7	1292.178	2	3655	1292.075424	0.607736317
1296.206	2.4	1296.305	3.1	3593	1296.202097	0.995603318
1298.7322	1.9	1298.834	1.2	3419	1298.730896	0.580263086
1300.226	4	1300.340      3	2747	1300.236776	2.155294335
1300.9807	2	1301.089	0.5	3504	1300.985717	2.43360954
1302.252	3	1302.349	2	2989	1302.245617	1.770327526
1303.320      4	1303.428	2	1227	1303.324531	1.013238473
1303.976	5	1304.082	2	3194	1303.978479	0.460417554
1304.813	3	1304.912	1	1472	1304.808414	1.450366826
1304.884	3	1304.986	1	1590	1304.882408	0.503541133
1305.704	3	1305.808	1	1802	1305.704342	0.108279915
1305.941	4	1306.038	1	2987	1305.934324	1.61913065
1306.620      15	1306.730      -	2445	1306.626269	0.417948066
1307.5993	1.9	1307.706	1	1384	1307.602192	1.346819551
1307.779	9	1307.881	2	2244	1307.777178	0.19763966
1308.660      4	1308.759	1	2554	1308.655108	1.186446681
1309.137	5	1309.236	4	3167	1309.13207	0.769891383
1309.660      3	1309.755	1	2157	1309.651029	2.836850642
1310.0784	1.9	1310.182	1.1	3055	1310.077995	0.184383854
1310.2561	1.9	1310.360      2	2723	1310.255981	0.043114082
1311.111	1.9	1311.218	1	2135	1311.113913	1.356698304
1311.5569	1.9	1311.663	2	3747	1311.558878	0.716890251
1312.656	12	1312.758	1	1294	1312.653791	0.183471832
1312.728	12	1312.839	1	1412	1312.734784	0.56340351
1312.740      4	1312.839	1	1997	1312.734784	1.264998723
1313.3056	2.1	1313.412	1	2056	1313.307739	0.919537982
1314.304	14	1314.439	0.7	2096	1314.334657	2.187072546
1314.690      10	1314.790      1.2	4004	1314.685629	0.433946507
1314.916	10	1315.024	3	3908	1314.919611	0.345854509
1314.995	7	1315.101	1.9	2429	1314.996605	0.221240012
1315.6679	1.9	1315.774	1.6	3216	1315.669551	0.664784571
1316.897	7	1317.007	1.8	2278	1316.902453	0.75451309
1316.912	5	1317.007	1.8	3000	1316.902453	1.796452769
1317.060      3	1317.153	5.4	3087	1317.048442	1.871049001
1317.4449	1.9	1317.552	0.7	3366	1317.44741	1.239674479
1318.151	1.9	1318.256	2	2903	1318.151354	0.128420821
1318.354	4	1318.456	0.7	3268	1318.351338	0.655442174
1318.5112	1.9	1318.618	1	2643	1318.513326	0.989957244
1319.752	2.4	1319.849	2	2949	1319.744228	2.487817663
1320.7152	1.9	1320.823	0.7	2378	1320.71815	1.457143711
1320.901	3	1321.005	3	2650	1320.900136	0.203636471
1321.251	4	1321.364	2	2976	1321.259108	1.812902287
1323.547	3	1323.655	1.1	2816	1323.549926	0.915617673
1323.557	4	1323.655	1.1	1322	1323.549926	1.705274024
1323.975	9	1324.094	0.9	2966	1323.988891	1.535766134
1324.464	5	1324.577	3	3533	1324.471852	1.346691146
1324.487	9	1324.577	3	3663	1324.471852	1.596687614
1324.733	4	1324.839	3	3491	1324.733832	0.166338643
1325.094	5	1325.201	2	2439	1325.095803	0.3348007
1332.928	4	1333.04	1.2	3830	1332.934181	1.480004878
1333.9644	1.8	1334.069	1.2	3464	1333.963099	0.601388483
1334.180      5	1334.288	2.7	2689	1334.182082	0.366324529
1334.205	11	1334.288	2.7	2308	1334.182082	2.023427478
1336.1601	1.8	1336.265	1	2944	1336.158925	0.570788266
1336.548	5	1336.664	3.9	4603	1336.557893	1.560130919
1337.007	3	1337.114	2	1978	1337.007857	0.237766622
1337.074	3	1337.183	3	3192	1337.076852	0.67217627
1339.065	5	1339.161	2	3308	1339.054695	1.913630405
1340.3529	2	1340.456	3	3092	1340.349592	0.917478212
1341.0684	1.8	1341.179	1.8	3093	1341.072535	1.624221177
1342.1803	1.8	1342.292	1.2	3283	1342.185446	2.37885028
1343.8962	1.8	1343.998	2	3758	1343.891311	1.817051524
1347.740      4	1347.844	2	2988	1347.737006	0.669588008
1351.4142	2	1351.519	2	2187	1351.411714	0.879010549
1355.553	4	1355.666	4	3150	1355.558385	0.951869276
1366.671	7	1366.799	2.8	612	1366.690501	2.586580406
1379.212	4	1379.311	2.8	2550	1379.201508	2.148927355
1390.917	6	1391.009	3.6	3877	1390.898579	2.632647711
1392.494	4	1392.612	3.4	1774	1392.501452	1.419443449
\end{lstlisting}

%% file: SMtables/G191_1.25.txt
\begin{lstlisting}
lam_K       error    lam_obs     error    q      lam_rest      match
1159.018	4	1159.115	3	3275	1159.022987	0.997437588
1159.036	5	1159.115	3	3368	1159.022987	2.231678857
1167.102	4	1167.194	2	2415	1167.101346	0.146269895
1171.114	11	1171.207	2	2842	1171.114027	0.002441909
1174.186	2.5	1174.276	2	3152	1174.182784	1.004610164
1182.606	3	1182.703	3	2791	1182.609115	0.734148062
1185.944	3	1186.045	2	3588	1185.950849	1.899690625
1195.1891	1.6	1195.283	2	3065	1195.188116	0.384147835
1197.994	4	1198.075	3	2374	1197.979894	2.821106587
1198.519	1.6	1198.619	2	3739	1198.523851	1.8941079
1202.034	3	1202.139	2	3605	1202.043572	2.654755979
1202.427	3	1202.516	2	2890	1202.420542	1.791145846
1204.7271	1.7	1204.821	2	2833	1204.725359	0.663284782
1229.402	5	1229.503	2	3317	1229.4054	0.631299352
1229.410	5	1229.503	2	3896	1229.4054	0.854263353
1230.440	4	1230.546	1.9	3139	1230.448317	1.87810716
1231.0942	2.4	1231.202	3	2161	1231.104265	2.619757883
1232.8074	1.7	1232.910	1	3231	1232.812129	2.397797931
1232.970	5	1233.073	1	3225	1232.975116	1.003380644
1233.115	3	1233.211	1	3272	1233.113105	0.599155871
1233.324	5	1233.429	1	1995	1233.331088	1.390070677
1234.3974	1.8	1234.496	1.7	2256	1234.398003	0.243669349
1234.5839	2	1234.678	2	3628	1234.579989	1.382800538
1235.8376	1.8	1235.939	1	3429	1235.840889	1.597157611
1236.277	4	1236.377	1.6	2067	1236.278854	0.430344191
1236.292	5	1236.377	1.6	4227	1236.278854	2.50411747
1236.668	4	1236.761	3	2421	1236.662823	1.035300631
1236.703	5	1236.803	1	3297	1236.70482	0.356963294
1237.2054	1.8	1237.308	3	2590	1237.20978	1.251961085
1241.0412	1.8	1241.142	3	2195	1241.043476	0.650472556
1241.059	5	1241.142	3	3797	1241.043476	2.662391282
1241.059	5	1241.173	2	3797	1241.074473	2.87331293
1241.333	4	1241.436	2	2656	1241.337452	0.995583824
1241.429	5	1241.534	1.6	3295	1241.435445	1.227599824
1241.440	4	1241.534	1.6	2535	1241.435445	1.057394105
1241.6405	1.9	1241.742	1.1	3610	1241.643428	1.333710879
1241.9881	1.5	1242.086	2	2687	1241.987401	0.279684822
1243.5042	1.7	1243.603	2	2377	1243.50428	0.030616826
1244.026	3	1244.127	2	2597	1244.028239	0.620922907
1244.183	5	1244.288	0.6	3662	1244.189226	1.236328032
1244.1847	2.2	1244.288	0.6	4017	1244.189226	1.984777424
1244.788	3	1244.877	2	3512	1244.778179	2.723790715
1244.9596	2.5	1245.065	1.6	2554	1244.966164	2.211571931
1244.965	3	1245.065	1.6	3372	1244.966164	0.342443855
1245.0666	1.6	1245.169	2.7	3958	1245.070156	1.133052822
1245.184	6	1245.286	1.1	3610	1245.187147	0.515863229
1245.448	17	1245.557	1.6	3815	1245.458125	0.59298256
1245.5389	2.5	1245.643	1.9	2581	1245.544118	1.661885506
1246.365	3	1246.461	5.6	3457	1246.362053	0.463801405
1246.549	5	1246.646	2.5	2296	1246.547039	0.350829005
1246.816	6	1246.905	2	1829	1246.806018	1.578253834
1246.836	14	1246.905	2	3325	1246.806018	2.120030134
1246.836	14	1246.944	2	3325	1246.845015	0.637467399
1247.612	5	1247.718	2	2707	1247.618954	1.291271305
1248.4946	2.5	1248.592	2.2	2590	1248.492884	0.515190976
1249.5321	1.9	1249.634	0.9	2831	1249.534802	1.285026244
1250.045	3	1250.151	1.3	2890	1250.051761	2.067733775
1250.065	5	1250.151	1.3	2983	1250.051761	2.562683183
1250.342	4	1250.444	1	2781	1250.344737	0.663896061
1250.396	22	1250.444	1	3954	1250.344737	2.327718683
1250.396	22	1250.501	-	3954	1250.401733	0.26058131
1250.405	3	1250.501	-	3634	1250.401733	1.089070396
1251.8353	1.9	1251.939	1	3583	1251.839619	2.011390041
1252.075	5	1252.160	2	2727	1252.060601	2.67380969
1252.075	5	1252.189	2	2727	1252.089599	2.710927632
1252.146	16	1252.269	1	4012	1252.169592	1.471656068
1252.173	11	1252.269	1	2638	1252.169592	0.308505849
1252.281	3	1252.382	1.3	3105	1252.282583	0.484307685
1253.015	5	1253.120	3	3231	1253.020525	0.947510386
1253.017	3	1253.120	3	2159	1253.020525	0.830823946
1253.202	3	1253.311	1.8	1970	1253.21151	2.718174121
1253.222	5	1253.311	1.8	2063	1253.21151	1.97403344
1253.301	3	1253.410	2	3597	1253.310502	2.635343665
1253.491	11	1253.603	2.2	2329	1253.503487	1.113096883
1253.500	11	1253.603	2.2	2616	1253.503487	0.310803603
1253.663	3	1253.772	1	3570	1253.672473	2.995666878
1253.983	8	1254.100      1	1387	1254.000447	2.16404557
1254.016	11	1254.100      1	2675	1254.000447	1.408094023
1254.1965	1.7	1254.302	1.5	4221	1254.202431	2.616077541
1254.420	6	1254.525	1	3424	1254.425413	0.889950217
1254.429	5	1254.525	1	1323	1254.425413	0.703398792
1254.442	6	1254.525	1	2084	1254.425413	2.726827503
1254.8657	2.4	1254.957	2	2171	1254.857379	2.663467083
1255.747	3	1255.849	2	2900	1255.749308	0.64019451
1255.818	5	1255.908	3	2232	1255.808304	1.662923924
1256.024	6	1256.136	2.6	2136	1256.036285	1.878767249
1256.222	4	1256.326	3	3042	1256.22627	0.854077788
1256.371	3	1256.472	2.5	2988	1256.372259	0.322345437
1256.9159	1.9	1257.020      2.2	3531	1256.920215	1.484507595
1257.002	3	1257.109	2.7	4097	1257.009208	1.785945766
1257.637	4	1257.738	0.7	3765	1257.638158	0.285240597
1258.0243	1.6	1258.124	2	3088	1258.024128	0.067287392
1258.537	1.7	1258.646	3.3	2845	1258.546086	2.4477033
1259.130      4	1259.232	2	3270	1259.13204	0.4560919
1259.592	3	1259.687	2	3728	1259.587004	1.385755859
1259.727	3	1259.829	1.8	2197	1259.728992	0.569465042
1260.407	3	1260.507	-	2145	1260.406938	0.020502335
1261.118	3	1261.229	3	3038	1261.128881	2.564718539
1261.219	10	1261.334	2	3512	1261.233873	1.458402354
1261.341	5	1261.451	1	3353	1261.350864	1.934402566
1261.437	5	1261.544	2	3533	1261.443856	1.273159533
1261.758	3	1261.862	1	3254	1261.761831	1.211446578
1262.552	5	1262.654	2	2123	1262.553768	0.328320489
1262.558	8	1262.654	2	2971	1262.553768	0.513198113
1262.739	5	1262.852	3.1	3915	1262.751752	2.167650012
1263.3452	1.8	1263.451	3	2029	1263.350705	1.573440216
1264.264	4	1264.352	4.6	1955	1264.251633	2.02869635
1264.436	10	1264.533	2	3119	1264.432619	0.331544009
1264.517	8	1264.637	-	2559	1264.536611	2.451330691
1264.755	5	1264.853	3	2678	1264.752593	0.412711512
1265.667	5	1265.770      1	3779	1265.669521	0.494351075
1265.733	3	1265.832	-	1948	1265.731516	0.494738634
1265.8763	1.9	1265.977	2	3068	1265.876504	0.074049164
1266.403	3	1266.510	-	3127	1266.409462	2.153987711
1267.305	5	1267.398	1	4177	1267.297391	1.492155113
1268.1818	1.8	1268.281	2	1979	1268.180321	0.549525652
1268.8826	1.8	1268.990      1.8	1935	1268.889265	2.618296945
1269.384	2.4	1269.488	2	4191	1269.387226	1.032477774
1270.205	1.9	1270.309	1.7	3220	1270.20816	1.239607374
1271.259	3	1271.369	1.9	4337	1271.268076	2.555928649
1271.448	3	1271.543	2	1772	1271.442062	1.646784578
1272.626	4	1272.736	2	3112	1272.634968	2.005245633
1272.656	10	1272.736	2	2567	1272.634968	2.062383645
1272.656	10	1272.771	2	2567	1272.669965	1.369376278
1272.731	6	1272.835	4	1440	1272.73396	0.410460434
1273.188	10	1273.315	1	2626	1273.213922	2.579312414
1273.209	8	1273.315	1	2719	1273.213922	0.610470305
1273.817	3	1273.922	1	4084	1273.820874	1.224934846
1274.2628	1.6	1274.367	2.3	2380	1274.265838	1.084401185
1276.4314	2	1276.535	1	3368	1276.433666	1.01345714
1276.6323	2.4	1276.725	3	4042	1276.623651	2.251224869
1278.872	3	1278.979	2	2862	1278.877472	1.517701219
1279.7132	2	1279.820	1	2212	1279.718405	2.327920884
1280.108	5	1280.214	1.5	2808	1280.112374	0.837928053
1280.120	4	1280.214	1.5	3490	1280.112374	1.78508516
1280.298	5	1280.415	2	3671	1280.313358	2.851938173
1281.285	5	1281.388	3	3586	1281.286281	0.2196757
1282.212	7	1282.307	2	2831	1282.205208	0.932957562
1282.281	7	1282.385	1	2949	1282.283202	0.311377955
1282.732	3	1282.842	1.9	3571	1282.740165	2.299455899
1283.1857	2.3	1283.285	2	2667	1283.18313	0.843081091
1284.475	5	1284.581	1	2751	1284.479027	0.789848295
1284.566	5	1284.670	1.5	3148	1284.56802	0.387035931
1285.806	10	1285.894	1.8	3493	1285.791923	1.385412882
1287.5713	2	1287.675	1	5009	1287.572782	0.66270083
1287.628	7	1287.731	3	511	1287.628777	0.102077452
1287.635	8	1287.731	3	3412	1287.628777	0.728300393
1287.807	3	1287.917	1	2642	1287.814763	2.454760305
1289.852	8	1289.955	2	2368	1289.852601	0.07286416
1290.3967	1.8	1290.502	2	2290	1290.399557	1.061955989
1292.002	7	1292.109	3.4	2054	1292.00643	0.569242755
1292.071	7	1292.179	2	3655	1292.076424	0.745085977
1296.206	2.4	1296.304	2.7	3593	1296.201097	1.357279479
1298.7322	1.9	1298.835	1	3419	1298.731896	0.141614155
1300.226	4	1300.340      3	2747	1300.236776	2.155294335
1300.9807	2	1301.089	0.5	3504	1300.985717	2.43360954
1302.252	3	1302.349	2	2989	1302.245617	1.770327526
1303.32	    4	1303.427	2	1227	1303.323531	0.789649425
1303.976	5	1304.082	2	3194	1303.978479	0.460417554
1304.813	3	1304.912	1	1472	1304.808414	1.450366826
1304.884	3	1304.986	1	1590	1304.882408	0.503541133
1305.704	3	1305.808	1	1802	1305.704342	0.108279915
1305.941	4	1306.038	1	2987	1305.934324	1.61913065
1306.620	15	1306.730	-	2445	1306.626269	0.417948066
1307.5993	1.9	1307.706	1	1384	1307.602192	1.346819551
1307.779	9	1307.881	2	2244	1307.777178	0.19763966
1308.660	4	1308.759	1	2554	1308.655108	1.186446681
1309.137	5	1309.235	4	3167	1309.13107	0.926052748
1309.660	3	1309.756	1	2157	1309.652029	2.520647979
1310.0784	1.9	1310.182	1.1	3055	1310.077995	0.184383854
1310.2561	1.9	1310.359	2	2723	1310.254981	0.405585023
1311.111	1.9	1311.219	1	2135	1311.114913	1.822407765
1311.5569	1.9	1311.663	2	3747	1311.558878	0.716890251
1312.656	12	1312.758	1	1294	1312.653791	0.183471832
1312.728	12	1312.839	1	1412	1312.734784	0.56340351
1312.740	4	1312.839	1	1997	1312.734784	1.264998723
1313.3056	2.1	1313.412	1	2056	1313.307739	0.919537982
1314.304	14	1314.439	0.7	2096	1314.334657	2.187072546
1314.690	10	1314.790	1.2	4004	1314.685629	0.433946507
1314.916	10	1315.024	3	3908	1314.919611	0.345854509
1314.995	7	1315.101	1.8	2429	1314.996605	0.222022117
1315.6679	1.9	1315.774	1.5	3216	1315.669551	0.682141867
1316.897	7	1317.007	1.7	2278	1316.902453	0.757053423
1316.912	5	1317.007	1.7	3000	1316.902453	1.807689738
1317.4449	1.9	1317.552	0.7	3366	1317.44741	1.239674479
1318.151	1.9	1318.257	2	2903	1318.152354	0.490891762
1318.354	4	1318.456	0.8	3268	1318.351338	0.652481269
1318.5112	1.9	1318.618	1	2643	1318.513326	0.989957244
1319.752	2.4	1319.849	2	2949	1319.744228	2.487817663
1320.7152	1.9	1320.823	0.7	2378	1320.71815	1.457143711
1320.901	3	1321.005	3.1	2650	1320.900136	0.200271259
1321.251	4	1321.363	2	2976	1321.258108	1.58931324
1323.547	3	1323.655	1.1	2816	1323.549926	0.915617673
1323.557	4	1323.655	1.1	1322	1323.549926	1.705274024
1323.975	9	1324.094	1	2966	1323.988891	1.533985861
1324.464	5	1324.577	3	3533	1324.471852	1.346691146
1324.487	9	1324.577	3	3663	1324.471852	1.596687614
1324.733	4	1324.838	3	3491	1324.732832	0.03364548
1325.094	5	1325.201	3	2439	1325.095803	0.309204566
1332.802	3	1332.908	3.5	2830	1332.802191	0.041468478
1332.928	4	1333.040	2	3830	1332.934181	1.38204247
1333.9644	1.8	1334.069	1	3464	1333.963099	0.631822525
1334.180	5	1334.288	2.4	2689	1334.182082	0.37532445
1334.205	11	1334.288	2.4	2308	1334.182082	2.035602289
1336.1601	1.8	1336.265	1	2944	1336.158925	0.570788266
1336.548	5	1336.664	3.9	4603	1336.557893	1.560130919
1337.007	3	1337.114	2	1978	1337.007857	0.237766622
1337.074	3	1337.183	3	3192	1337.076852	0.67217627
1339.065	5	1339.161	2	3308	1339.054695	1.913630405
1340.3529	2	1340.456	3	3092	1340.349592	0.917478212
1341.0684	1.8	1341.179	1.8	3093	1341.072535	1.624221177
1342.1803	1.8	1342.292	1.2	3283	1342.185446	2.37885028
1343.8962	1.8	1343.999	2	3758	1343.892311	1.445433952
1347.740	4	1347.844	2	2988	1347.737006	0.669588008
1351.4142	2	1351.519	2	2187	1351.411714	0.879010549
1355.553	4	1355.666	3	3150	1355.558385	1.076917152
1366.671	7	1366.799	3	612	1366.690501	2.560583963
1390.917	6	1391.009	3.8	3877	1390.898579	2.593736856
1392.494	4	1392.611	2.5	1774	1392.500452	1.367783003
\end{lstlisting}